\documentclass[prc,twocolumn,nofootinbib,floatfix]{revtex4}

\usepackage{graphicx,color}
\usepackage{amsmath,amssymb,bm}
\usepackage{epstopdf}
\usepackage{epsfig}
\usepackage{graphicx}

\def\Xint#1{\mathchoice
{\XXint\displaystyle\textstyle{#1}}%
{\XXint\textstyle\scriptstyle{#1}}%
{\XXint\scriptstyle\scriptscriptstyle{#1}}%
{\XXint\scriptscriptstyle\scriptscriptstyle{#1}}%
\!\int}
\def\XXint#1#2#3{{\setbox0=\hbox{$#1{#2#3}{\int}$ }
\vcenter{\hbox{$#2#3$ }}\kern-.5\wd0}}

\def\dashint{\Xint{\bf-}}

\begin{document}

\title{Symmetries of the Similarity Renormalization Group for Nuclear Forces}

\author{V. S. Tim\'oteo} \email{varese@ft.unicamp.br}

\affiliation{Faculdade de Tecnologia, Universidade Estadual de Campinas - UNICAMP,
13484-332, Limeira - SP - Brazil}

\author{S. Szpigel} \email{szpigel@mackenzie.br}

\affiliation{Faculdade de Computa\c{c}\~ao e Inform\'atica, Universidade Presbiteriana Mackenzie,
01302-907, S\~ao Paulo - SP - Brazil}

\author{E. Ruiz
  Arriola}\email{earriola@ugr.es}
  \affiliation{Departamento de F\'isica At\'omica, Molecular y Nuclear
  and Instituto Carlos I de Fisica Te\'orica y Computacional, Universidad de Granada, E-18071 Granada, Spain}
\date{\today}

\begin{abstract}
We analyze the role played by Long Distance Symmetries within the context of the
Similarity Renormalization Group (SRG) approach, which is based on phase-shift
preserving continuous unitary transformations that evolve hamiltonians with a
cutoff on energy differences. We find that there is a SRG cutoff for
which almost perfect fulfillment of Wigner symmetry is found. We
discuss the possible consequences of such finding.
\end{abstract}

\maketitle

\section{Introduction}

The use of effective interactions in Nuclear Physics is rather old
(for a review see e.g.~\cite{Bogner:2003wn,Coraggio:2008in} and
references therein).  The basic idea is to emphasize the role of the
physically relevant degrees of freedom, which in the case of nucleons
in finite nuclei depends very much on the relevant energy scale or
equivalently on the shortest de Broglie wavelength sampling the
interactions. Moreover, the explicit effects of the (hard) core which
traditionally induce strong short distance correlations actually
depend on detailed and accurate knowledge of the interaction at rather
short distances (see e.g. Ref.~\cite{Pieper:2001mp} and references
therein).

While the issue of scale dependence is best formulated within a
renormalization framework~\cite{Wilson:1973jj}, Wilsonian methods have
only been seriously considered as an insightful technique in the study
of nuclear forces about a decade ago (for a balanced review see
\cite{Birse:2010fj} and references therein).

Some years ago Glazek and Wilson~\cite{Glazek:1993rc,Glazek:1994qc}
and independently Wegner~\cite{wegner1994flow} showed how
high-momentum degrees of freedom can decouple while keeping
scattering equivalence via the so-called similarity renormalization
group (SRG). The SRG is a renormalization method based on a series of
continuous unitary transformations that evolve hamiltonians with a
cutoff on energy differences. Such transformations are the group
elements that give the method its name. Viewing the hamiltonian as a
matrix in a given basis, the similarity transformations suppress
off-diagonal matrix elements as the cutoff is lowered, forcing the
hamiltonian towards a band-diagonal form. An important feature of the
SRG is that all other operators are consistently evolved by the same
unitary transformation.  Quite generally the transformation may
involve any number of particles and allows to generate well-behaved multi-particle
renormalized interactions. In Refs. ~\cite{Szpigel:1999gf,Szpigel:2000gf}, the main features of the SRG formalism were illustrated by Perry and one
of the present authors (S.S.) through simple examples from quantum mechanics, namely the Schr{\"o}dinger equations for non-relativistic two-body systems in one and two dimensions with Dirac-delta contact potentials.

Recently, the SRG has been applied to evolve several nucleon-nucleon ($NN$) potentials to
phase-shift equivalent softer forms~\cite{Bogner:2006pc} which become extremely
handy for many-body calculations in Nuclear Physics (for a review see
e.g. \cite{Bogner:2009bt}). The role played by bound-states onto the
SRG evolution has also been analyzed~\cite{Glazek:2008pg}, and it was
found that the SRG generator and evolution SRG scale may be suitably
chosen to avoid that undesirable singularities sneak in ruining the
low-energy properties of the SRG-evolved interaction. This has been
recently applied to prevent the spurious bound-states which normally
appear in meson exchange potentials~\cite{Wendt:2011qj}. The
complicated neutron-nucleus scattering becomes far simpler within a
SRG perspective~\cite{Navratil:2010jn}. The general discussion of
operator evolution and in particular deuteron properties has also been
carried out where a link to the Operator Product Expansion has been
envisaged~\cite{Anderson:2010aq}.

A great advantage of the SRG method over other approaches such as the
$V_{\rm low k}$~\cite{Bogner:2003wn} or the Unitary Correlation
Operator Method (UCOM) method~\cite{Roth:2010bm}, is the
straightforward application to the scale dependence of few-body
forces~\cite{Jurgenson:2009qs,Jurgenson:2010wy} which consistently
treat two-body induced as well as initially introduced few-body interactions (see
Ref.~\cite{Jurgenson:2008jp} for a discussion on simple
one-dimensional models). A further main advantage of SRG is the
tremendous reduction of the many-body problem, since effectively the
two-body interaction becomes almost diagonal and consequently the
corresponding phase space gets enormously reduced. This of course is
done at the cost of allowing three or even four-body forces which
precisely due to the high-energy decoupling of the SRG remain shorter
range than the two-body effective interaction but with a different and
scale dependent strength. The connection to the UCOM configuration
space method was discussed in Ref.~\cite{Hergert:2007wp,Roth:2008km}
where it was shown how a suitable rescaling of the radial coordinate
is in fact equivalent to choose a static generator. The relation of
the SRG to the $V_{\rm low k}$ method was discussed in
Ref.~\cite{Anderson:2008mu} as a block diagonal cutoff representation
with a Hilbert space with high- and low-momentum components where the
mixing between off-diagonal elements becomes negligible along the SRG
evolution.

The momentum-space $V_{\rm low k}$ approach~\cite{Holt:2003rj} (see
e.g. Ref.~\cite{Bogner:2009bt} for a review) takes a Wilsonian point
of view of integrating out high-energy components. This allows to
obtain a self-adjoint $V_{\rm low k}$ potential from a bare potential,
V. Given a symmetry group with a generic generator $X$, a standard
symmetry means that $[ V,X] =0 $ implies $[ V_{\rm low k},X] =0 $. The
reverse, however, is not true.  We define a {\it long distance symmetry} as a symmetry of the effective interaction, i.e. $[ V_{\rm
    low k},X] =0 $ but $[ V ,X] \neq 0 $. From a renormalization
viewpoint that corresponds to a symmetry of the potential broken only
by short distance counterterms. This is discussed at length in
Refs.~\cite{CalleCordon:2008cz,CalleCordon:2009ps,RuizArriola:2009bg}
and summarized in Ref.~\cite{Arriola:2010qk}. On the other hand, when
interaction is itself evolved to low-energies the symmetry of the long-range part of the potential reappears explicitly for the $V_{\rm low
  k}$ potential.

In a recent work it has been made clear that the momentum-space
$V_{\rm low k}$ approach~\cite{Holt:2003rj} displays a remarkable
symmetry pattern which in the case of Wigner SU(4) spin-isospin
symmetry with nucleons in the fundamental representation may be
related unambiguously to large $N_c$ dynamical features of
QCD~\cite{Kaplan:1996rk} as analyzed in detail in
Refs.~\cite{CalleCordon:2008cz,CalleCordon:2009ps,RuizArriola:2009bg,Arriola:2010qk}. This
possibility of linking a very distinct pattern of QCD to an observable
feature of the $NN$ interaction is extremely intriguing. However, the
symmetries are very well satisfied and a constructive point of view is
to select what definitions of the effective interaction comply to the
symmetries. Of course, we do not expect them to be perfect, but given
the fact that in the $V_{\rm low k}$ (the diagonal elements) approach
they work so well, it is worth testing other definitions. We want to
analyze whether the currently used SRG complies to the symmetry
pattern. In a previous work by one of the present authors
(E.R.A.)~\cite{Arriola:2010hj} the running scale dependence of the
effective interaction was carried out at low momenta with the sole
input of low-energy scattering parameters such as scattering lengths
(volumes) and effective ranges. It was found that the mixing induced by
the tensor force was essential to achieve the Wigner symmetry
condition. It was also found that Wigner symmetry sets in at a scale
predicted {\it just} from the lowest threshold parameters.
Also, it is clear from Ref.~\cite{CalleCordon:2008cz} that the difference in the
$^1S_0$ and $^3S_1$ phase-shifts comes from difference of the singlet and triplet scattering lengths
while the effective interactions are similar in both channels. The realization of the Wigner symmetry
does not arise from the size of the scattering length but rather from the long distance properties
of the effective interactions. In Ref.~\cite{MSW:2009ca}, the origin of the Wigner symmetry is unveiled
in the framework of pionless effective field theory at leading order, but no mention is made
concerning what happens when pion exchange interactions and the tensor coupling are included.

The purpose of the present paper is to outline under what conditions
can these long distance symmetries be displayed. We will show that this
is indeed the case and moreover for the SRG cutoff of about $600 {\rm
  MeV}$ an extremely accurate fulfillment of Wigner SU(4) symmetry is
found. Taking into account that in the SRG we are at any rate preserving
phase equivalence this result actually suggests a representation of
the interaction based explicitly on the symmetry. This of course will
have some implications for Nuclear Structure and Nuclear Reactions
which deserve further study.

The paper is organized as follows. In Section~\ref{eq:srg-sym} we
review the SRG approach and provide the definition of a long distance
symmetry within such a framework. One good way to unveil a symmetry is
to construct a set of sum rules where the symmetry is linearly
broken. This is done in Section~\ref{sec:pw}. In Section~\ref{sec:num}
we show our numerical methods and the corresponding results. Finally in
Section~\ref{sec:concl} we present our main conclusions and outlook
for further work. In Appendix~\ref{eq:su4} we review the meaning of
both the Wigner and Serber symmetries within the $NN$ context. In
Appendix~\ref{sec:appa} we analyze a fixed-point solution of the SRG
equations and its stability properties.

\section{SRG and symmetries}
\label{eq:srg-sym}

The formulation of the Similarity Renormalization Group (SRG) is well
known~\cite{Wegner200177,Perry200133,Szpigel:2000gf}. However, the
equations are rather complicated (integro-differential non-linear
coupled equations) and hence very little general properties have so
far been deduced, so much insight is provided by numerical
calculations or the study of simplified models. A textbook
presentation is available~\cite{Kehrein:2006ti} and a rigorous
discussion has been carried out only recently~\cite{bach2010rigorous}
although the interesting case of unbounded operators which is the
standard situation in Nuclear Physics has been left out. Therefore
much understanding of these equations relies on numerical approaches
and the use of a discretized momentum-space basis on a finite Hilbert
space (see however \cite{Szpigel:1999gf,Szpigel:2010bj} for some
analytical models).  The continuum limit will be analyzed in some
detail in Appendix~\ref{sec:appa}.

The SRG makes a transformation which actually drives the system into a
diagonal basis, suppressing exponentially off-diagonal elements, as we
review here.  Let us consider the evolution equation at the operator
level in the SRG approach induced by the unitary transformation $H_s =
e^{\eta_s} H e^{-\eta_s}$
\begin{eqnarray}
\frac{d H_s}{ds} = \left[ \eta_s , H_s \right] \, ,
\end{eqnarray}
with the anti-hermitian generator $\eta_s = [ G_s, H_s ]$, where $G_s$
is the SRG hermitian generator which will be specified shortly. If
$H_{s=0}= H$ is the initial Hamiltonian one can easily see that when $
[G_s , H_s ] =0$ then the SRG equation has a stationary
point. Conversely, stationary points of the SRG equation fulfill $ [
  [G_s , H_s ],H_s] =0$. This means that the invariant subspaces of
$H_s$ eigenvectors are also invariant subspaces of $G_s$, whence a
band-diagonal structure follows. Due to the commutator character of
the equations one has an infinite set of conservation laws. Indeed,
using the cyclic property of the trace we get
\begin{eqnarray}
&&\frac{d }{ds}  {\rm Tr} \left(H_s^n \right) = n  {\rm Tr} \left( H_s^{n-1} \frac{d H_s}{ds} \right) \nonumber \\
&&=
n  {\rm Tr} \left( H_s^{n-1} [\eta_s,H_s]]  \right) =0 \, .
\end{eqnarray}
In this paper we will use as initially suggested by Glazek and
Wilson~\cite{Glazek:1993rc,Glazek:1994qc} the kinetic energy as the
SRG generator, $G_s=T$. Using $H_s=T+V_s$ we get for the potential
energy the evolution equation
\begin{eqnarray}
\frac{d V_s}{ds} = \left[ [T,V_s] , V_s \right] \, .
\end{eqnarray}
For this choice one has that for any $n$ and $s$
\begin{eqnarray}
\frac{d}{ds} {\rm Tr} \left(V_s^n \right) = 0 \, .
\end{eqnarray}
In appendix~\ref{sec:appa} we will also show that the diagonal
matrix-elements of the standard scattering reaction R-matrix is a
stationary point of the previous equation when the potential is diagonal.

More generally, for an arbitrary operator $O$ we get the equation
\begin{eqnarray}
\frac{d O_s}{ds} = \left[ [T,V_s] , O_s \right] \, .
\end{eqnarray}
The purpose of the present paper is to outline under what conditions
can the so-called long distance symmetries be analyzed. In particular,
for a symmetry group generator $X$ we have
\begin{eqnarray}
\frac{d X_s}{ds} = \left[ [T,V_s] , X_s \right] \, .
\end{eqnarray}
A symmetry at a given scale $s$ must commute with both kinetic and
potential energy operators and thus fulfills, $[ X_s, T]=0$ and $[
  X_s, V_s]=0$. Using Jacobi's identity $[[A,B],C] + [[C,A],B] +
[[B,C],A] =0 $ we get
\begin{eqnarray}
\frac{d X_s}{ds} = -\left[ [X_s,T] , V_s \right]- \left[ [V_s,X_s] , T
  \right] =0 \, .
\end{eqnarray}
A standard symmetry corresponds to a vanishing of the l.h.s. whereas a
long distance symmetry is a fixed-point of the evolution along the
similarity renormalization group at a given point. We will see that
such a symmetry pattern appears within the SRG in regard to Wigner and
Serber symmetries (see Appendix~\ref{eq:su4} for a short
description). The variable $s$ has dimensions of ${\rm energy}^{-2}$ and it is
customary to introduce the SRG cutoff $\lambda= s^{-1/4}$ which has
dimensions of momentum.

As we have already mentioned, the previous manipulations are fully
justified in a finite Hilbert space and typically corresponds to a
discretized momentum-space basis. The discretization effects are
enhanced as the running Hamiltonian is driven towards the diagonal
form. So one expects to evolve to a scale where the high-energy
components are decoupled from the dynamics but also where the
discretization effects are also unimportant. A feature of the SRG is
that much of the SRG scale evolution happens already at the beginning and
slows down as one approaches the scales of interest in nuclear
applications.

We expect the long distance symmetry (if at all) to happen at a given
SRG scale, but this may depend on the SRG generator. For definiteness,
we use here the SRG equations with the kinetic energy as the
generator, which has $T |\vec p \rangle = E_p |\vec p \rangle $ with
$E_p = p^2/M$. In momentum-space the equations read
\begin{eqnarray}
&&\frac{d V_s (\vec p',\vec p) }{ds} = -(E_p-E_p')^2 V_s (\vec p',\vec
p) \nonumber \\ &&+ \int \frac{d^3 q}{(2\pi)^3} \left( E_p+E_{p'}-2
E_q \right) V_s (\vec p',\vec q)V_s (\vec q,\vec p) \, .
\end{eqnarray}
Note that if we take zero momentum states
\begin{eqnarray}
  &&\frac{d V_s (\vec 0,\vec 0) }{ds} = - 2 \int \frac{d^3
    q}{(2\pi)^3} E_q \langle \vec q | V_s V_s^\dagger | \vec q \rangle \le
  0 \, ,
\end{eqnarray}
which means that zero-momentum matrix elements of the SRG evolved
potentials always decrease. Note that this does not prevent that it
becomes infinite at some finite value of $s$. The conservation of
${\rm Tr} \left(V_s^n \right) $ means that the zero momentum strength
increases at the cost of depleting the high- and off-diagonal matrix
elements.

\section{Symmetries and Partial waves decomposition}
\label{sec:pw}

\subsection{Kinematical Symmetries}

The most general self-adjoint interaction in momentum-space which is
invariant under parity, time-reversal, isospin and Galilean invariance
has the following form~\cite{okubo1958velocity},
\begin{eqnarray}
V(\vec p',\vec p) &=& V_C + \vec \tau_1 \cdot \vec \tau_2\, W_C +
\left (V_S + \vec \tau_1 \cdot \vec \tau_2 \, W_S \right ) \vec \sigma_1 \cdot
\vec \sigma_2 \nonumber \\
&+& \left ( V_{LS} + \vec \tau_1 \cdot \vec \tau_2 \, W_{LS} \right )
i (\vec \sigma_1 + \vec \sigma_2 ) \cdot (\vec p' \times \vec p) \nonumber \\
&+&
\left ( V_T + \vec \tau_1 \cdot \vec \tau_2 \, W_T \right ) S_{12} (\vec p'-\vec p)
\nonumber \\
&+& \left ( V_{Q} + \vec \tau_1 \cdot \vec \tau_2 \, W_{Q} \right ) S_{12} (\vec p' \times \vec p) \nonumber \\
&+& \left ( V_{P} + \vec \tau_1 \cdot \vec \tau_2 \, W_{P} \right ) S_{12} (\vec p'+ \vec p) \, , \nonumber
\end{eqnarray}
where the tensor operator is defined as
\begin{eqnarray}
S_{12} (\vec q) = \left[\vec \sigma_1 \cdot \vec q \, \vec \sigma_2
  \cdot \vec q - \frac{1}3 q^2 \vec \sigma_1 \cdot \vec \sigma_2
  \right] \, ,
\end{eqnarray}
and $\vec \sigma_i$ and $\vec \tau_i$ are the Pauli matrix spin and
isospin operators for $i$-th particle respectively. The subscripts
refer to the central (C), spin-spin (S), tensor (T), spin-orbit (SL),
quadratic spin-orbit (Q) and quadratic velocity dependent (P)
components of the $NN$ interaction, each of which occurs in an isoscalar
(V) and an isovector (W) version.  Here $\vec \sigma_{1,2}$ and $\vec
\tau_{1,2}$ are the usual spin and isospin operators of the two
nucleons. Note that the so-defined condition of zero angular averaging
for the tensor operator $ \int d^2 \hat p S_{12} (\vec p) =0$ is
fulfilled and as a consequence our results will be stated in a rather
simple form. Note also that the central parts can also be deduced by
tracing the potential in the spin-isospin space after appropriate
multiplication of the operators ${\bf 1}$, ${\bf S}_i$, ${\bf T}_a$
and $ {\bf G}_{ia}$ (see Appendix~\ref{eq:su4})

An advantage of using the momentum-space basis is that the generalized Pauli principle can be
incorporated directly into the potential which is  antisymmetric under
the exchange of the final states which in the CM reads,
\begin{eqnarray}
V_{1',2'; 1,2} ( \vec p' , \vec p ) = - V_{2',1'; 1,2} ( -\vec p' , \vec p )
\label{eq:gen-pauli}
\end{eqnarray}
where the indices represent a full Pauli spinor-isospinor state.
These kinematical symmetries are preserved by the SRG equations.  This
means that if we have a starting potential $ V (\vec p',\vec p) $
which evolves into $ V_s (\vec p',\vec p) $ then $V ( R\vec p',R\vec
p)$ evolves into $V_s ( R\vec p',R\vec p)$. In particular, a potential
remains rotational invariant along the SRG evolution and the general
form is maintained throughout the evolution. This of course includes
the generalized Pauli principle~(\ref{eq:gen-pauli}).

\subsection{Long distance Symmetries}

The off-shell $T-$matrix has the same decomposition as for the
potential and hence contains 12 independent amplitudes but on-shell,
because of the condition $|\vec p'|=|\vec p|$, one just gets 10
different amplitudes since one has the identity $ \sigma_1 \cdot
\sigma_2 = \sigma_1 \cdot \hat n \sigma_2 \cdot \hat n + \sigma_1
\cdot \hat m \sigma_2 \cdot \hat m + \sigma_1 \cdot \hat l \sigma_2
\cdot \hat l $ where $\hat n$, $\hat m$, $\hat l$ are three
orthonormal vectors in three-dimensional space such as those
proportional to $\vec p'-\vec p , \vec p'+\vec p $ and $ \vec p' \times \vec p$.

Following the description in ref.~\cite{1971NuPhA.176..413E} (see also
\cite{Kaiser:1997mw,Epelbaum:1999dj}) one can undertake the projection
onto partial waves. The non-linear SRG equation can be simplified
using the conservation of angular momentum. If we just define for a
given spin the standard partial wave decomposition
\begin{eqnarray}
\langle  \vec p' | V_\lambda^S | \vec p \rangle =
N \sum_{JMLL'} {\cal Y}_{LS}^{JM} (\hat p') V_{LL'}^{JS} (p',p)
{{\cal Y}_{L'S}^{JM}}^\dagger (\hat p) \, ,
\end{eqnarray}
with $N= 4\pi^2/M$ we get an infinite set of coupled equations with
good total angular momentum. The three-dimensional structure contains
central and non-central forces. Inserting this into the SRG equation
we get the coupled-channel equations,
\begin{eqnarray}
&& \frac{d V_{s}(p,p')}{ds} = -(p^2-p'^2)^2 \; V_{s}(p,p') \nonumber \\
&&+\frac{2}{\pi} \int_{0}^{\infty}dq \; q^2\;
(p^2+p'^2-2 q^2)\; V_{s}(p,q)\; V_{s}(q,p'),
\label{flowNN}
\end{eqnarray}
\noindent
where $V_{s}(p,p')$ is used as a brief notation for the projected $NN$
potential matrix elements,
\begin{equation}
V^{(JLL'S;I)}_{s}(p,p')= \langle\;p{(LS)J;I}|V_{s}|\;p'{(L'S)J;I} \;\rangle \; ,
\end{equation}
\noindent
in a partial-wave relative momentum-space basis, $|\;q{(LS)J;I} \;\rangle$, with normalization such that
\begin{equation}
1=\frac{2}{\pi} \int_{0}^{\infty}dq \; q^2 \; |\;q{(LS)J;I} \;\rangle \;\langle \; q{(LS)J;I}\;|.
\label{PWnorm}
\end{equation}
\noindent
The superscripts $J$, $L(L')$, $S$ and $I$ denote respectively the
total angular momentum, the orbital angular momentum, the spin and the
isospin quantum numbers of the $NN$ state. For non-coupled channels
($L=L'=J$), the matrix elements $V_{s}(p,p')$ are simply given by
$V_{s}(p,p')\equiv V^{(JJJS;I)}_{s}(p,p')$. For coupled-channels
($L,L'=J \pm 1$), the $V_{s}(p,p')$ represent $2 \times 2$ matrices of
matrix-elements for the different combinations of $L$ and $L'$.  The
advantage of using the partial-wave decomposition is that every single
channel may be evolved independently from a given initial solution,
$V_{s=0} (p',p)$, to yield a unitarily equivalent potential
\begin{eqnarray}
V_{s} (p',p) = \int dq q U_s (q',p')^* V_{s=0} (q',q) U_s (q,p)
 \end{eqnarray}
A further property is that for low-momenta we have the threshold behaviour
\begin{eqnarray}
V_{s} (p',p) = C_{L,L'}(s) p^L (p')^{L'} \left[ 1 + {\cal O} (p^2,p'^2) \right]
 \end{eqnarray}
which means also that the matrix $C_{L,L'}(s)$ is a decreasing
function of the SRG cut-off, i.e. $C_{L,L'}'(s)$ is negative definite.

\subsection{Perturbation theory}

In the case of rotational invariance, the symmetry is preserved along
the SRG trajectory. In the case of a long distance symmetry the
symmetry breaking strength depends on the SRG scale.  Let us thus
consider the expansion around the central solution, i.e. let us split
\begin{eqnarray}
{\cal V}_{NN} = V_0 + V_1 \, ,
\label{eq:V0+V1}
\end{eqnarray}
where $[ \vec L, V_0] =0 $ whereas $[ \vec J, V_1] = 0 $ and $[ \vec
  L, V_1] \neq 0 $.  The zeroth-order potential commutes with $L,S,T$
and so the corresponding potential may be denoted as $V_L^{ST}$.  The
total potential commutes with the total angular momentum $J = L+S$ and
$S^2$. Of course the goodness of this separation depends on the SRG
scale $\lambda$, but we expect it to work better the lower the scale.
Inserting the decomposition~(\ref{eq:V0+V1}) into the SRG equation we
have the zeroth-order
\begin{widetext}
\begin{eqnarray}
\frac{d V_s^{(0)} (\vec p',\vec p) }{ds} = -(E_p-E_p')^2 V_s^{(0)}
(\vec p',\vec p)
+ \int \frac{d^3 q}{(2\pi)^3} \left( E_p+E_{p'}-2 E_q
\right) V_s^{(0)} (\vec p',\vec q)V_s^{(0)} (\vec q,\vec p) \, ,
\end{eqnarray}
whereas the first-order is
\begin{eqnarray}
\frac{d V_s^{(1)} (\vec p',\vec p) }{ds} &=& -(E_p-E_p')^2 V_s^{(1)}
(\vec p',\vec p) \nonumber \\
&+& \int \frac{d^3 q}{(2\pi)^3} \left( E_p+E_{p'}-2 E_q
\right) \left[ V_s^{(1)} (\vec p',\vec q) V_s^{(0)} (\vec q,\vec p)+
  V_s^{(0)} (\vec p',\vec q)V_s^{(1)} (\vec q,\vec p) \right]
\end{eqnarray}
\end{widetext}
Clearly, because $V_0$ is central we get that the linear combinations
are preserved as long as the non-central contribution remains a small
correction. Generally, we do not expect it to be the case, and in
particular for the initial potential.

\subsection{Sum rules}

Rather than analyzing the evolution of the generators we prefer to
check the running of a set of sum rules based on first-order
perturbation theory in non-central components of the potential, which
helped to disentangle some correlations in $NN$
fits~\cite{Holinde:1975vg,Nagels:1977ze} and have been shown to work
well at the level of phase-shifts and $V_{\rm low k}$ potentials in
Ref.~\cite{CalleCordon:2008cz,CalleCordon:2009ps}. The main idea is
to decompose the potential into central and a non-central pieces,
as in Eq.~(\ref{eq:V0+V1}) and assume the non-central piece
to be of the form
\begin{eqnarray}
V_1= L \cdot S V_{LS} + S_{12} V_T
 \, ,
\end{eqnarray}
One property fulfilled by the non-central interaction is that the
trace in the spin-isospin space vanishes. This suggests a set of sum
rules at the partial waves level. Using first-order perturbation
theory~\cite{CalleCordon:2008cz,CalleCordon:2009ps} we deduce the
following linear combinations for triplet P-waves,
\begin{eqnarray}
V_{^3P_C}    &=& \frac19   \big( V_{^3P_0} + 3 V_{^3P_1} + 5 V_{^3P_2}
\big)\, , \\
V_{^3P_T}    &=&-\frac5{72}\big(2V_{^3P_0} - 3 V_{^3P_1} +   V_{^3P_2}
\big)\, , \\
V_{^3P_{LS}} &=&-\frac1{12}\big(2V_{^3P_0} + 3 V_{^3P_1} - 5 V_{^3P_2}
\big)\, ,
\label{eq:linear-P}
\end{eqnarray}
for triplet D-waves
\begin{eqnarray}
V_{^3D_C}    &=& \frac1{15} \big(3 V_{^3D_1} + 5 V_{^3D_2} + 7
V_{^3D_3} \big)\, , \\
V_{^3D_T}    &=&-\frac7{120}\big(3 V_{^3D_1} - 5 V_{^3D_2} + 2
V_{^3D_3} \big)\, , \\
V_{^3D_{LS}} &=&-\frac1{60} \big(9 V_{^3D_1} + 5 V_{^3D_2} -
14V_{^3D_3} \big)\, ,
\label{eq:linear-D}
\end{eqnarray}
for triplet F-waves
\begin{eqnarray}
V_{^3F_C}    &=& \frac1{21} \big(5  V_{^3F_2} + 7 V_{^3F_3} + 9
V_{^3F_4} \big)\, , \\
V_{^3F_T}    &=&-\frac5{112}\big(4  V_{^3F_2} - 7 V_{^3F_3} + 3
V_{^3F_4} \big)\, , \\
V_{^3F_{LS}} &=&-\frac1{168}\big(20 V_{^3F_2} + 7 V_{^3F_3} - 27
V_{^3F_4} \big)\, ,
\label{eq:linear-F}
\end{eqnarray}
and for triplet G-waves
\begin{eqnarray}
V_{^3G_C}    &=& \frac1{27}      \big(7 V_{^3G_3} + 9 V_{^3G_4} + 11
V_{^3G_5} \big)\, , \\
V_{^3G_T}    &=&-\frac{77}{2160} \big(5 V_{^3G_3} - 9 V_{^3G_4} + 4
V_{^3G_5} \big)\, , \\
V_{^3G_{LS}} &=& \frac1{360}     \big(-35 V_{^3G_3} - 9 V_{^3G_4} + 44
V_{^3G_5} \big)\, .
\label{eq:linear-G}
\end{eqnarray}
In terms of the previous definitions Serber symmetry reads
\begin{eqnarray}
0= V_{^3P_C} = V_{^3F_C} = V_{^3H_C}  = \dots
\end{eqnarray}
whereas Wigner symmetry implies
\begin{eqnarray}
V_{^1S_0} = V_{^3S_C} \, , \quad
V_{^1D_2} = V_{^3D_C} \, ,  \quad
V_{^1G_4} = V_{^3G_C} \, .
\end{eqnarray}

\begin{figure*}[tbc]
\begin{center}
\includegraphics[height=4cm,width=5cm]{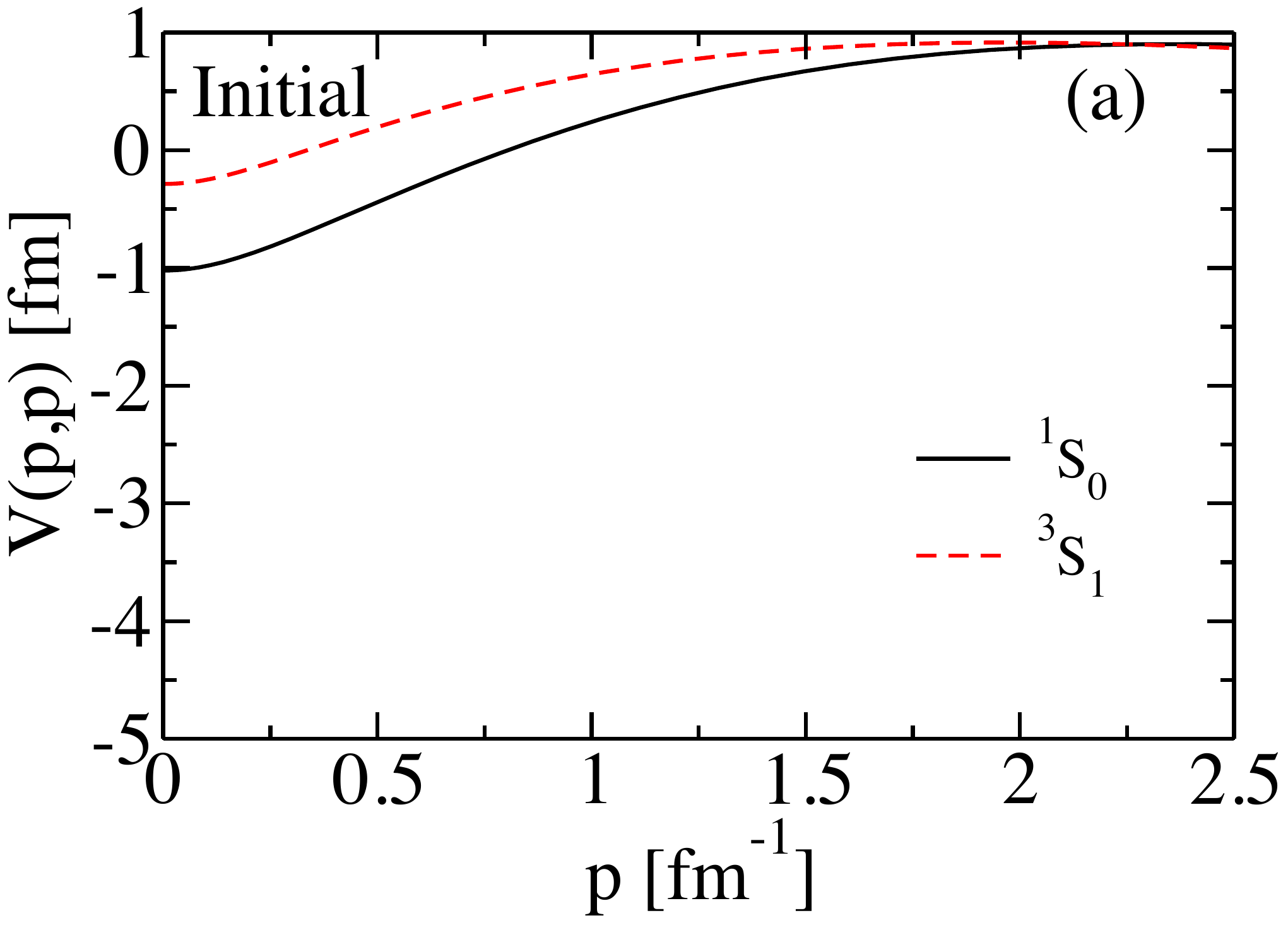}
\includegraphics[height=4cm,width=5cm]{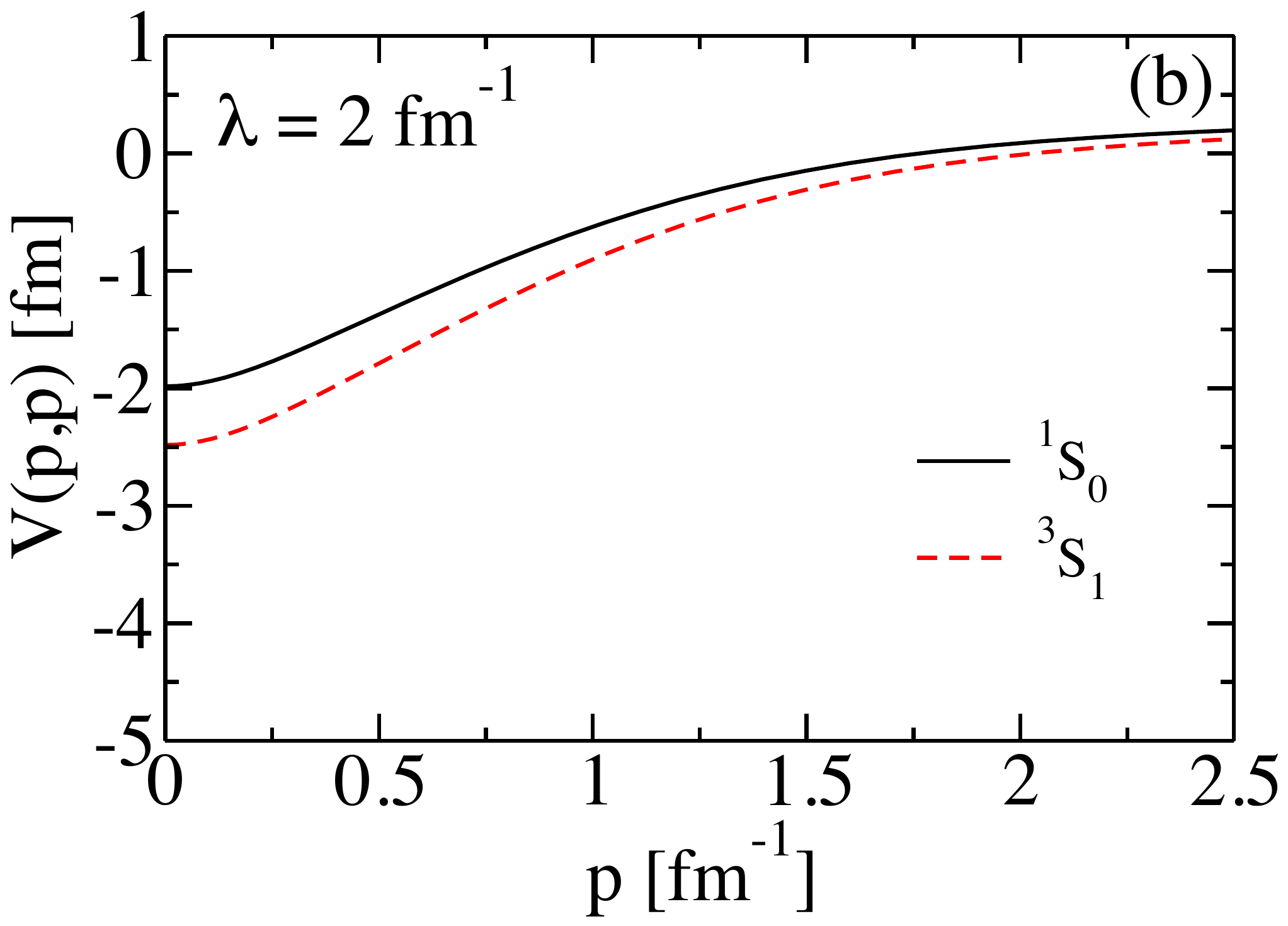}
\includegraphics[height=4cm,width=5cm]{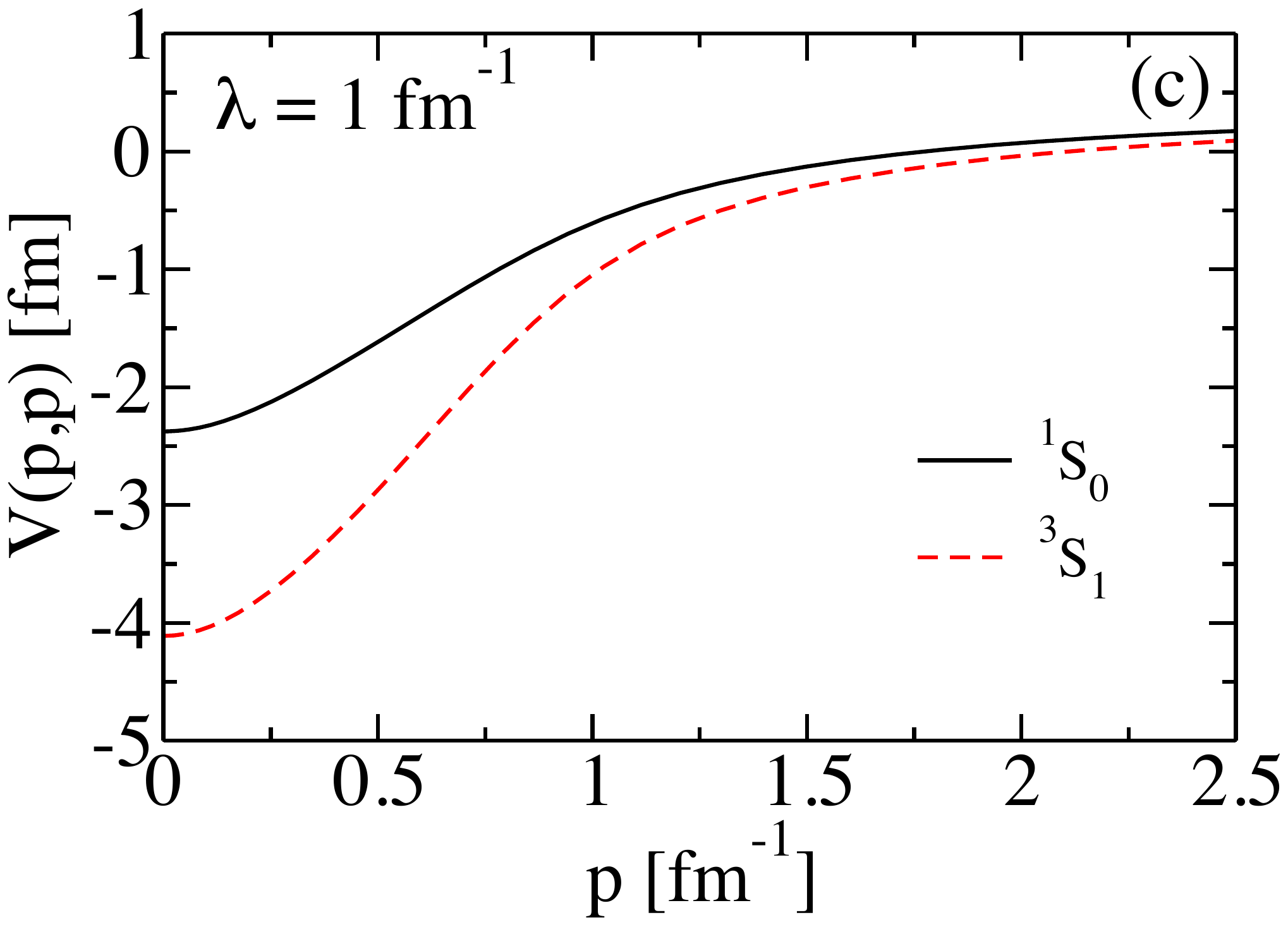} \\
\includegraphics[height=4cm,width=5cm]{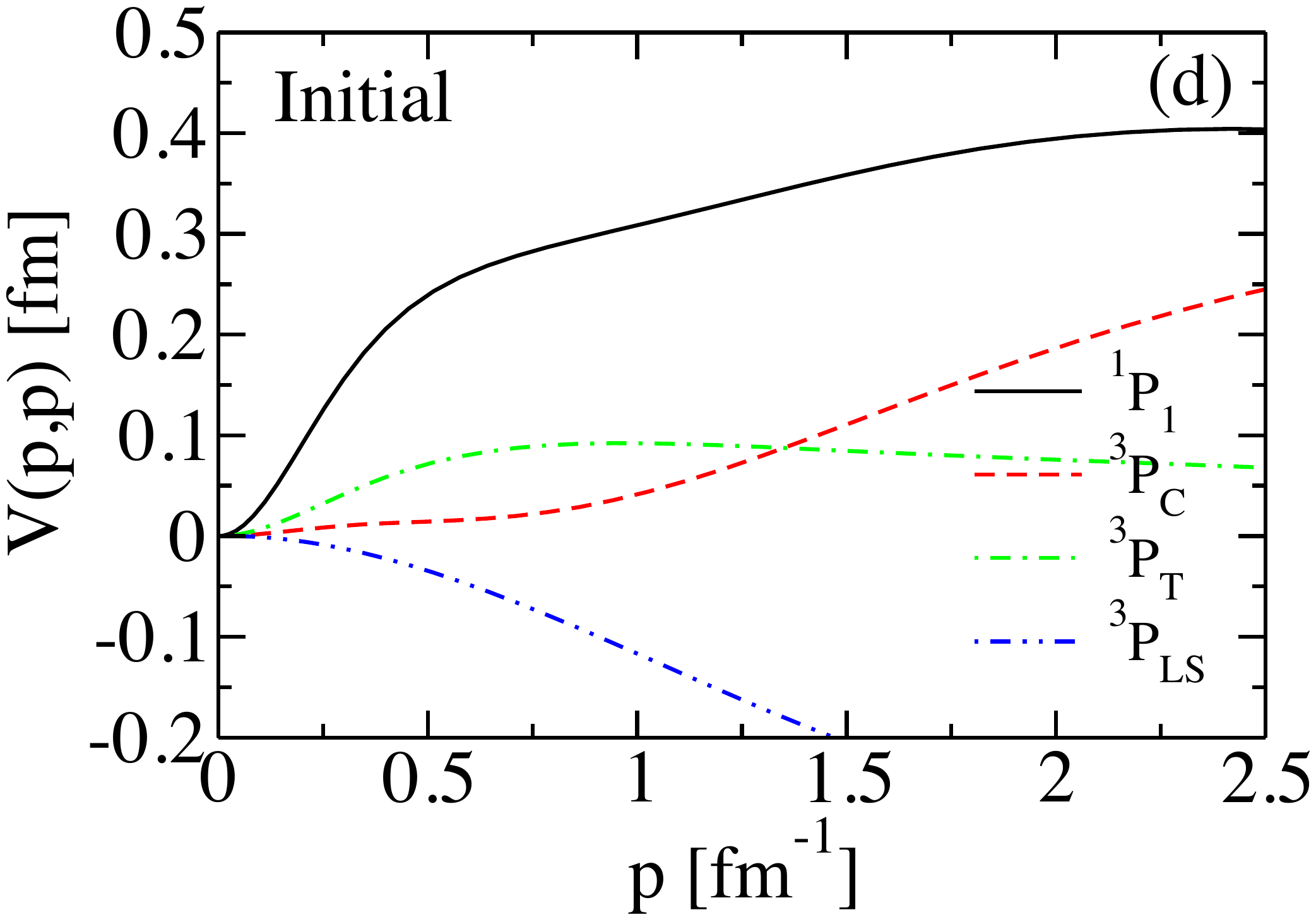}
\includegraphics[height=4cm,width=5cm]{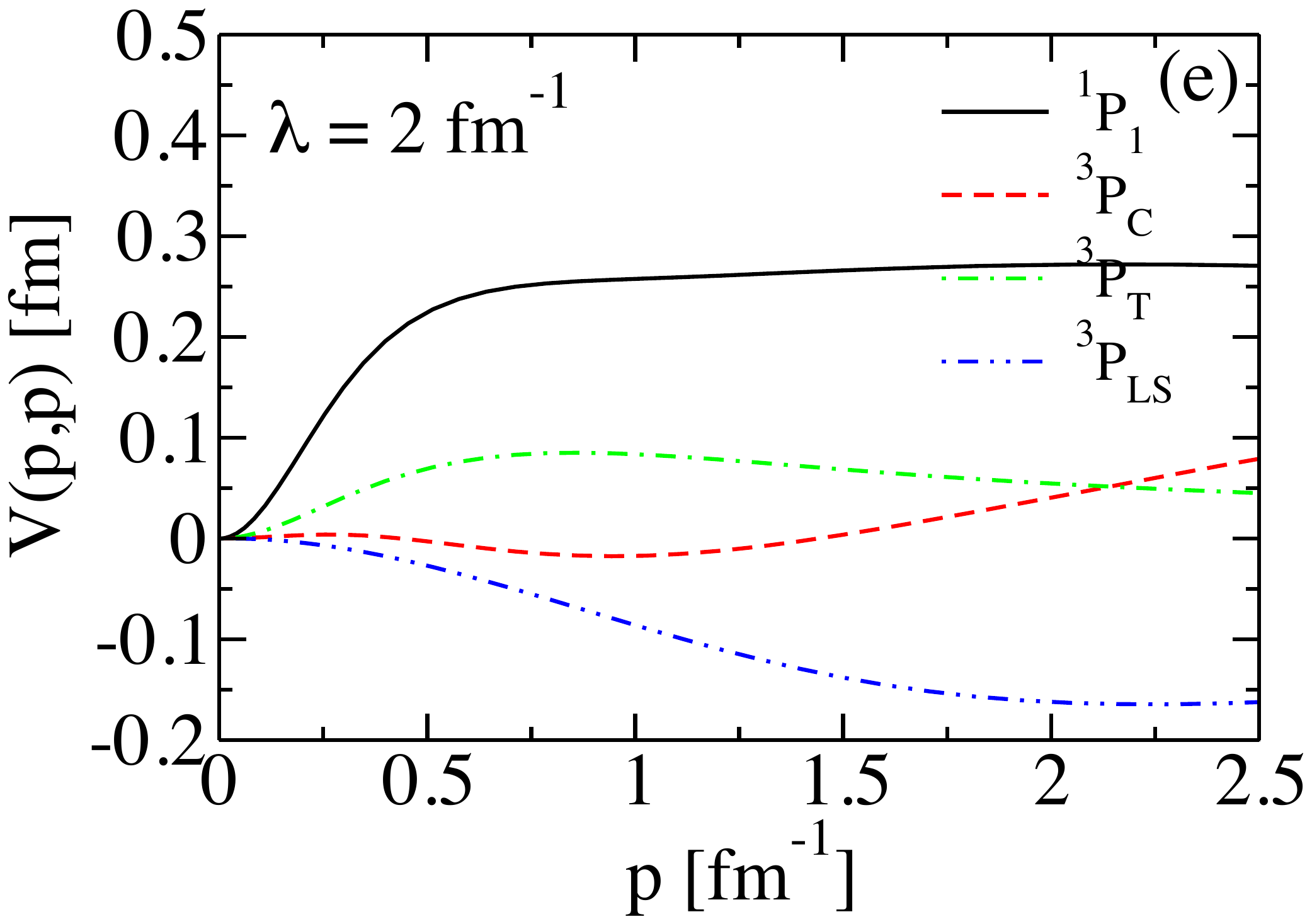}
\includegraphics[height=4cm,width=5cm]{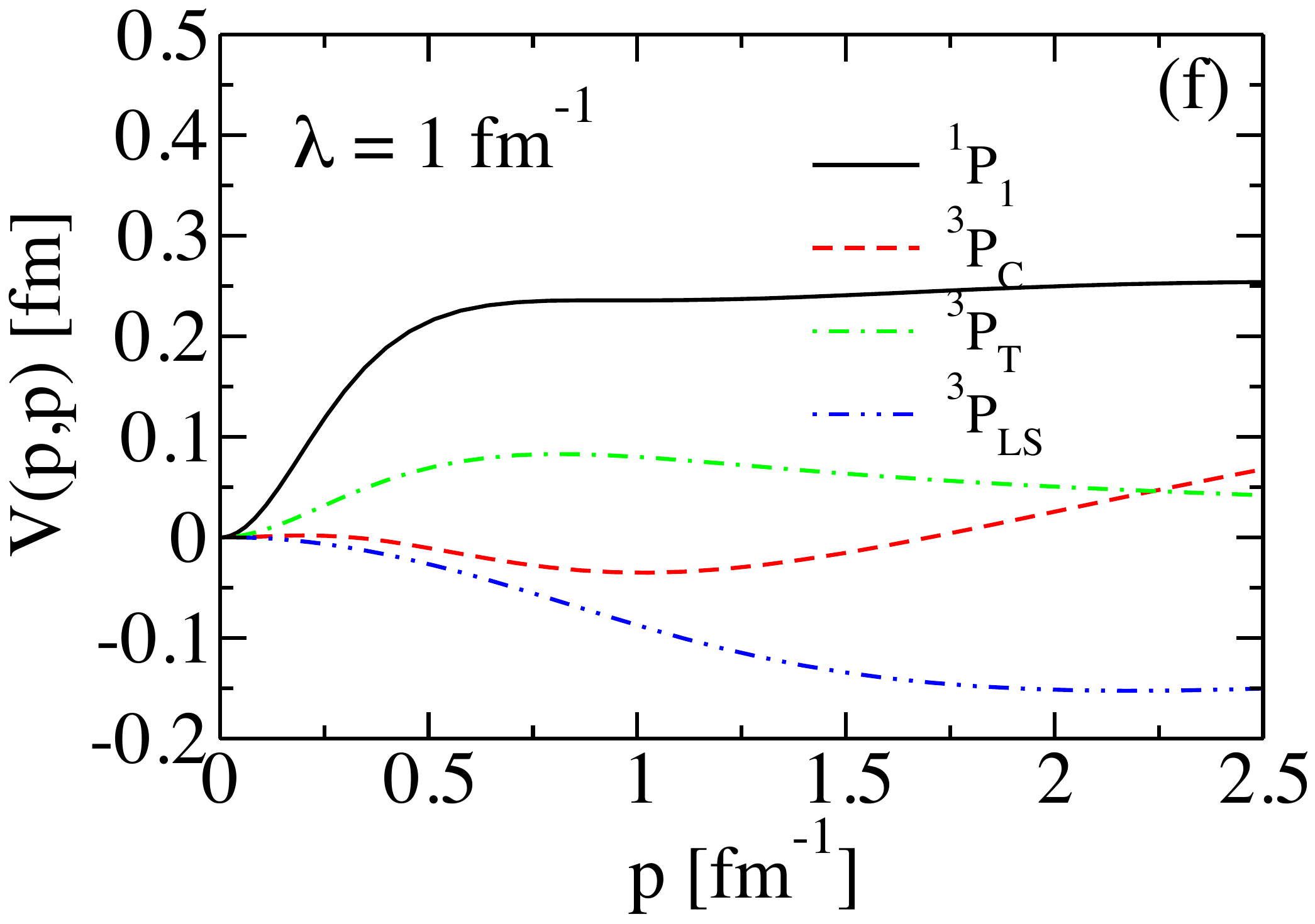} \\
\includegraphics[height=4cm,width=5cm]{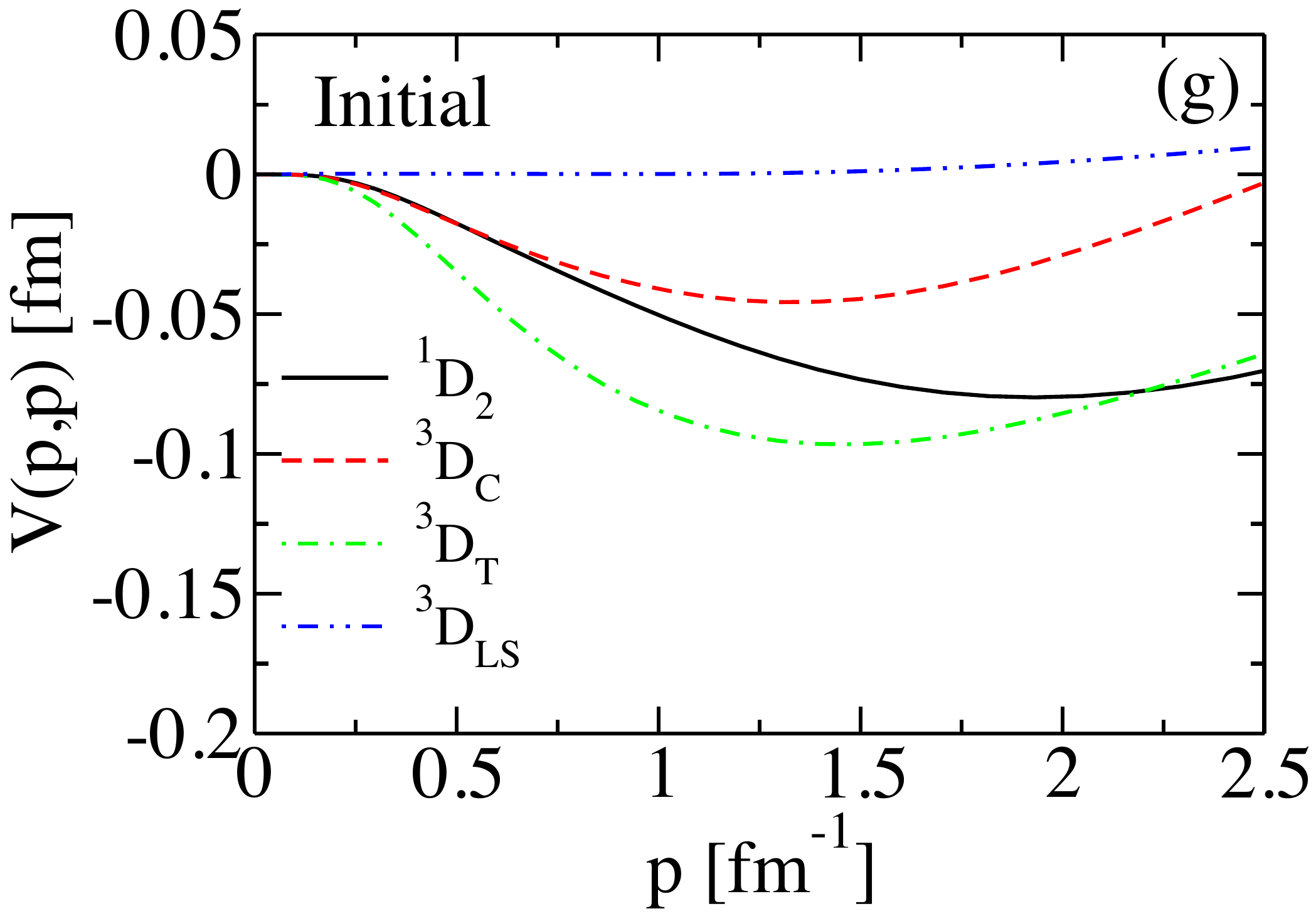}
\includegraphics[height=4cm,width=5cm]{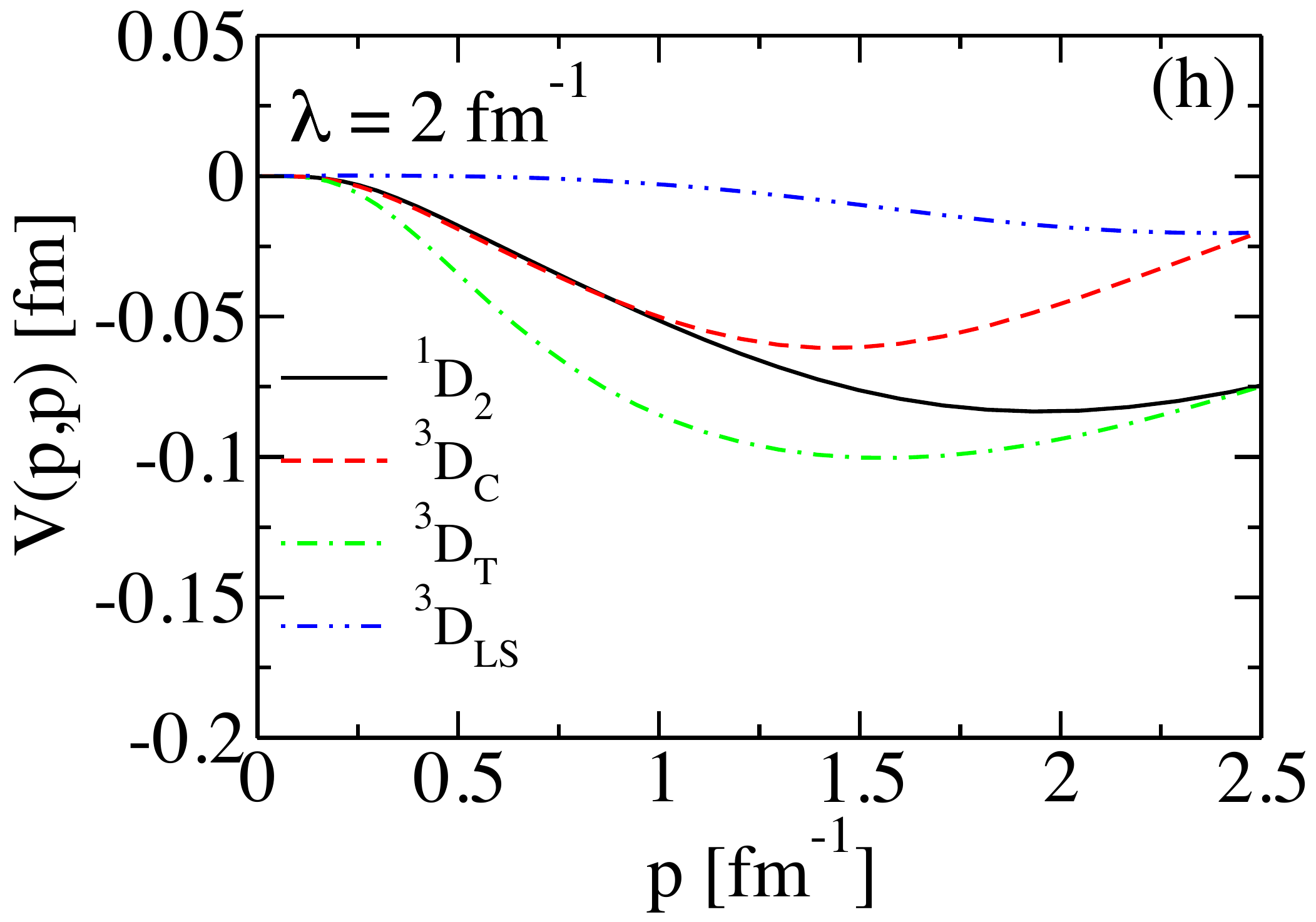}
\includegraphics[height=4cm,width=5cm]{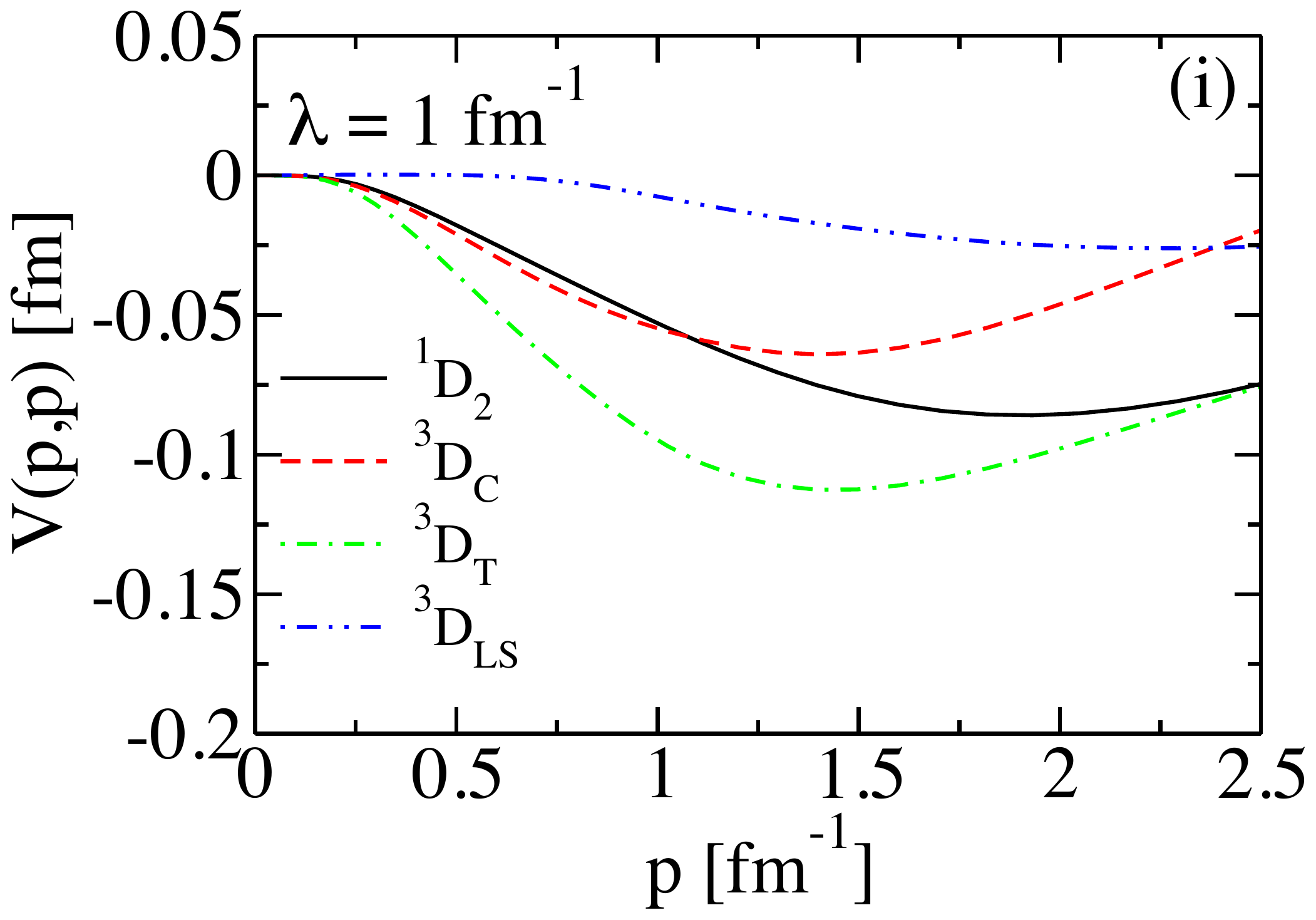} \\
\includegraphics[height=4cm,width=5cm]{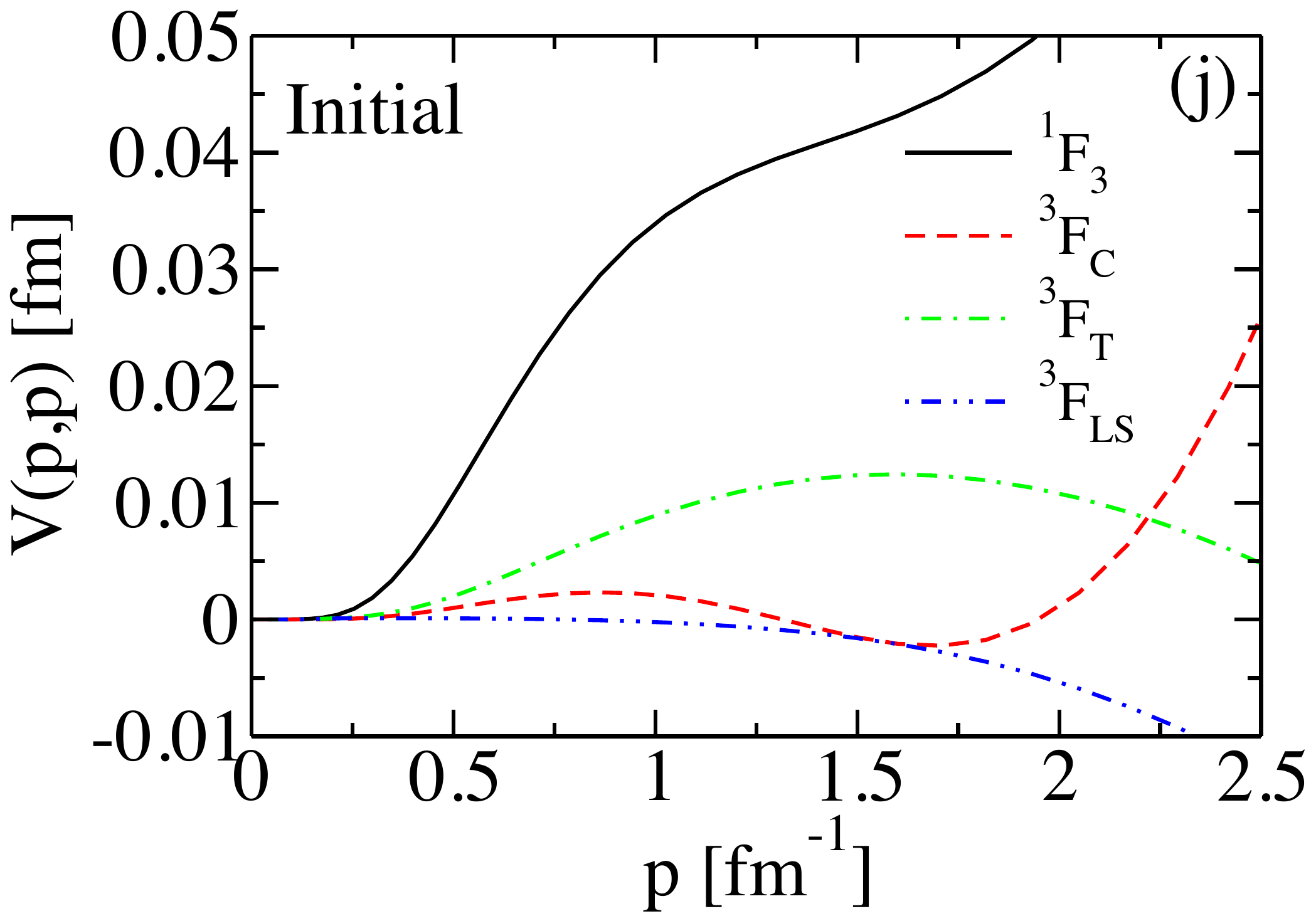}
\includegraphics[height=4cm,width=5cm]{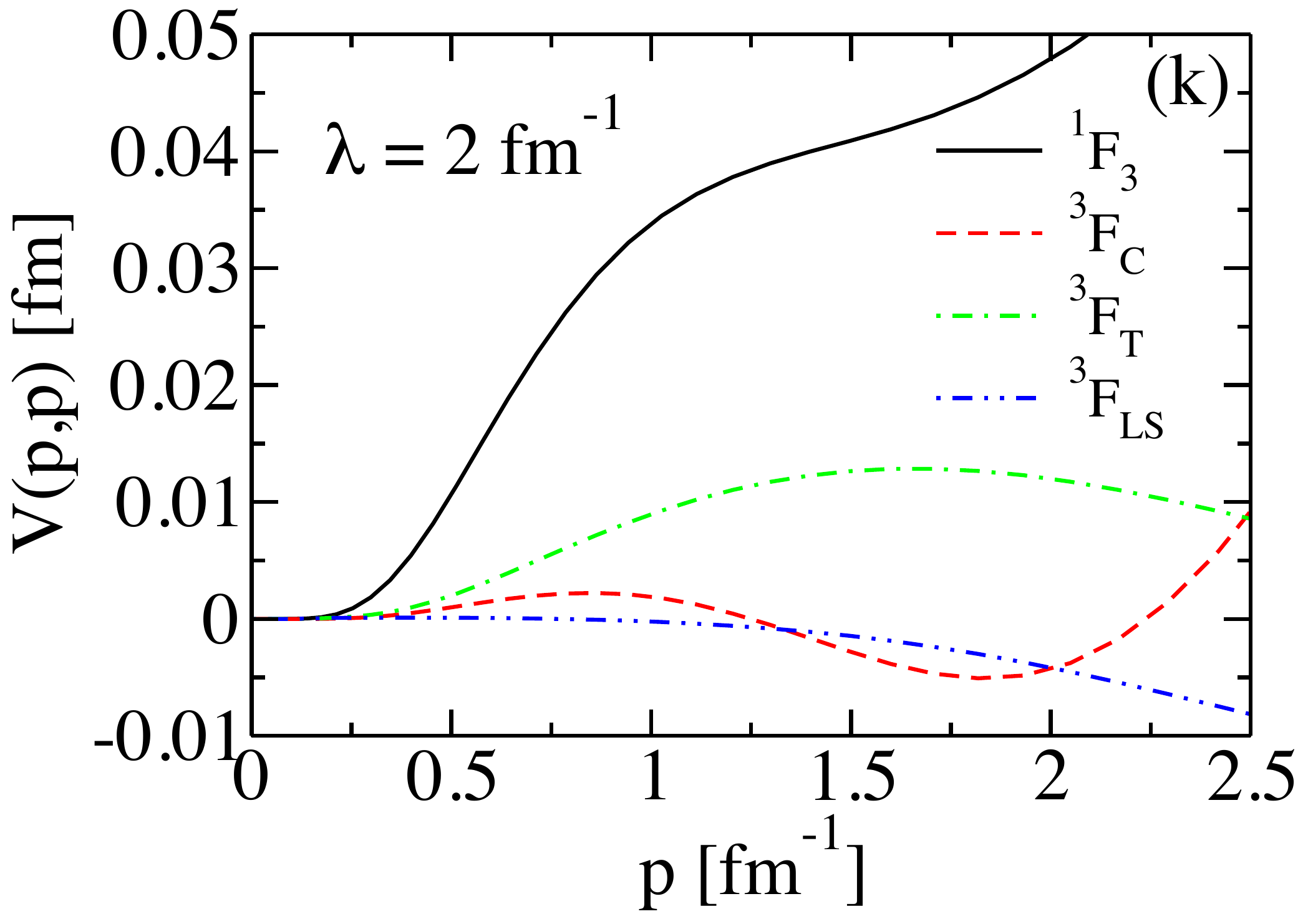}
\includegraphics[height=4cm,width=5cm]{fig1k} \\
\includegraphics[height=4cm,width=5cm]{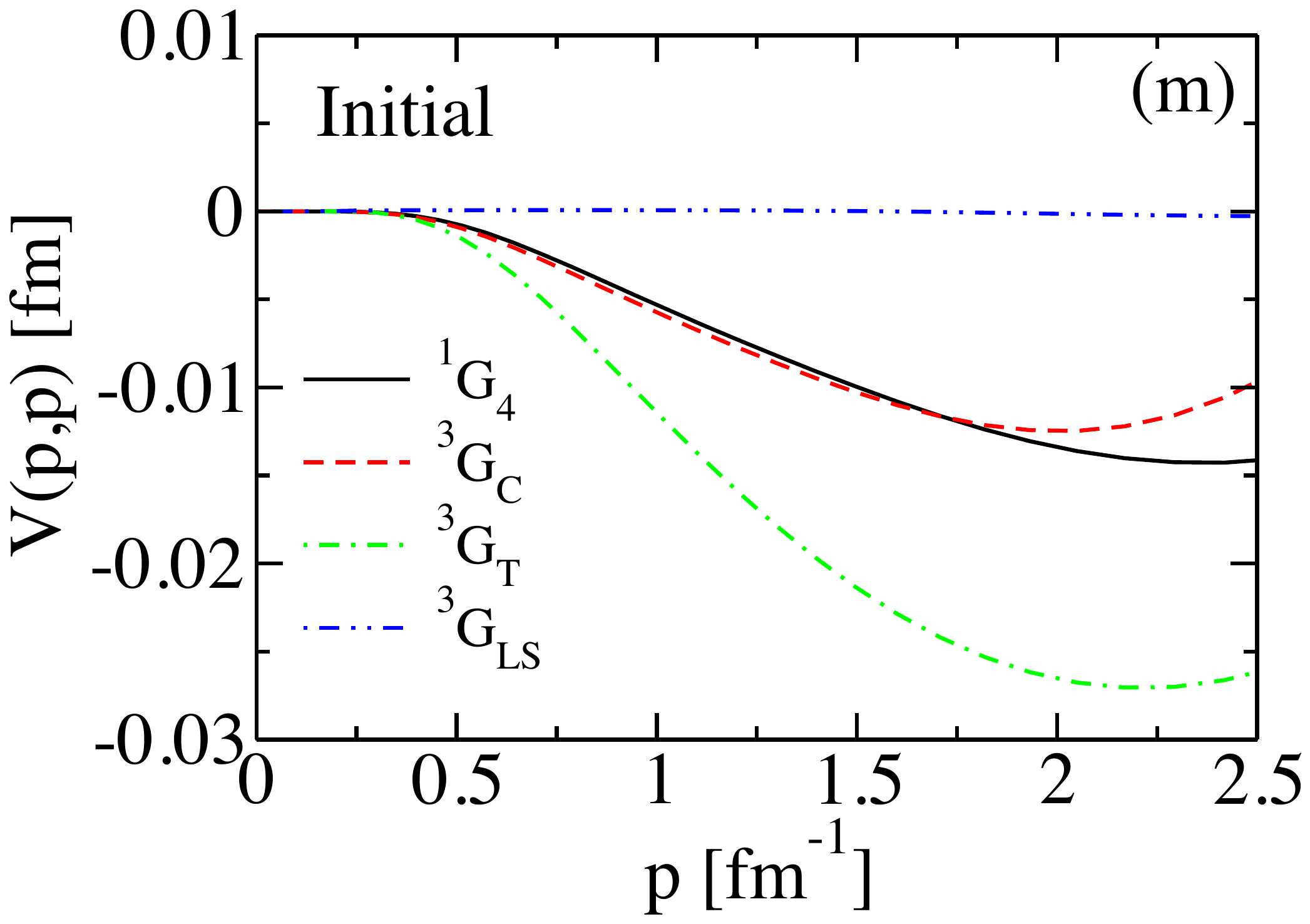}
\includegraphics[height=4cm,width=5cm]{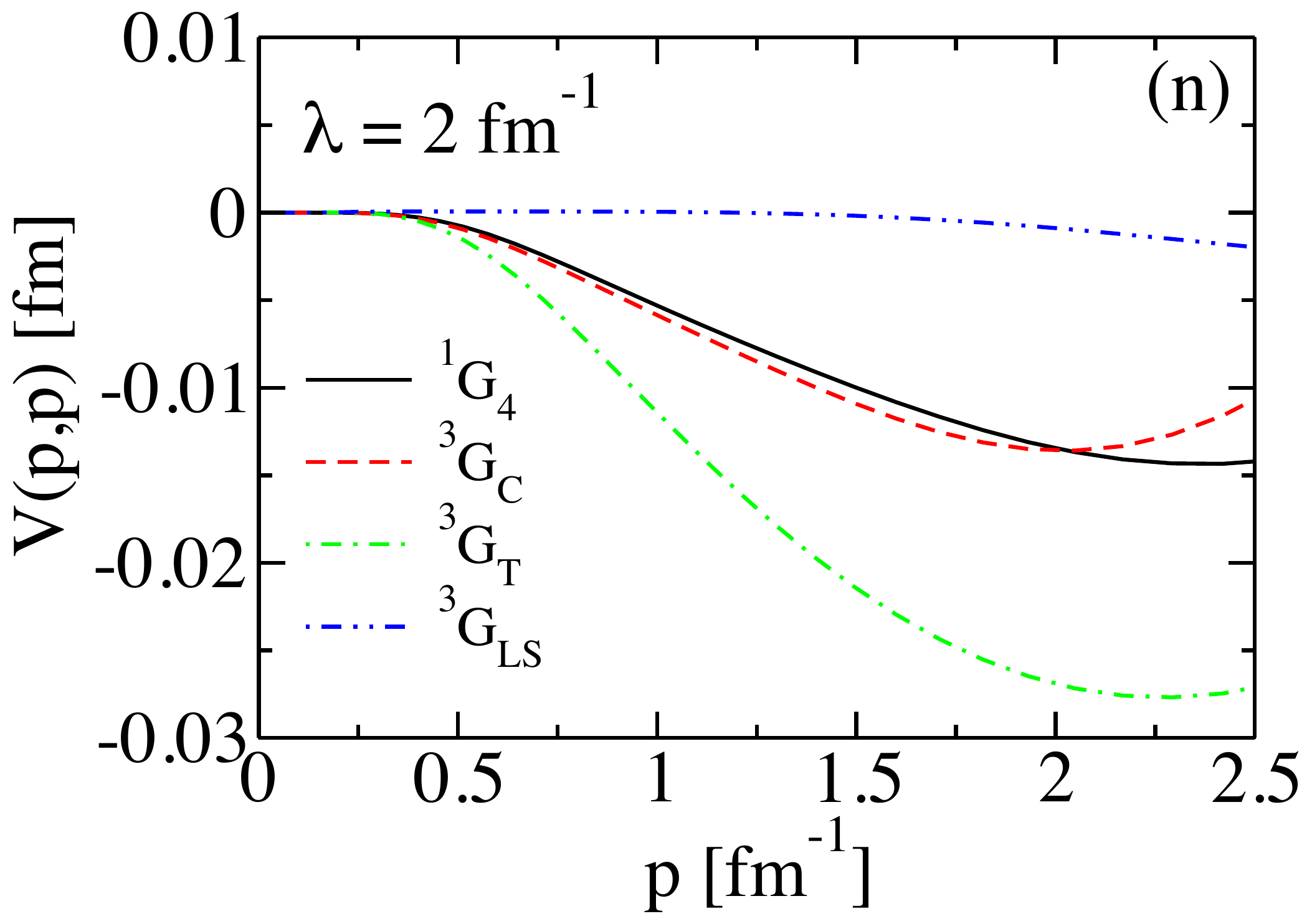}
\includegraphics[height=4cm,width=5cm]{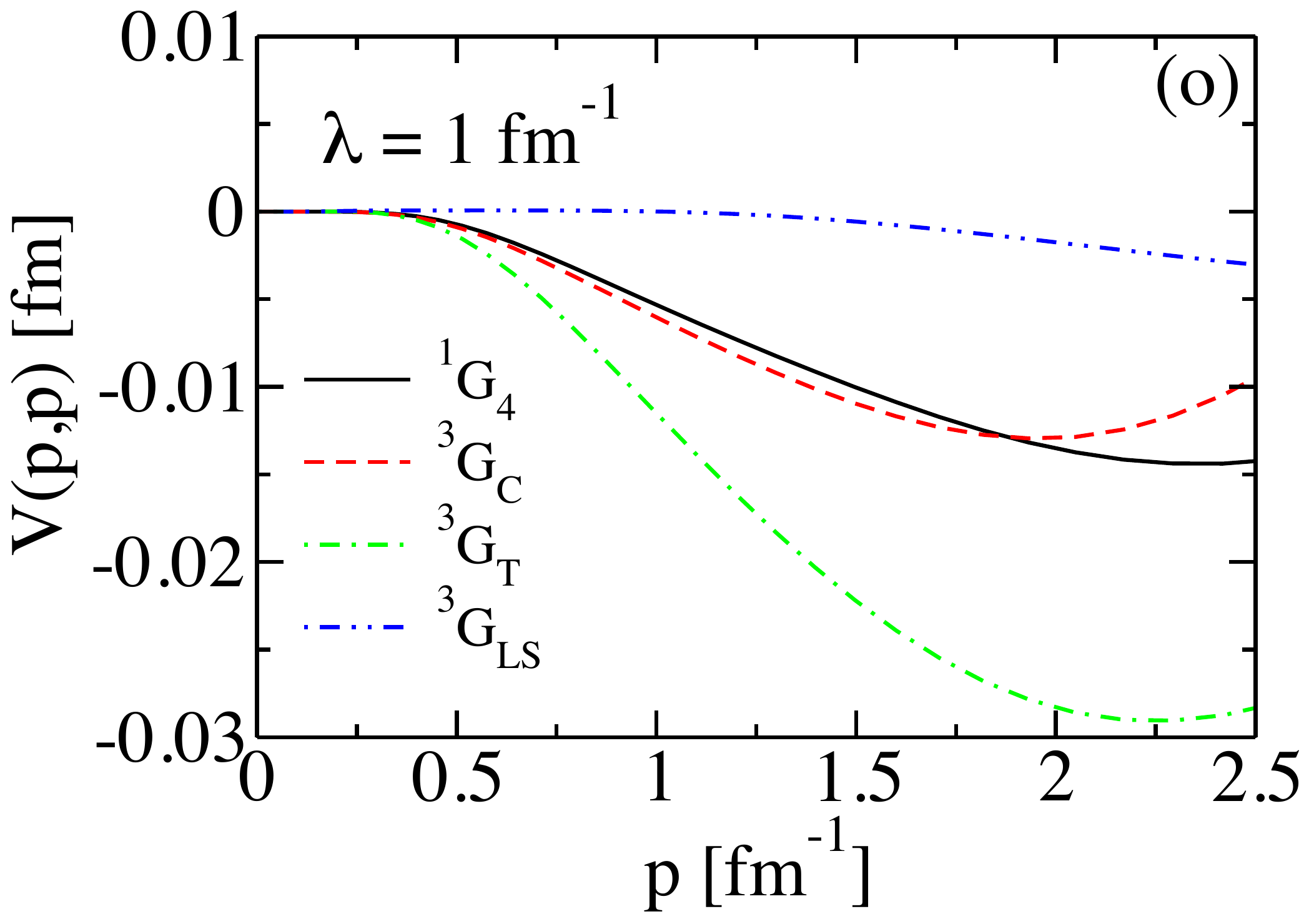}
\end{center}
\caption{(Color online) Diagonal matrix-elements of the SRG-evolved Argonne AV18 potential~\cite{Wiringa:1994wb} $V(p,p)$ (in {\rm fm}) as a function
  of the CM momentum (in ${\rm fm}^{-1}$) for the S, P, D, F and G partial-wave
  components for different values of the similarity cutoff
  $\lambda$. Left Panel: Initial potential ($\lambda=\infty$). Central Panel: $\lambda =
  2~{\rm fm}^{-1}$.  Right Panel: $\lambda = 1~{\rm fm}^{-1}$. }
\label{fig:AV18-diag}
\end{figure*}

\begin{figure*}[tbc]
\begin{center}
\includegraphics[height=4cm,width=5cm]{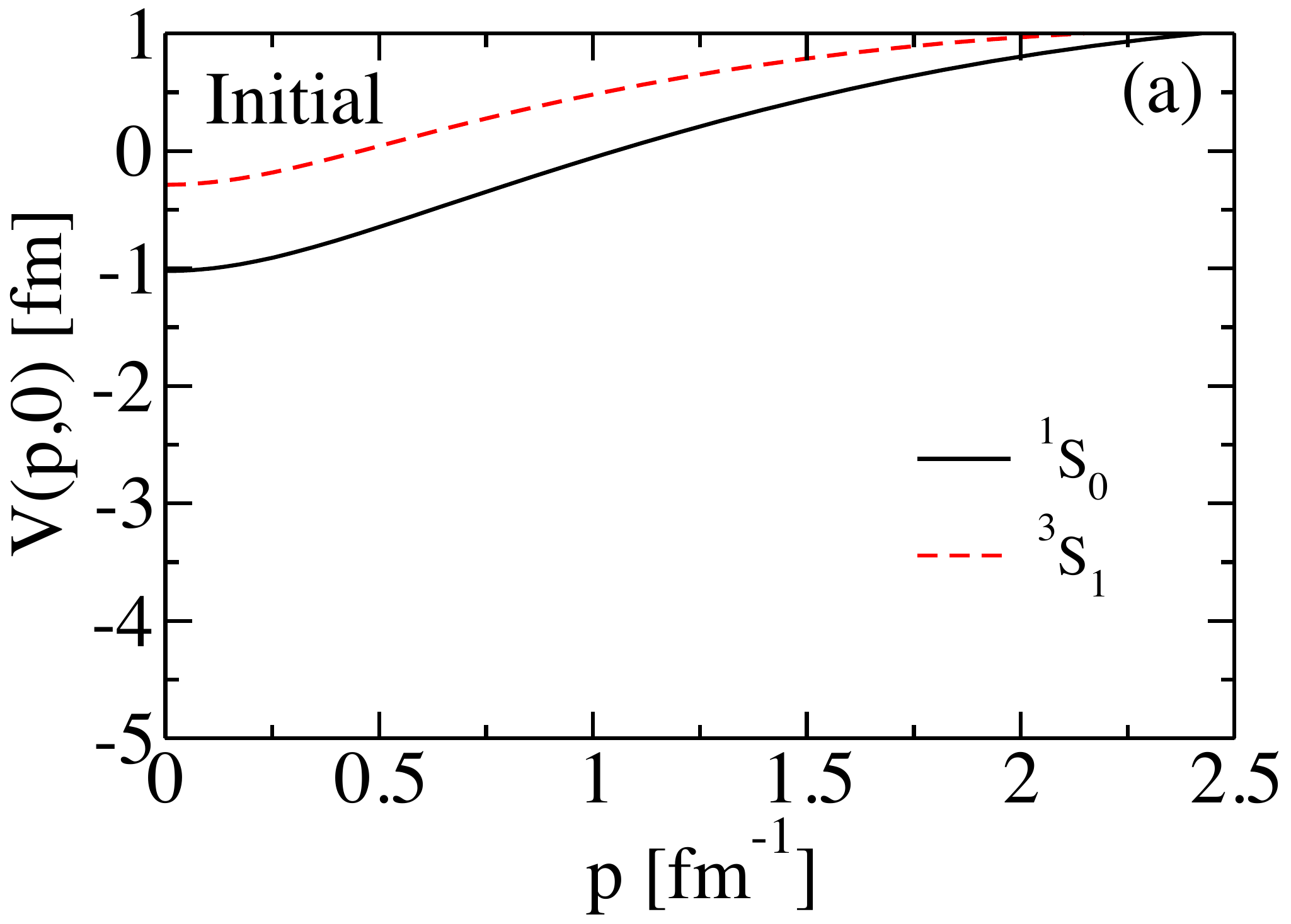}
\includegraphics[height=4cm,width=5cm]{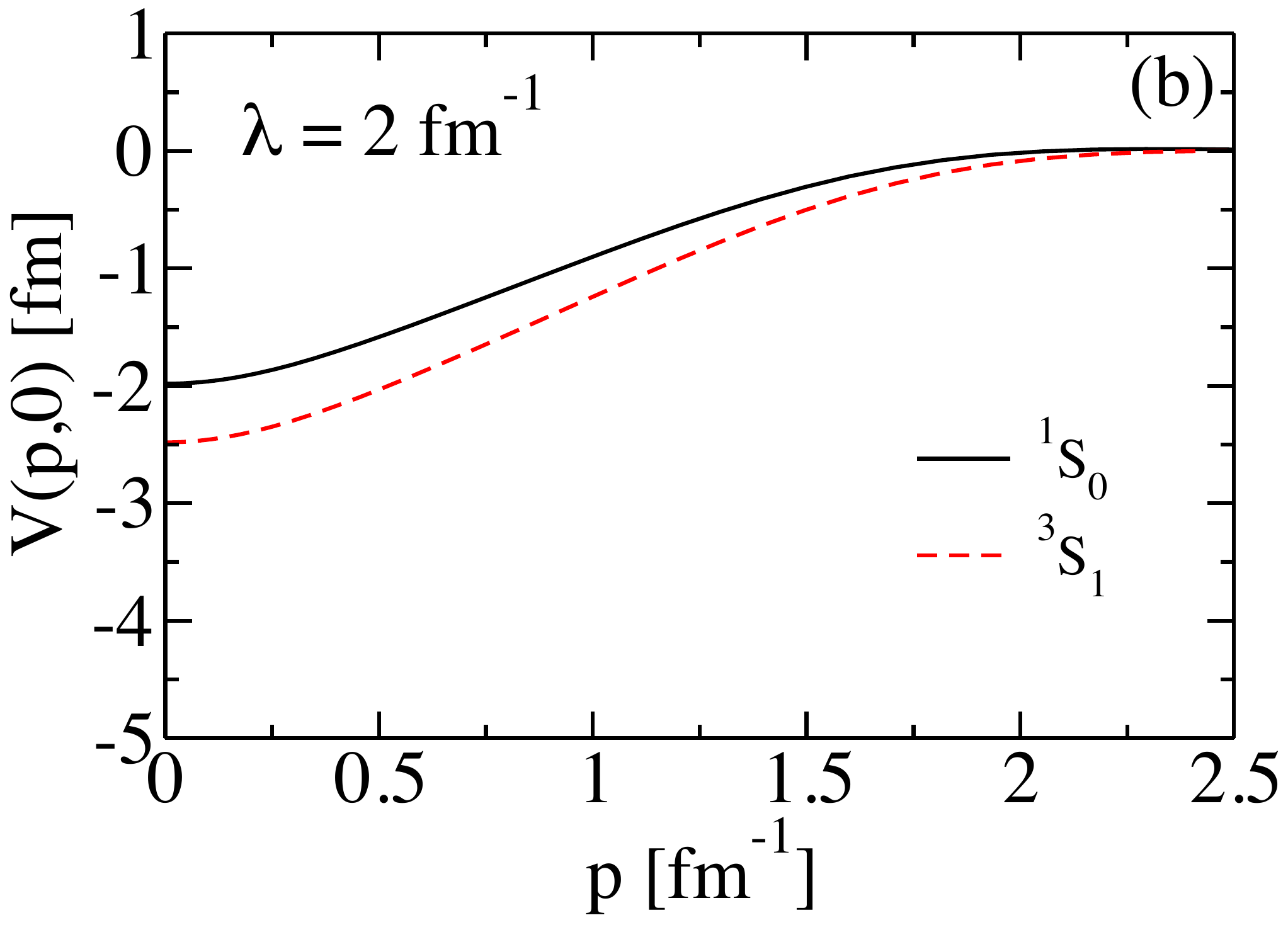}
\includegraphics[height=4cm,width=5cm]{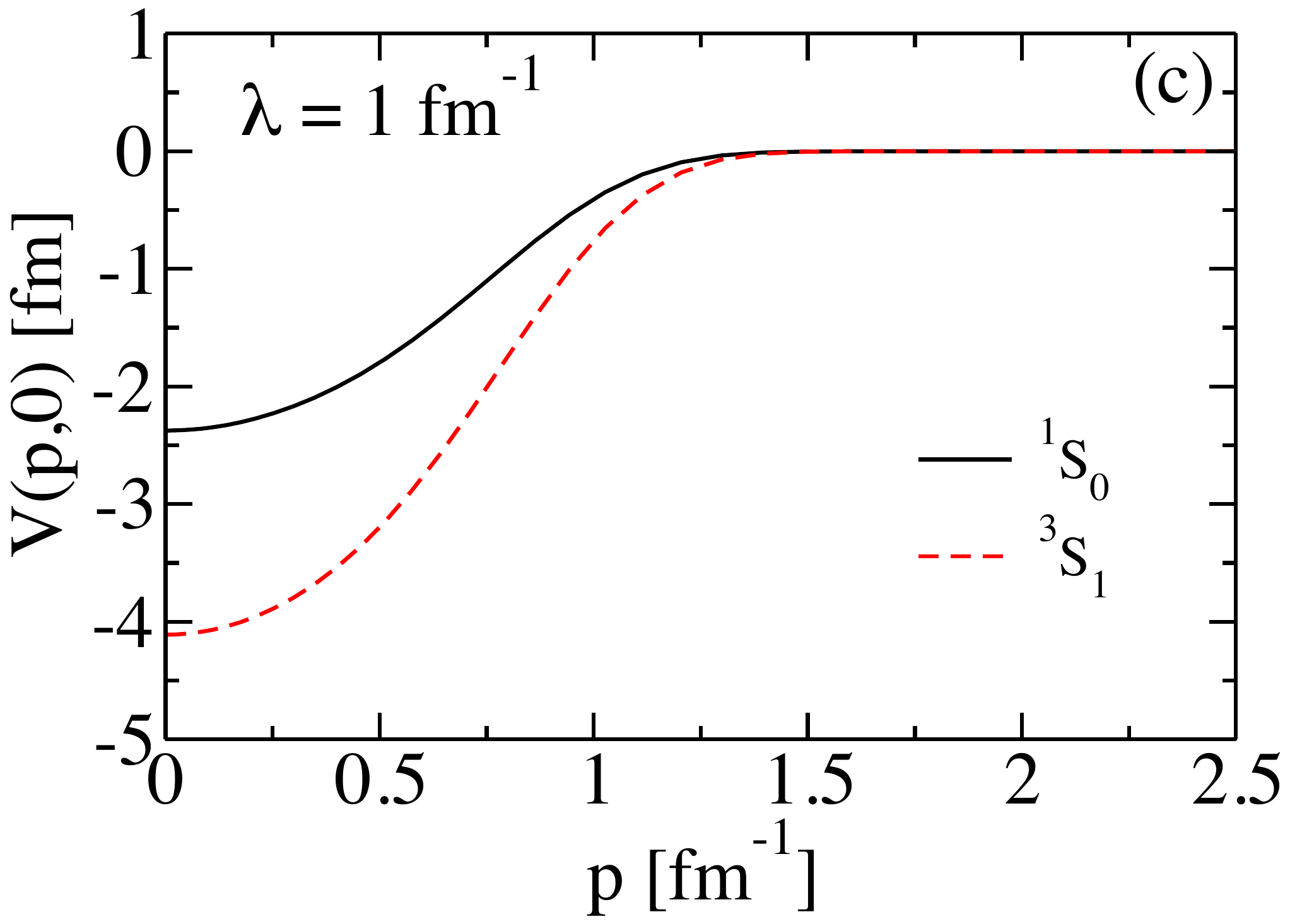} \\
\includegraphics[height=4cm,width=5cm]{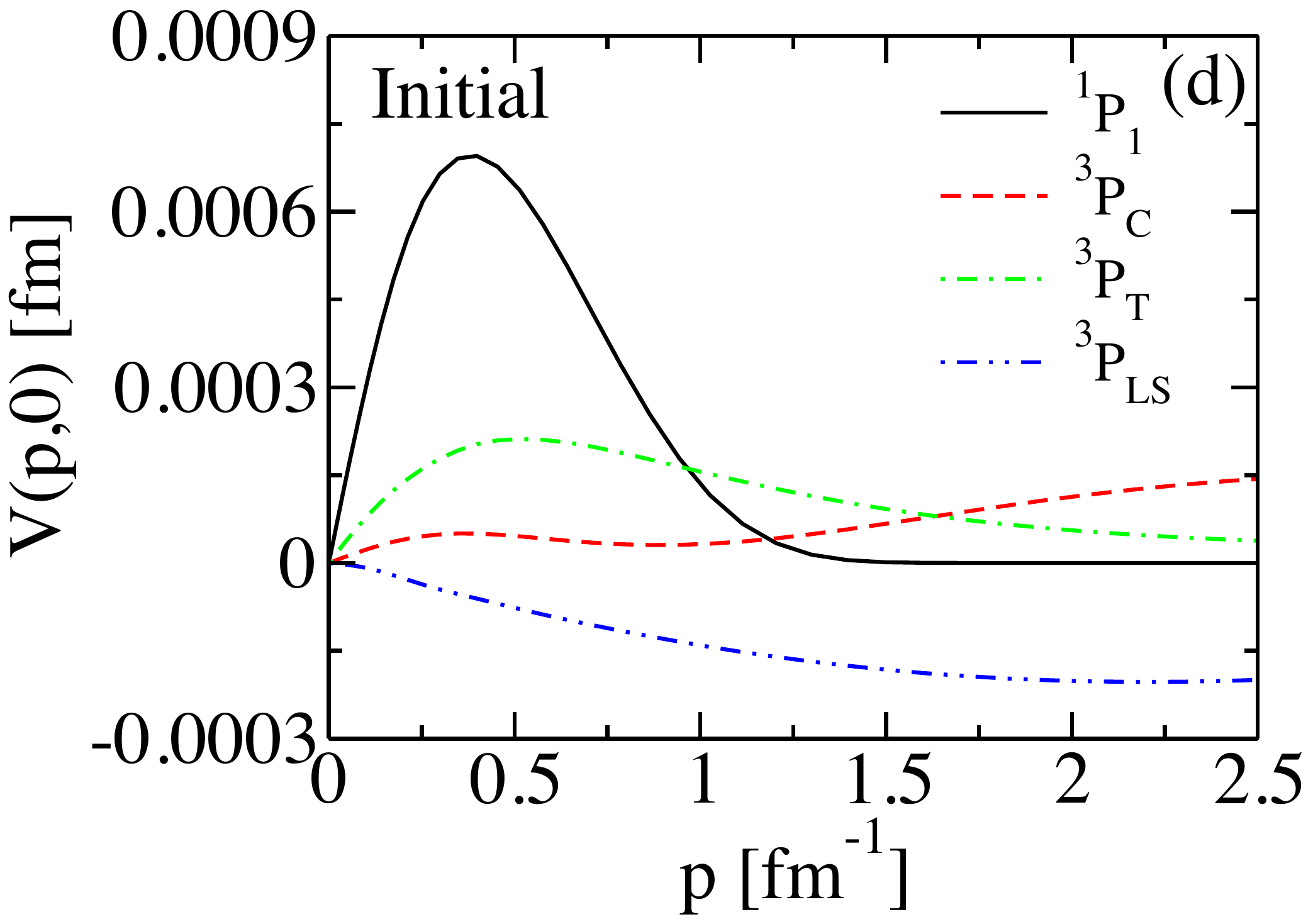}
\includegraphics[height=4cm,width=5cm]{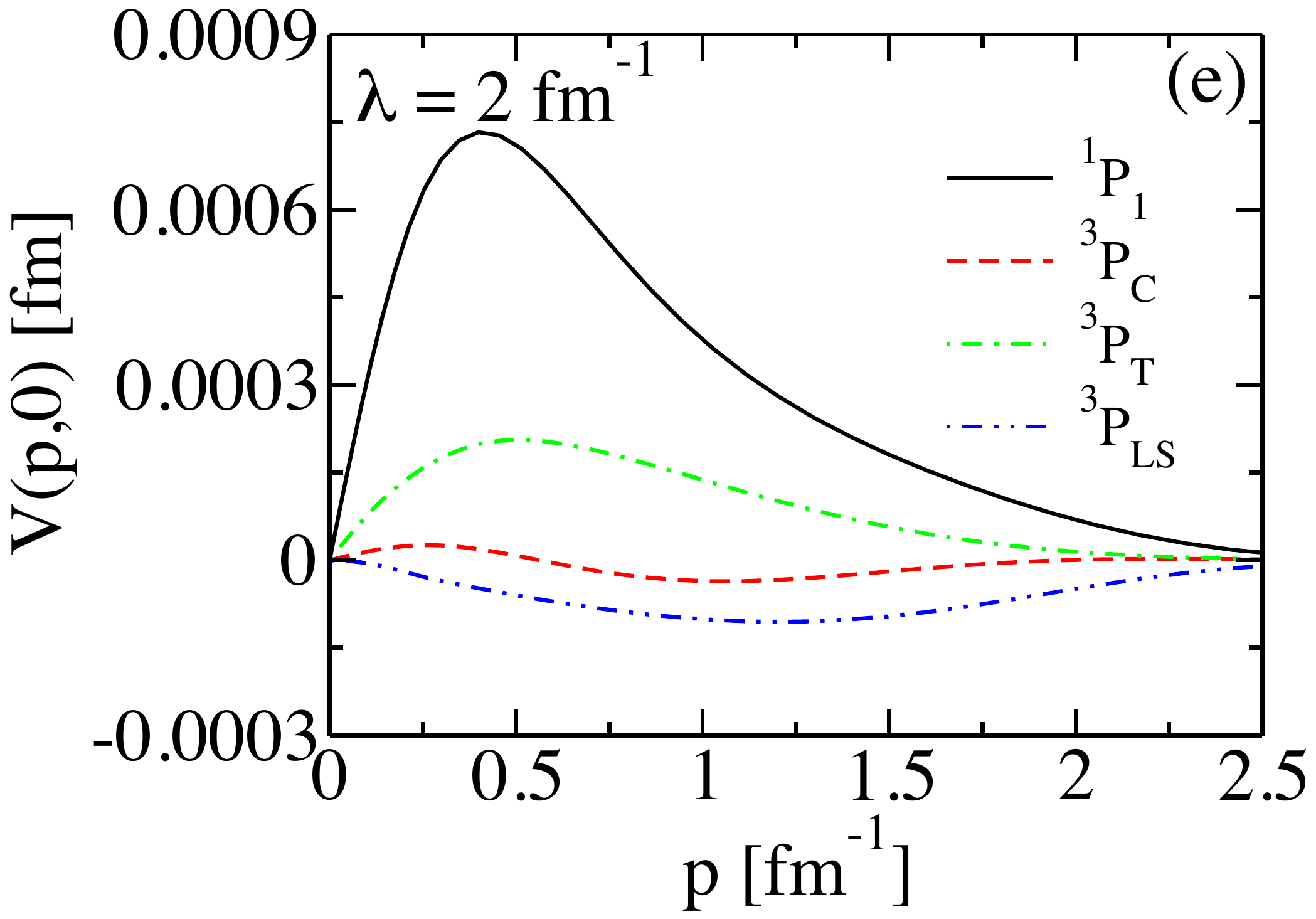}
\includegraphics[height=4cm,width=5cm]{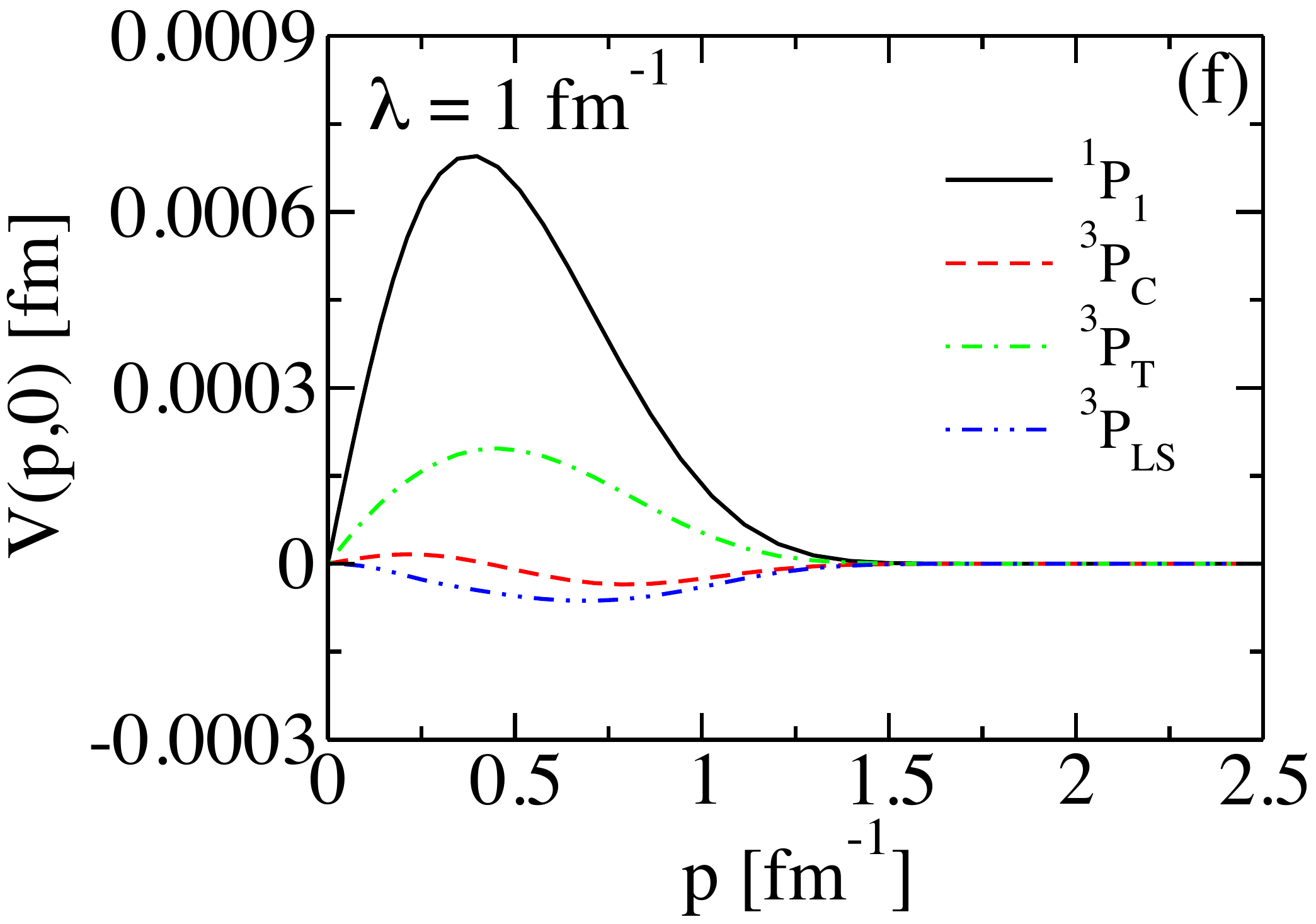} \\
\includegraphics[height=4cm,width=5cm]{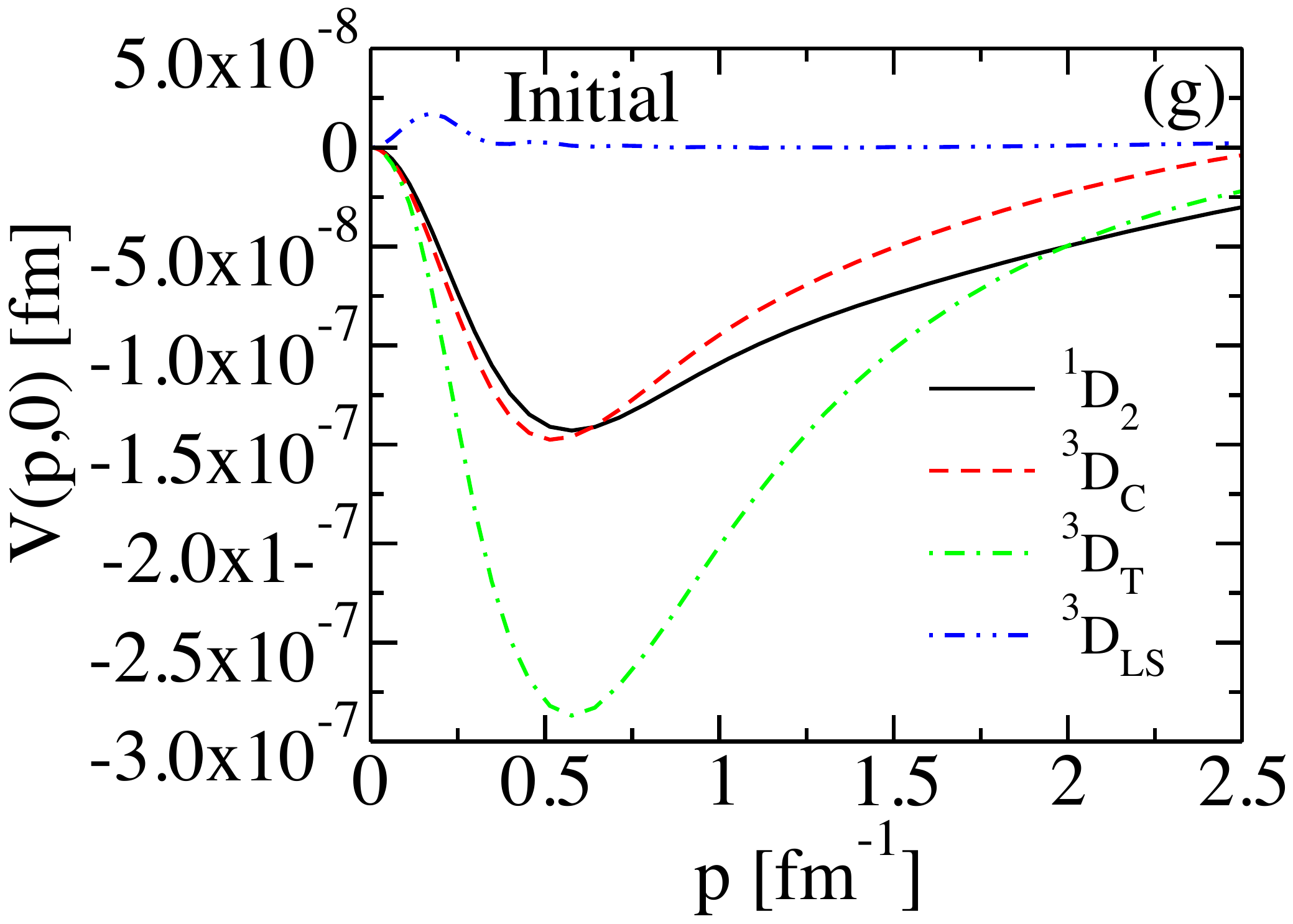}
\includegraphics[height=4cm,width=5cm]{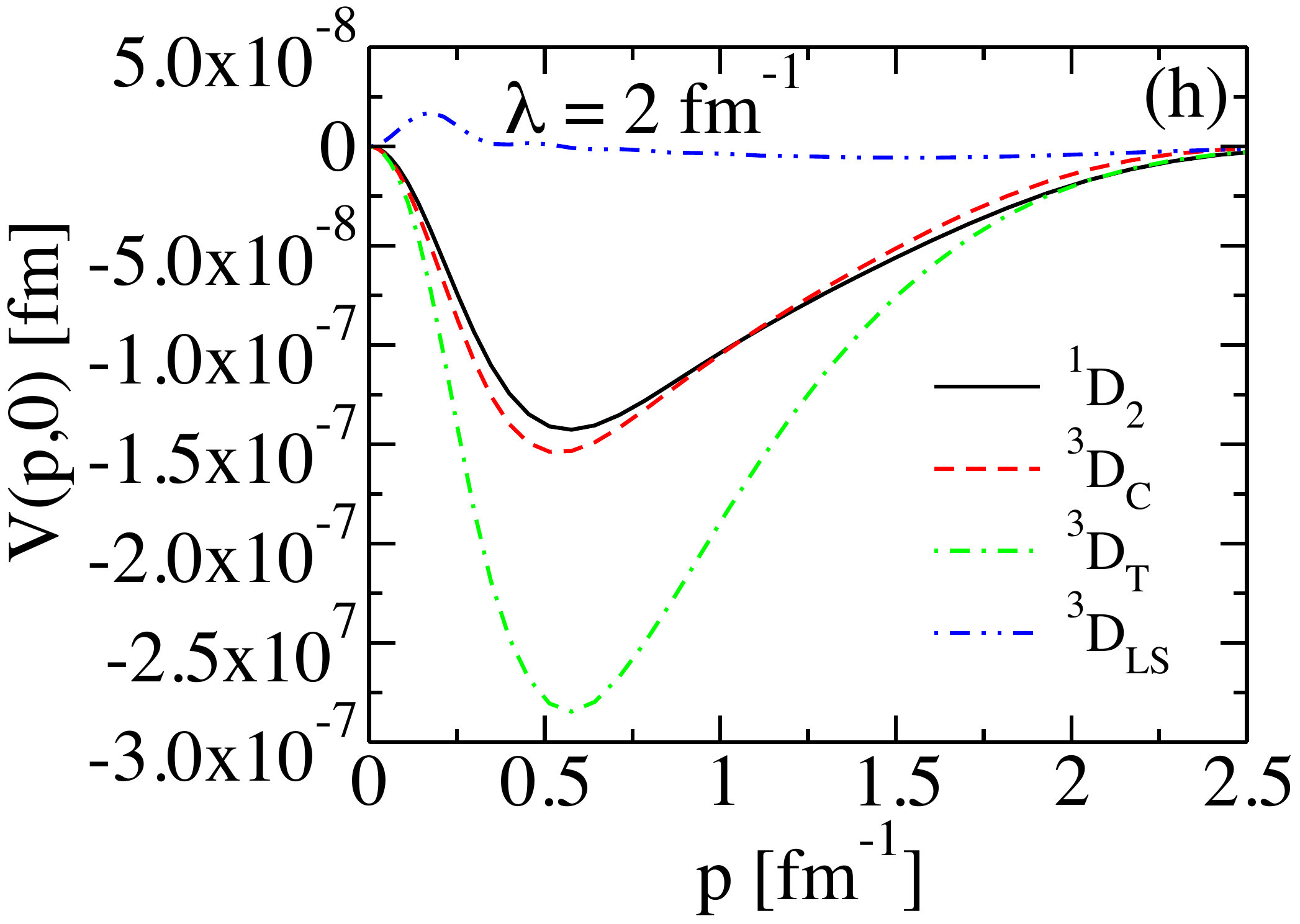}
\includegraphics[height=4cm,width=5cm]{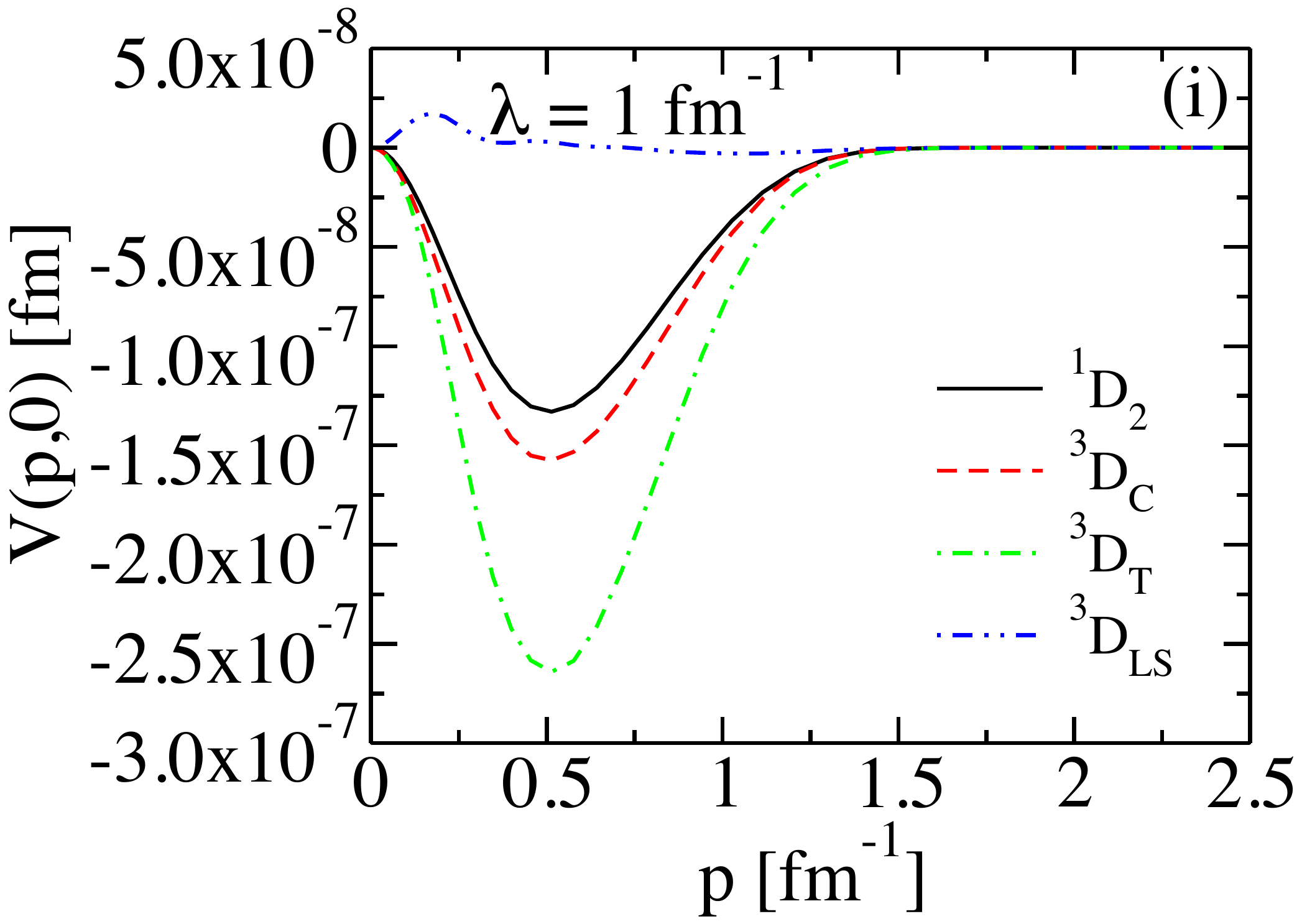} \\
\includegraphics[height=4cm,width=5cm]{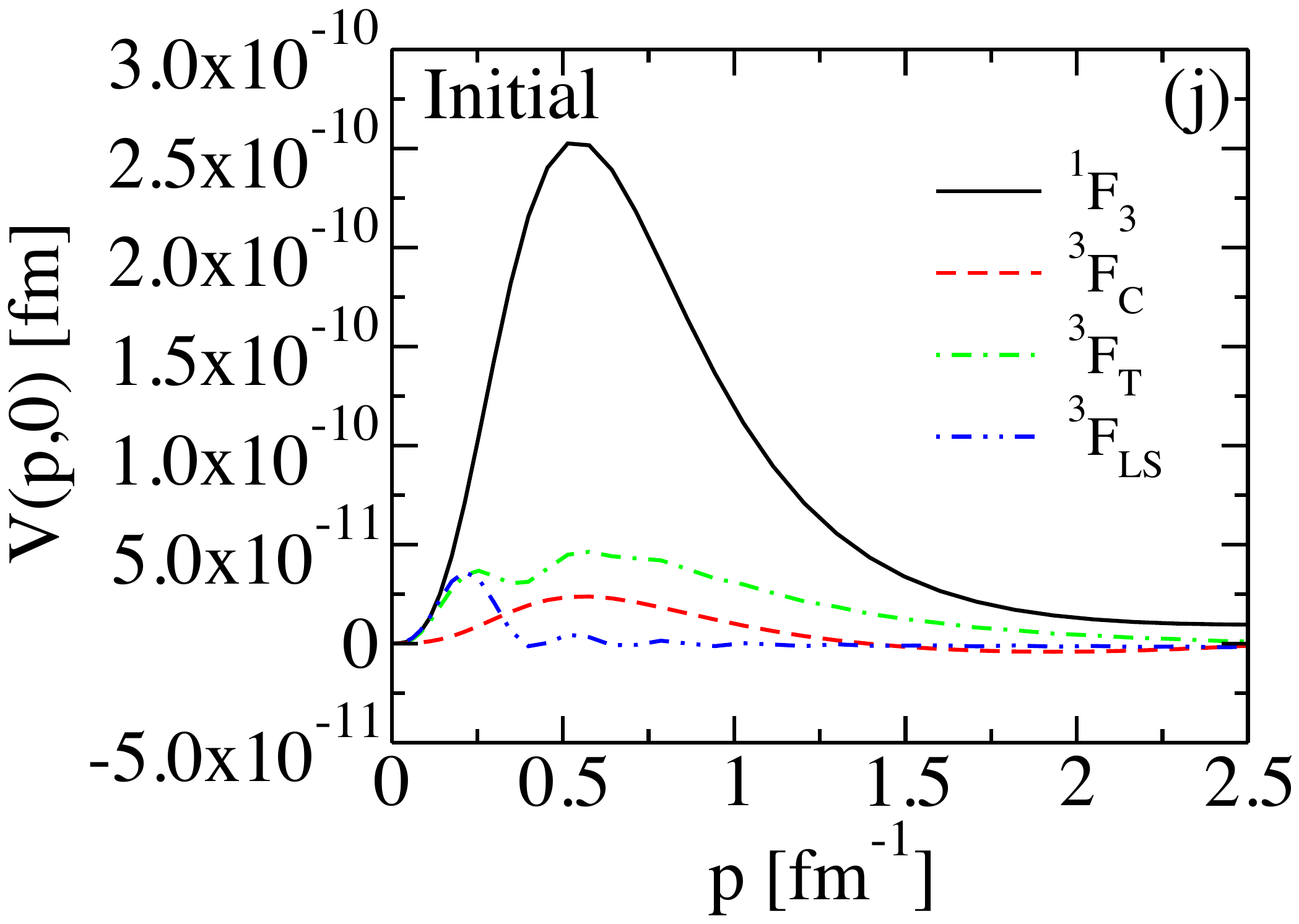}
\includegraphics[height=4cm,width=5cm]{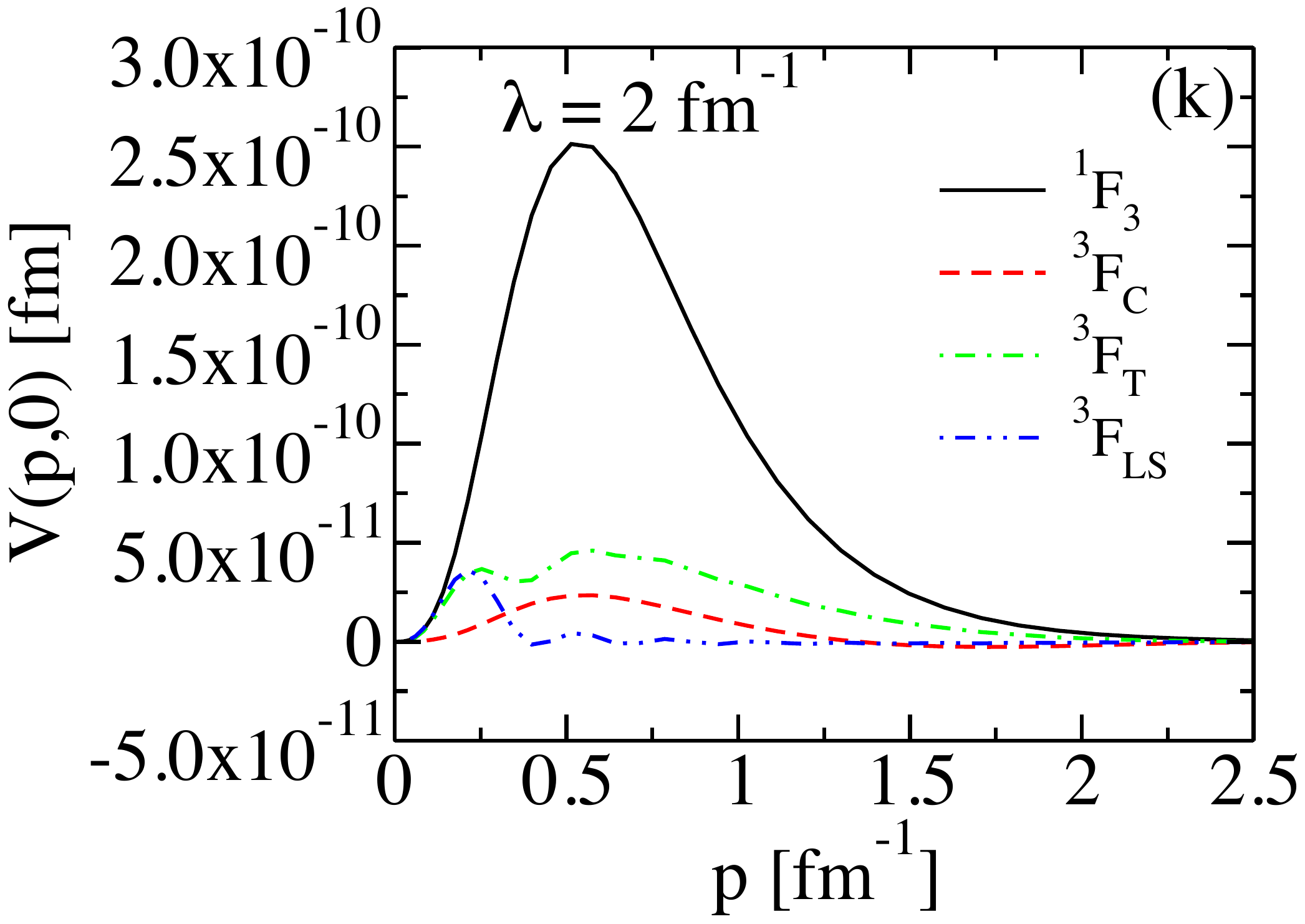}
\includegraphics[height=4cm,width=5cm]{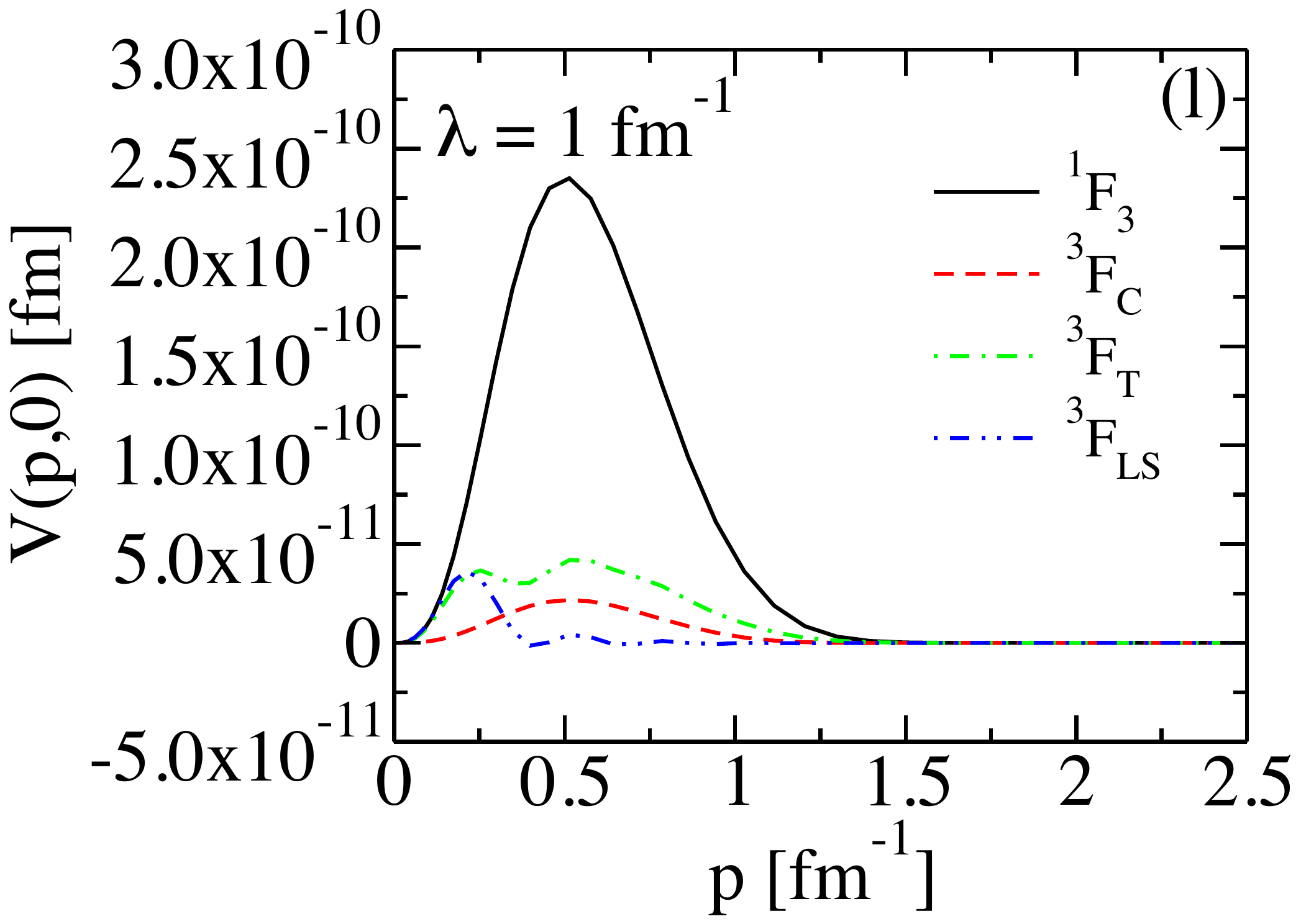} \\
\includegraphics[height=4cm,width=5cm]{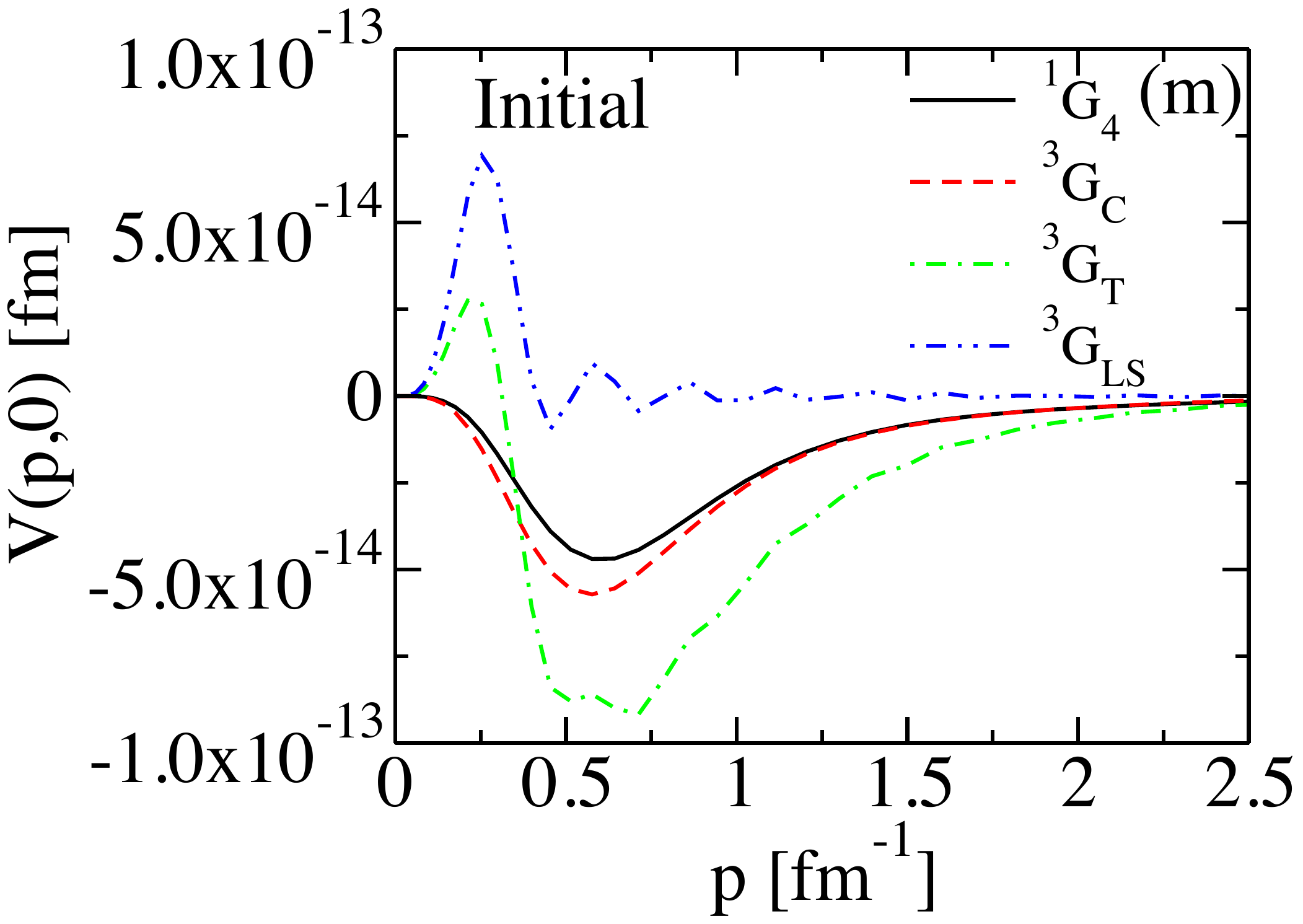}
\includegraphics[height=4cm,width=5cm]{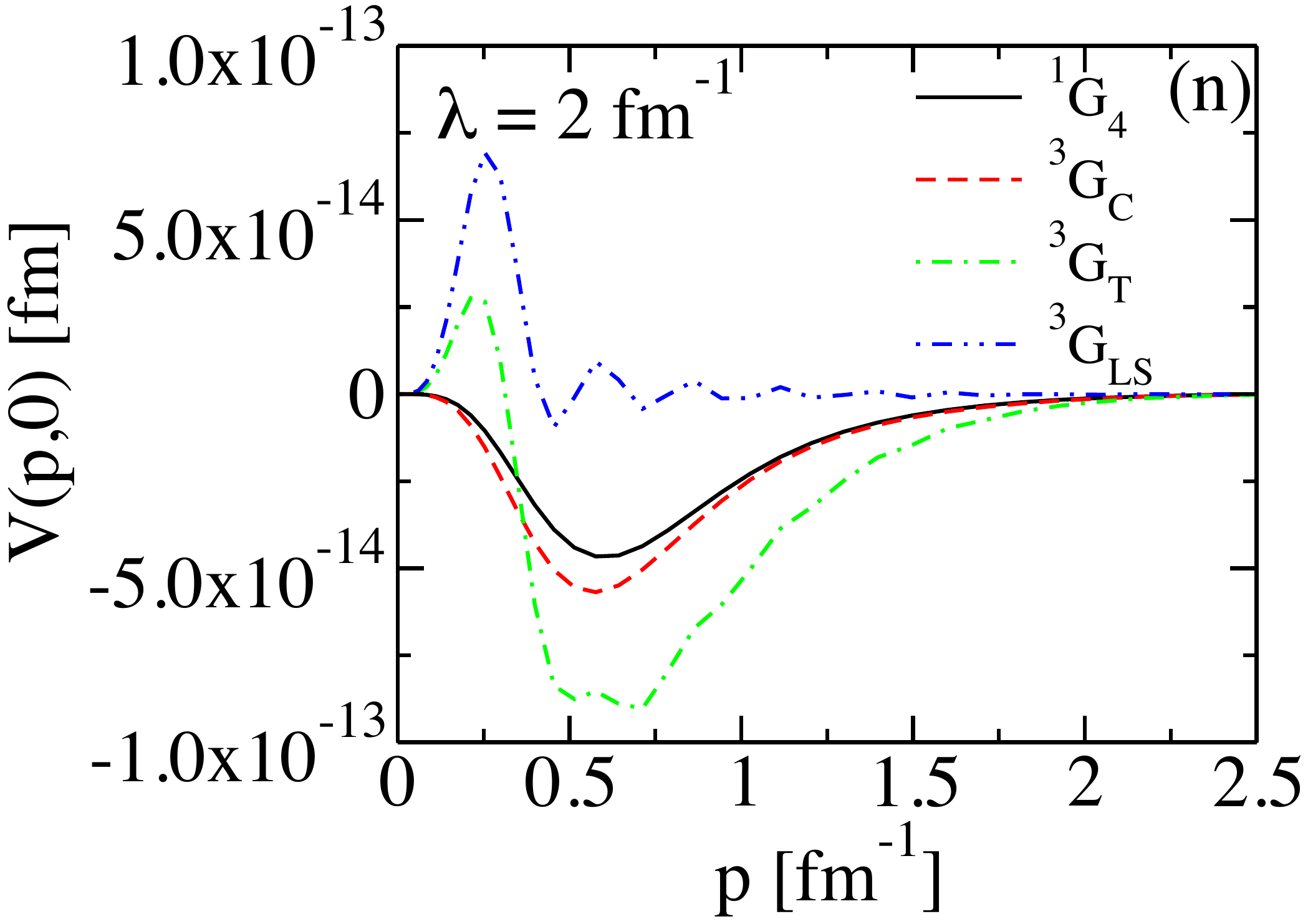}
\includegraphics[height=4cm,width=5cm]{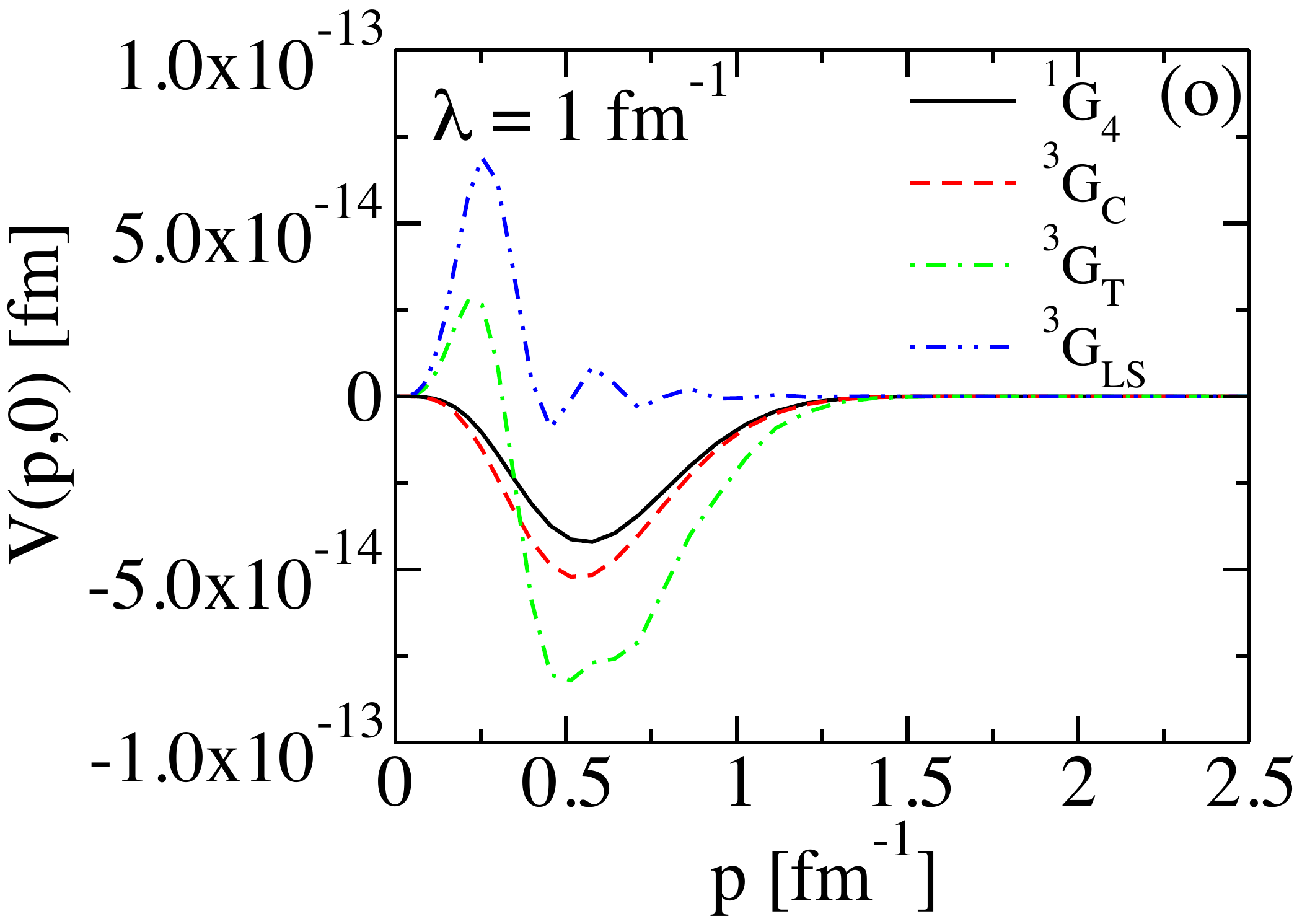}
\end{center}
\caption{(Color online) Fully off-diagonal matrix-elements of the SRG-evolved Argonne AV18 potential~\cite{Wiringa:1994wb} (in {\rm fm}) as a function
  of the CM momentum (in ${\rm fm}^{-1}$) for the S, P, D, F and G partial-wave components for different values of the similarity cutoff $\lambda$. Left Panel: Initial potential  ($\lambda=\infty$). Central Panel: $\lambda = 2~{\rm
    fm}^{-1}$.  Right Panel: $\lambda = 1~{\rm fm}^{-1}$. }
\label{fig:AV18-off-diag}
\end{figure*}

\begin{figure*}[tbc]
\begin{center}
\includegraphics[height=4cm,width=5cm]{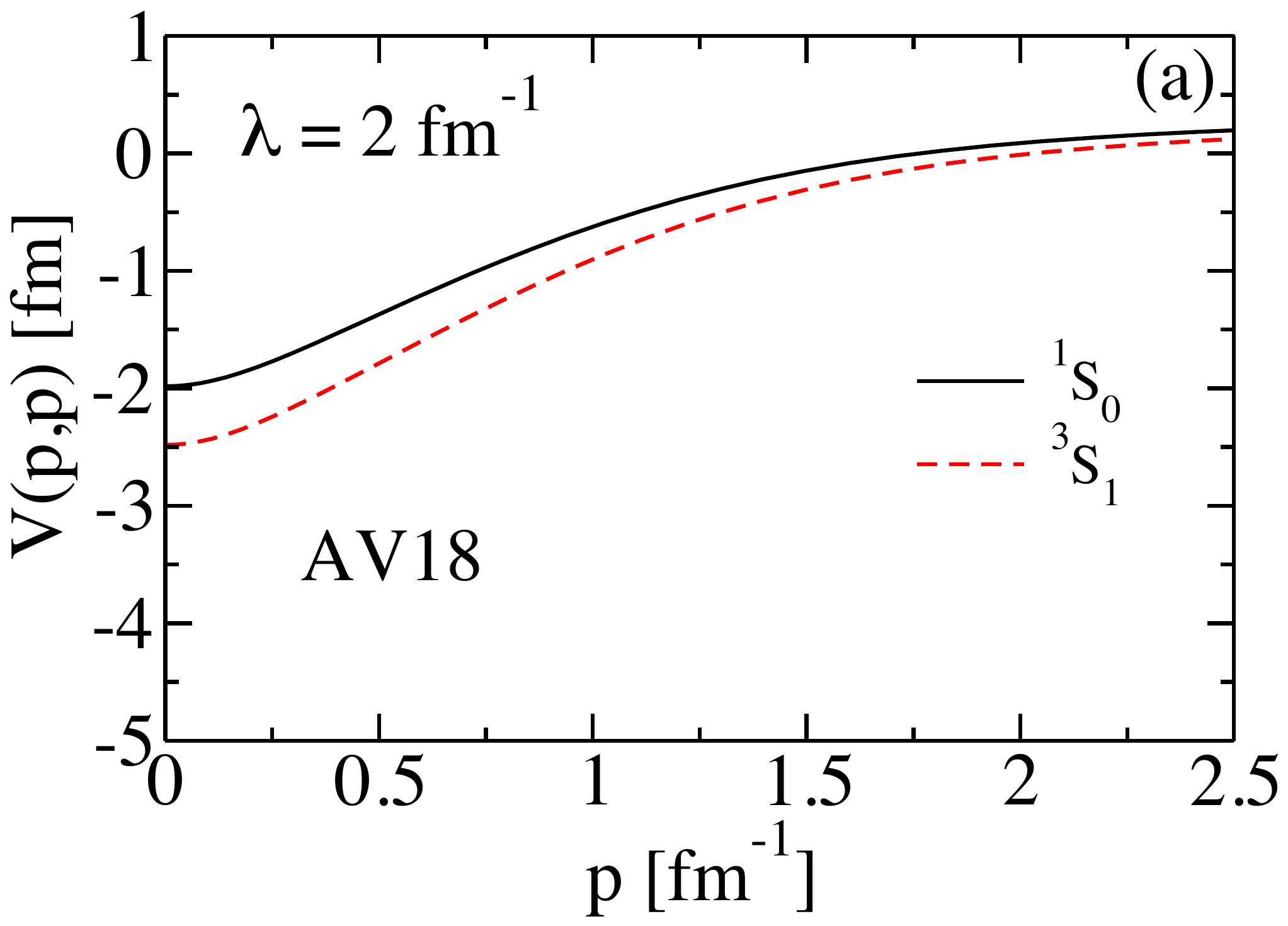}
\includegraphics[height=4cm,width=5cm]{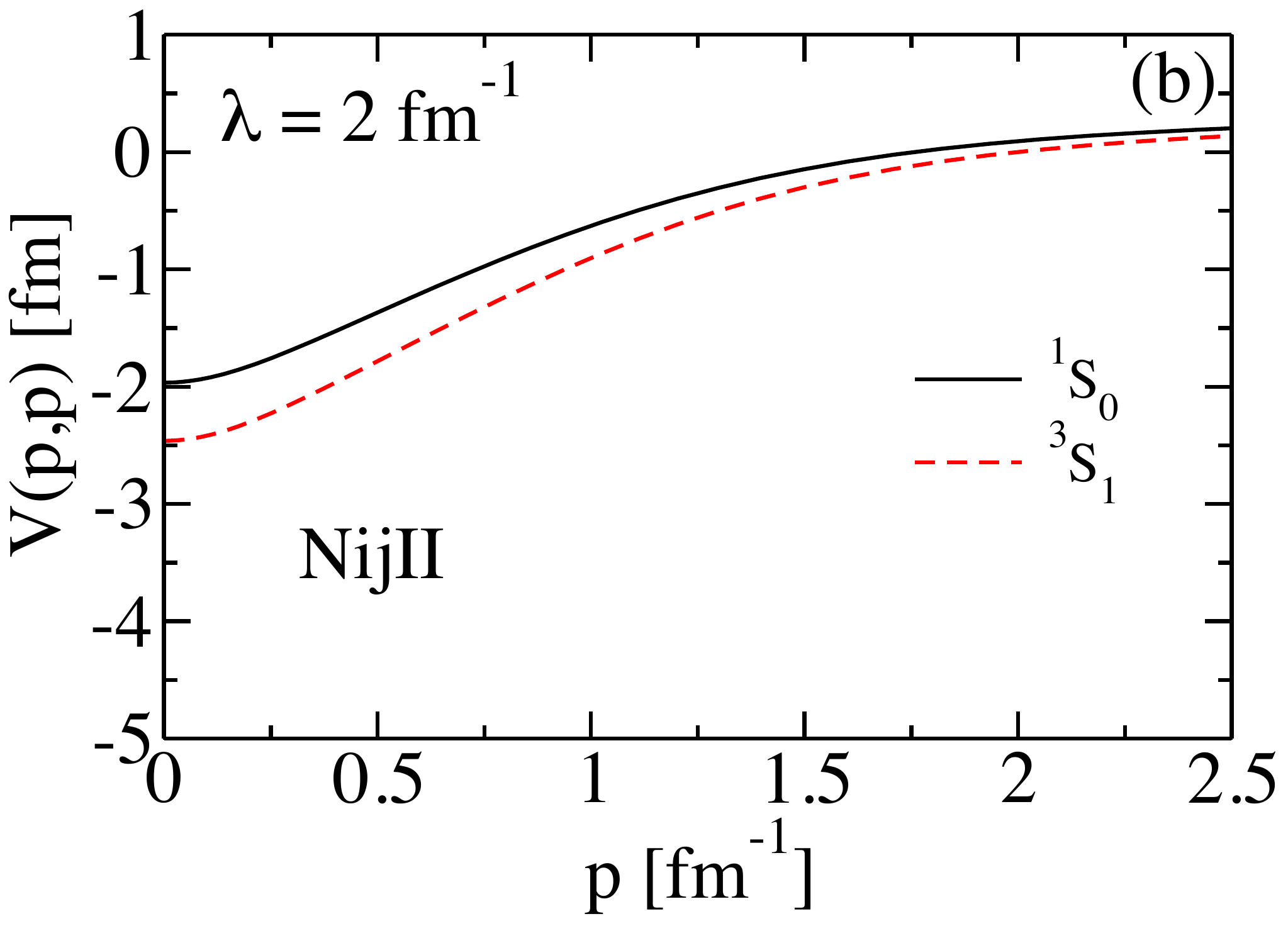}
\includegraphics[height=4cm,width=5cm]{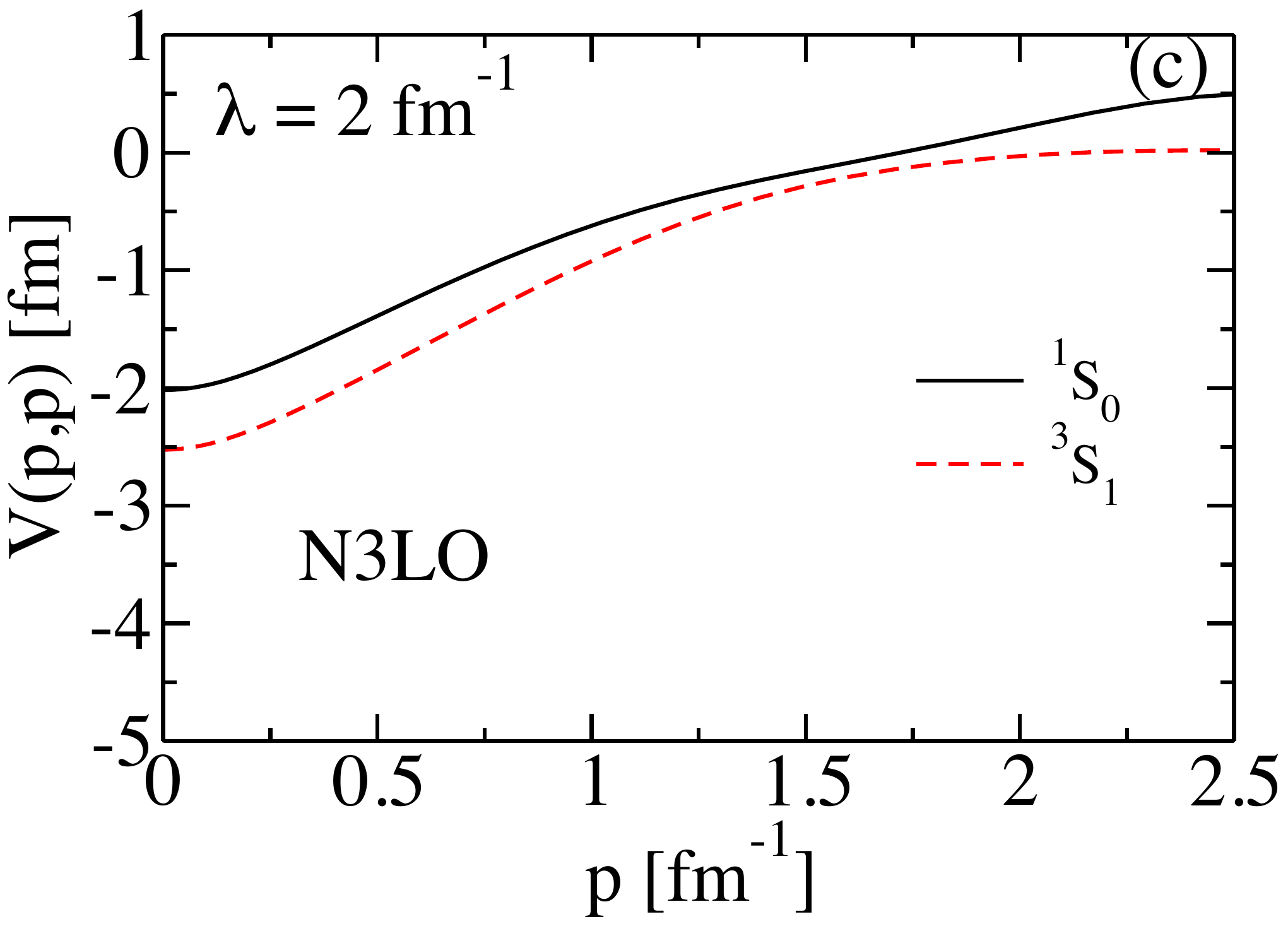} \\
\includegraphics[height=4cm,width=5cm]{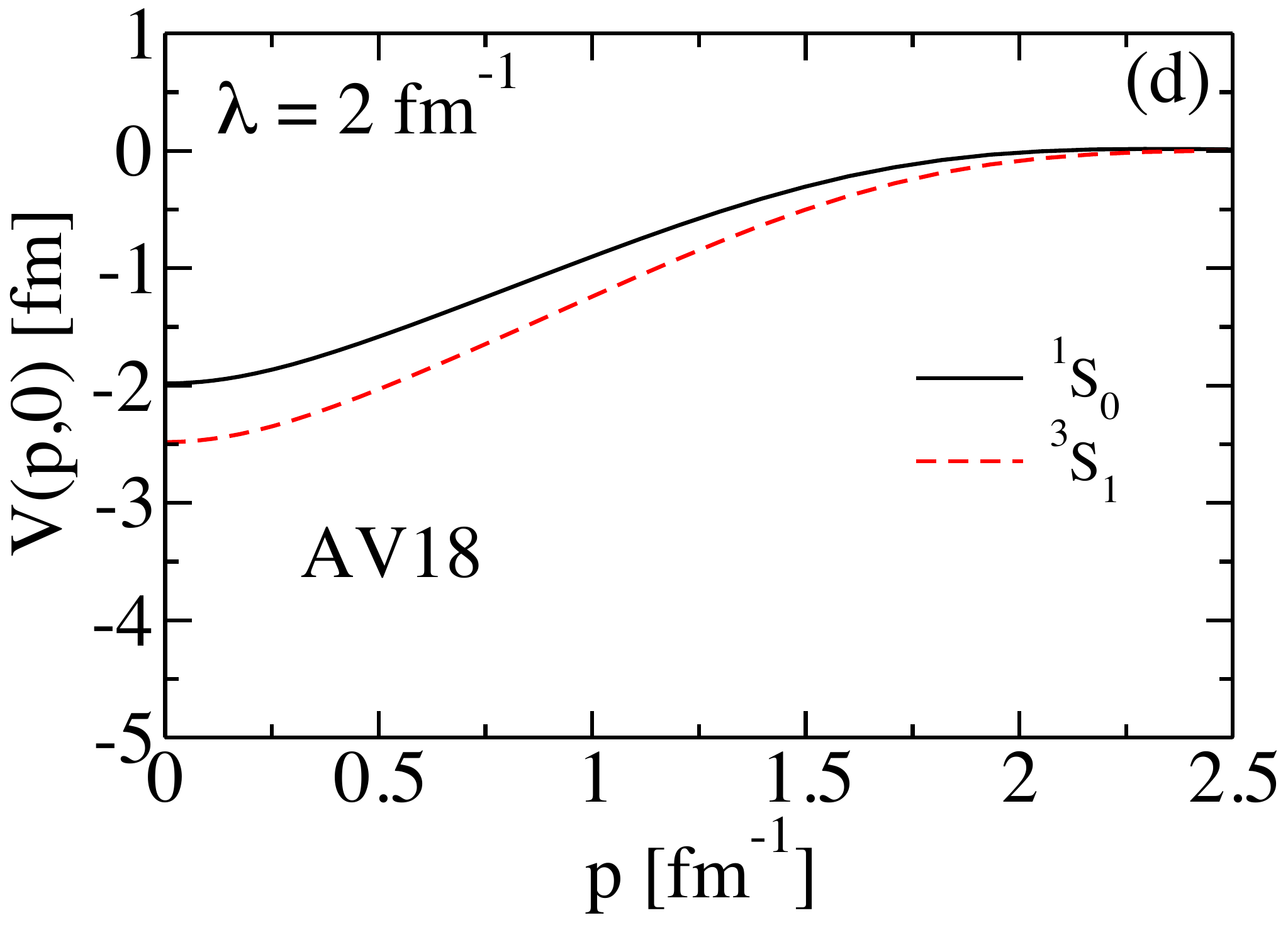}
\includegraphics[height=4cm,width=5cm]{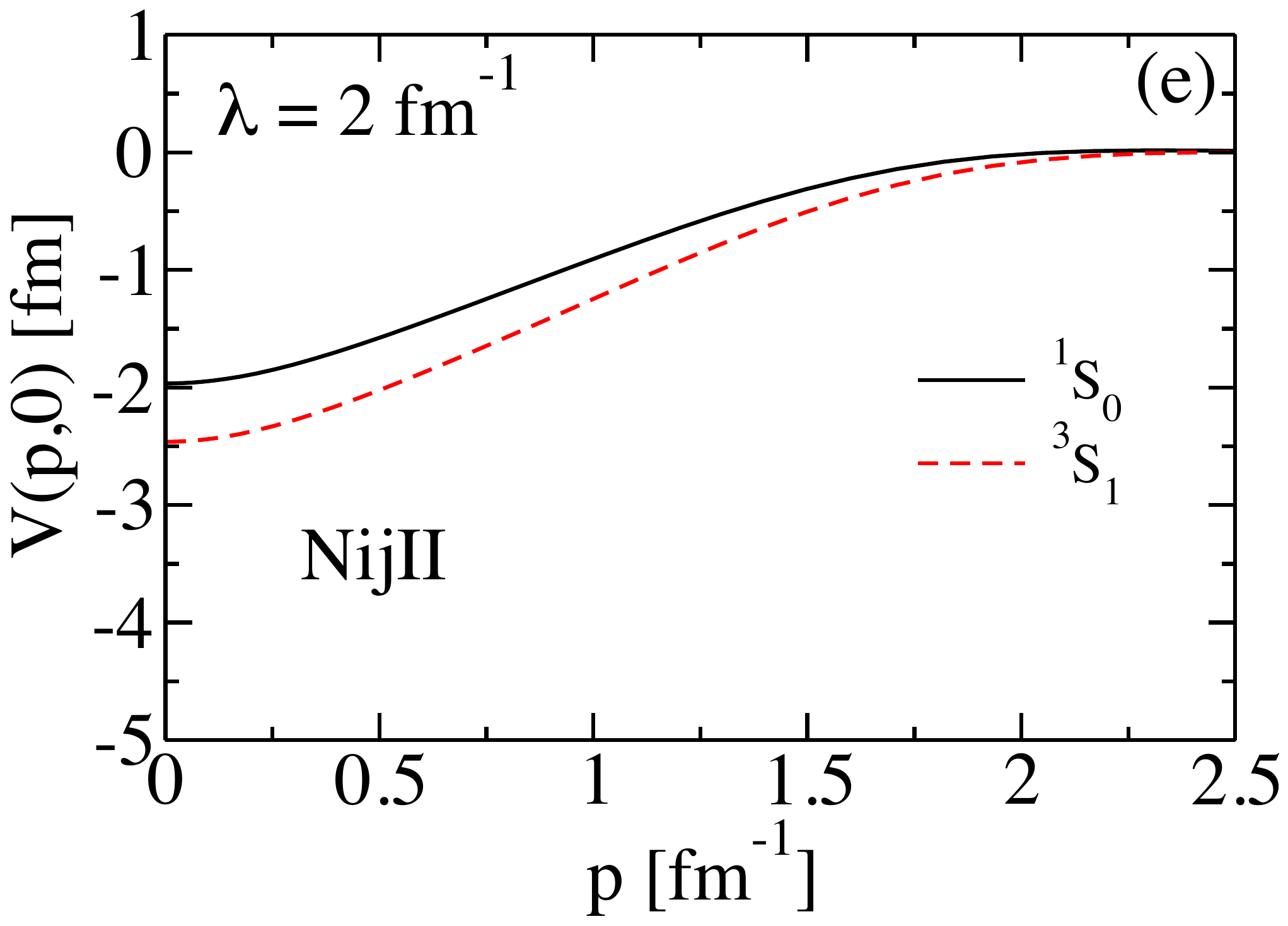}
\includegraphics[height=4cm,width=5cm]{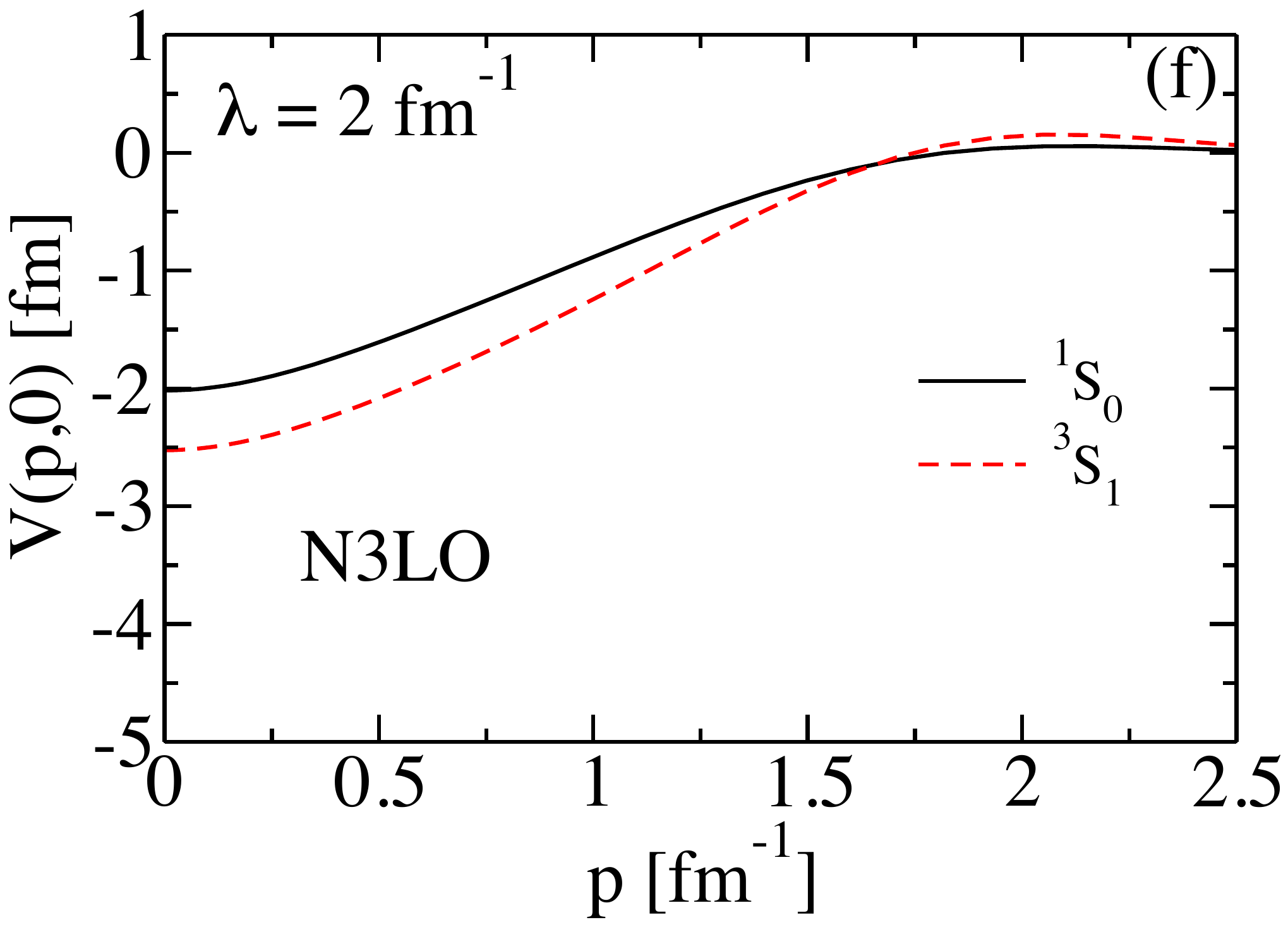} \\
\includegraphics[height=4cm,width=5cm]{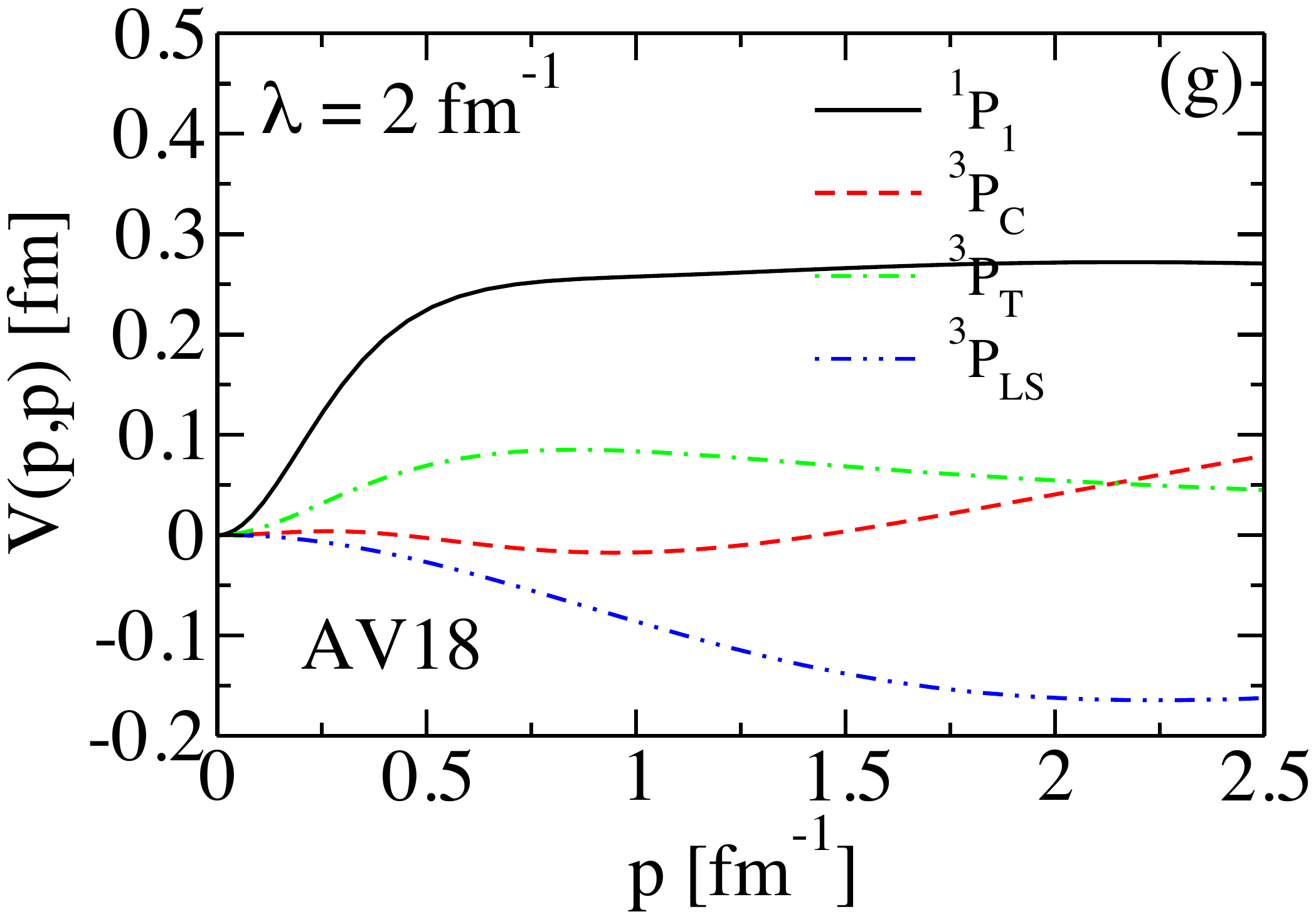}
\includegraphics[height=4cm,width=5cm]{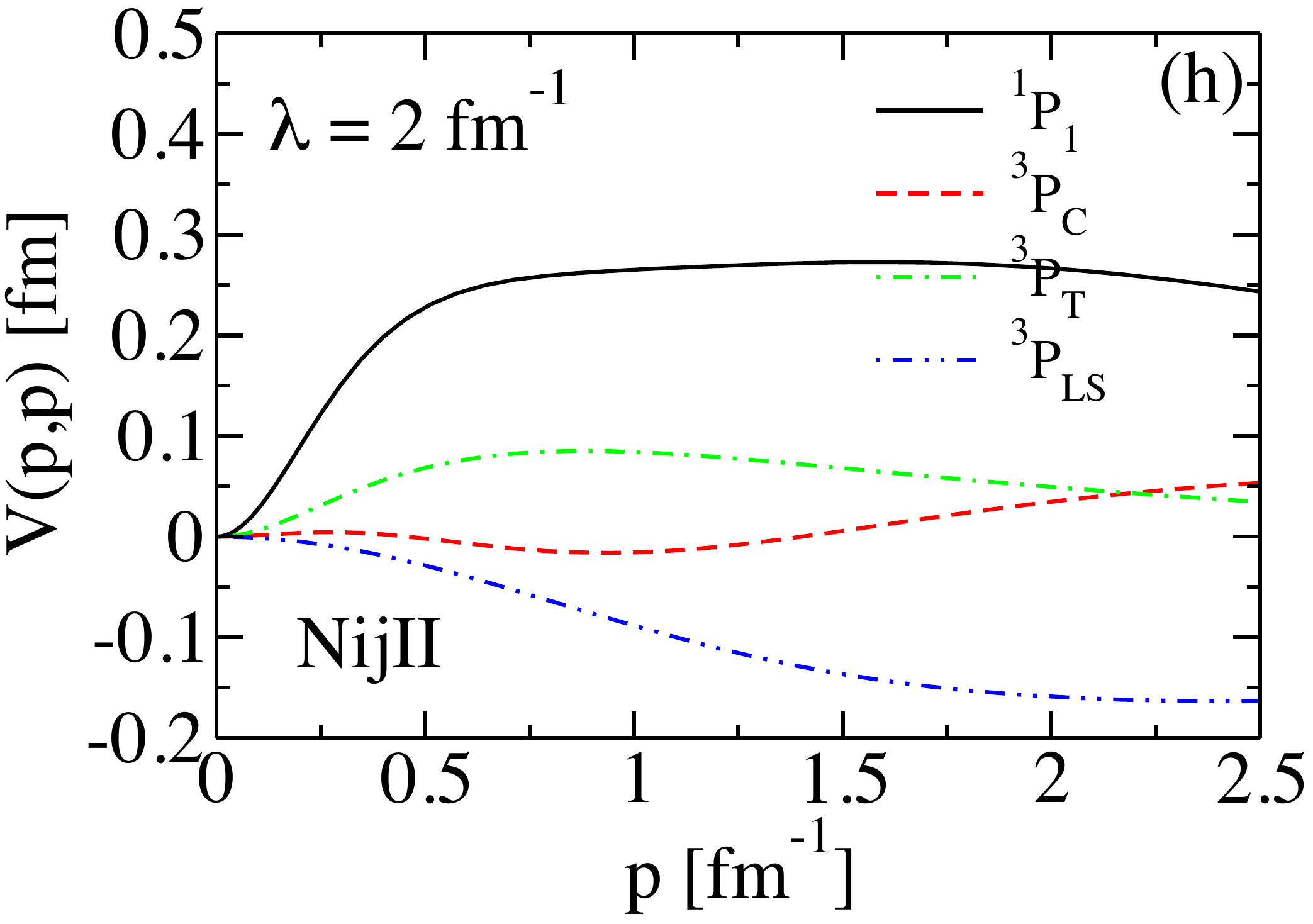}
\includegraphics[height=4cm,width=5cm]{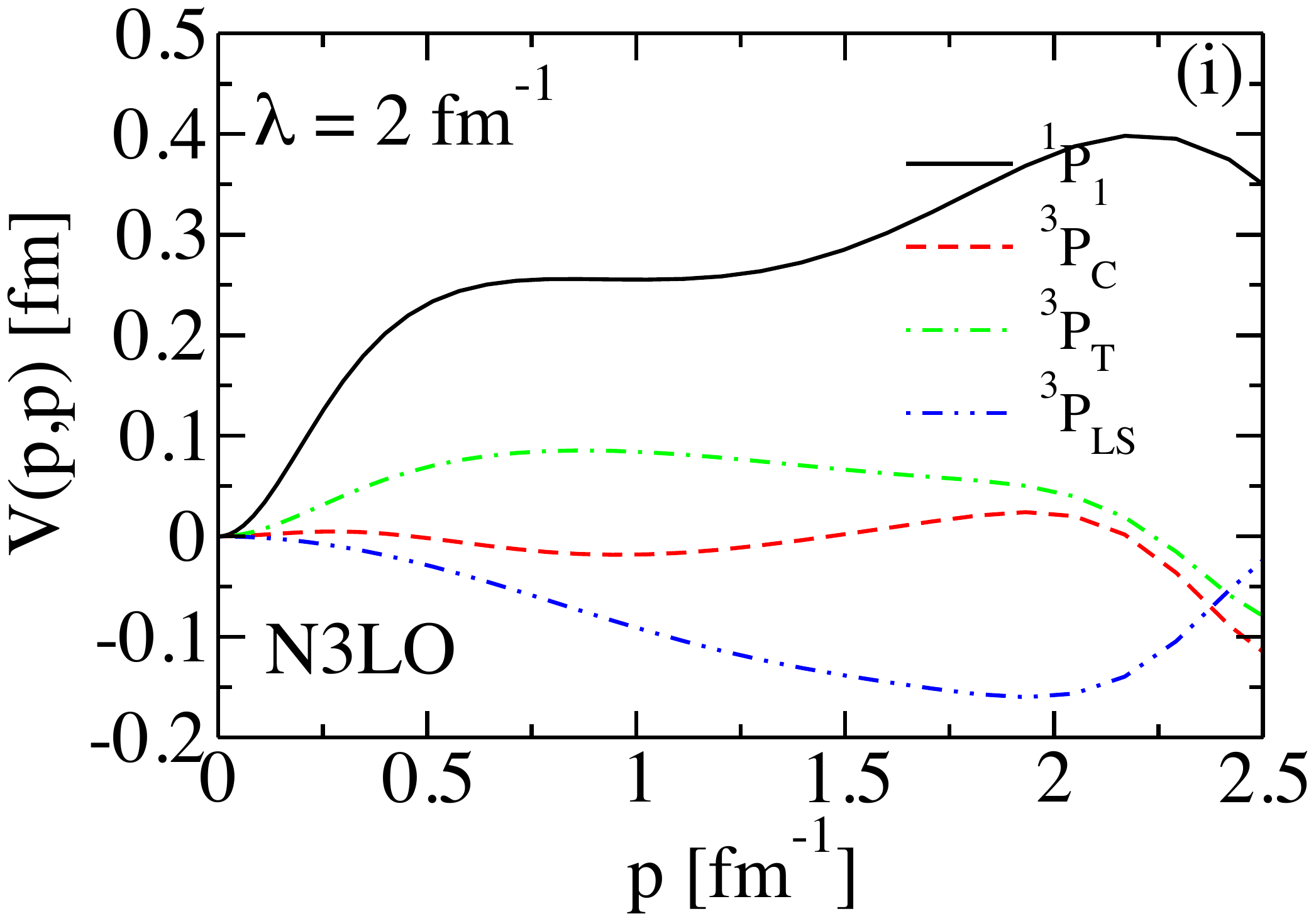} \\
\includegraphics[height=4cm,width=5cm]{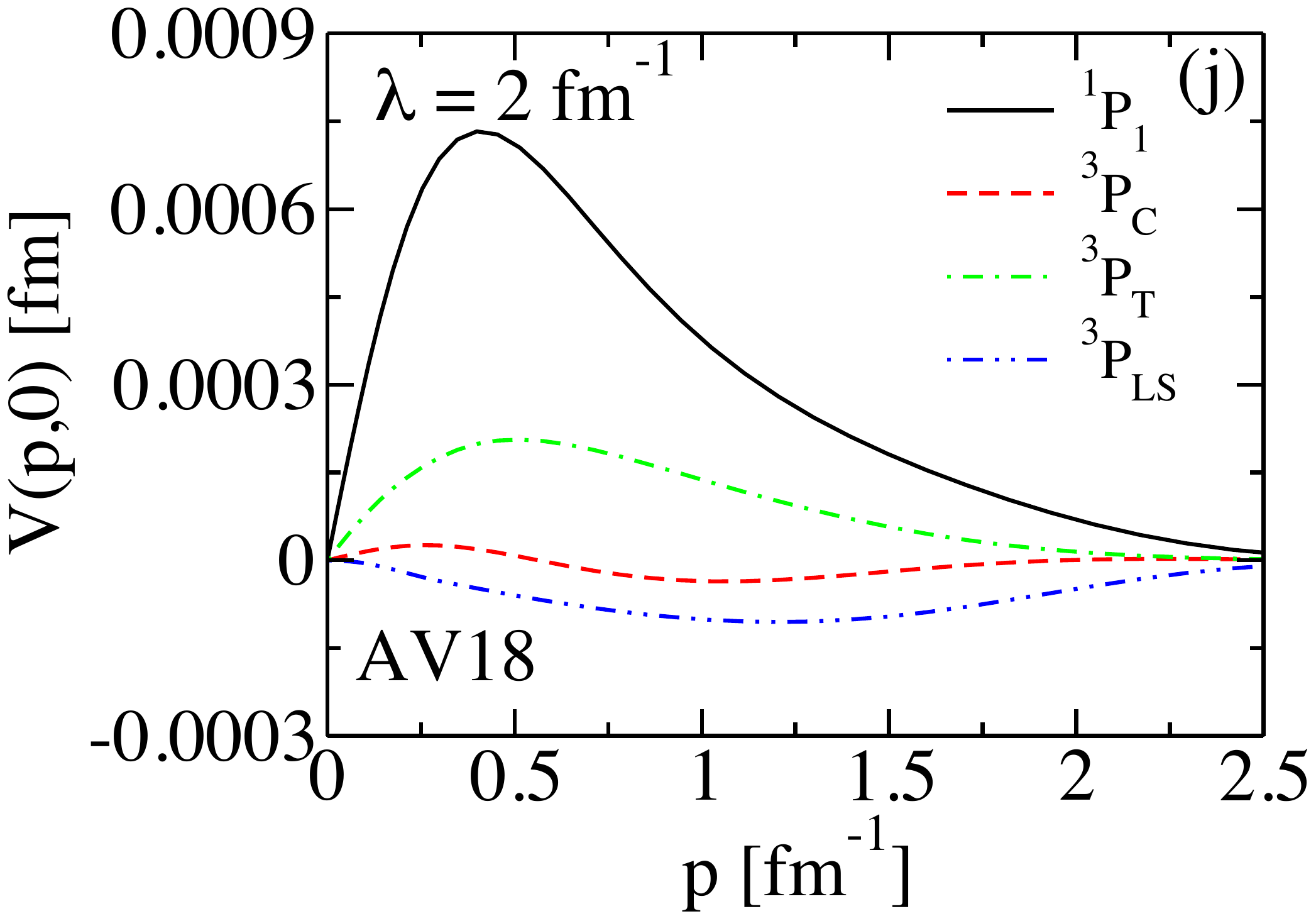}
\includegraphics[height=4cm,width=5cm]{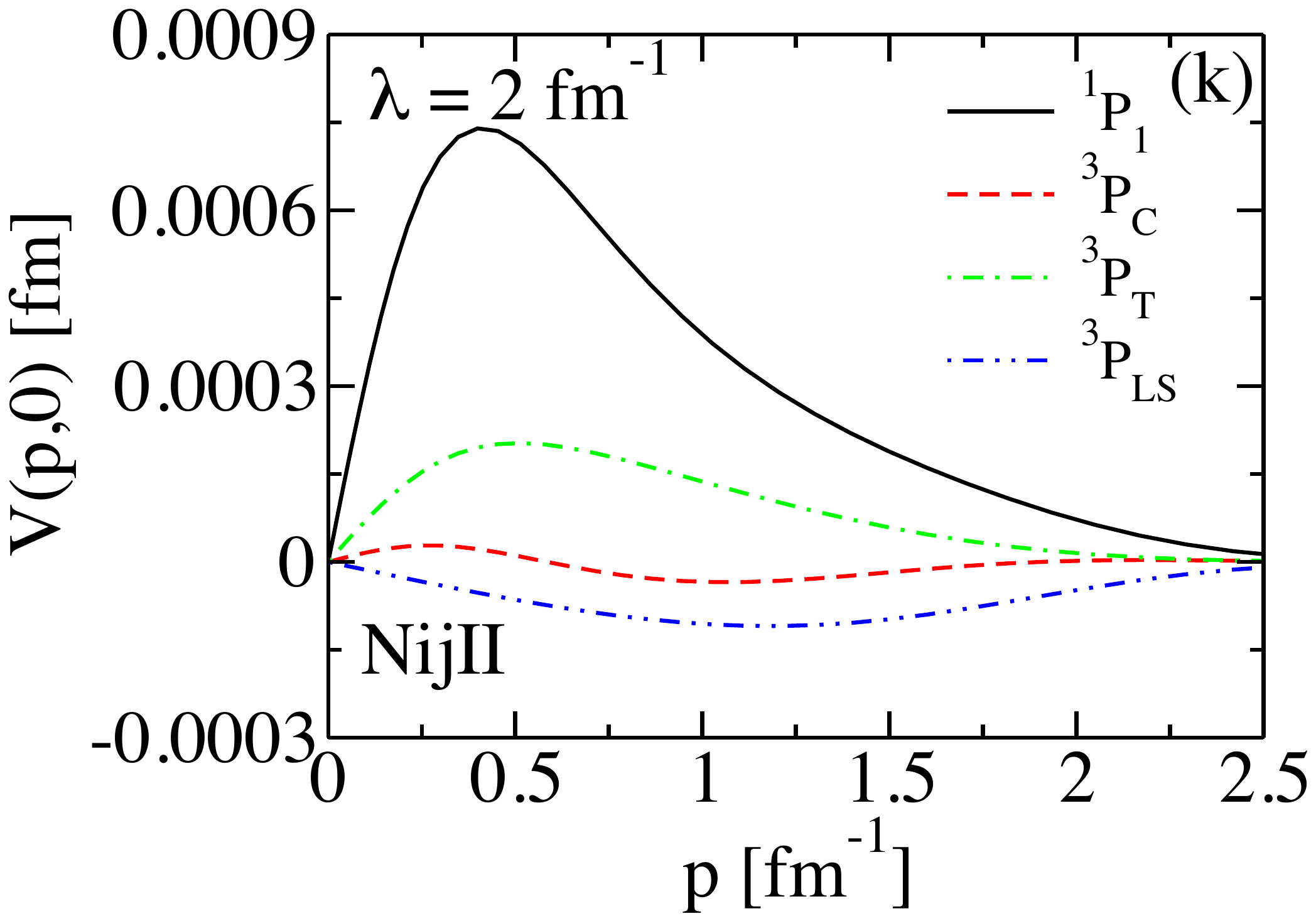}
\includegraphics[height=4cm,width=5cm]{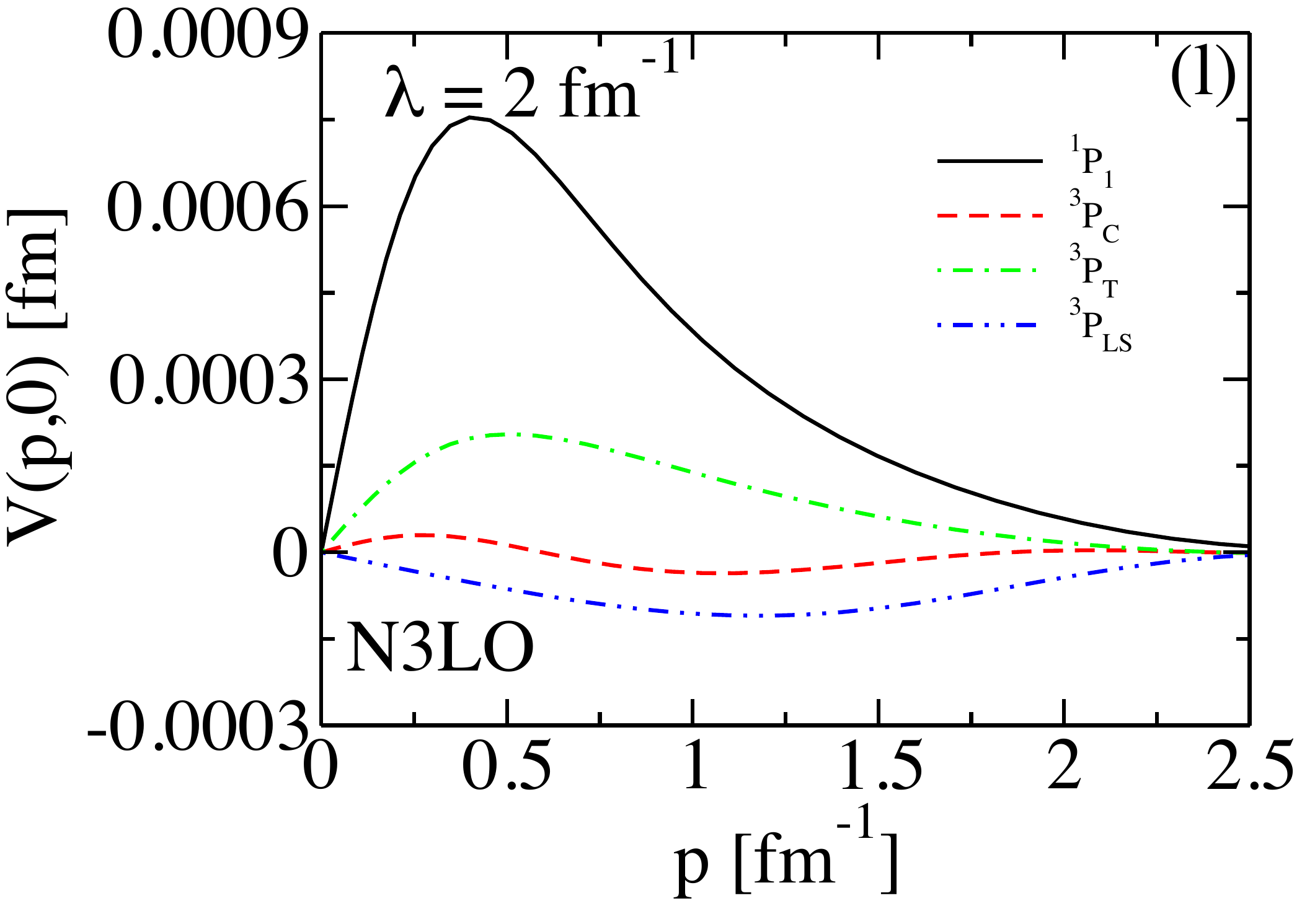}
\end{center}
\caption{(Color online) Comparison between the diagonal $V(p,p)$ and fully off-diagonal $V(p,0)$ matrix-elements of the SRG-evolved potentials for the S and P waves as a function of the CM momentum $p $(in ${\rm fm}^{-1}$) for a similarity cutoff $\lambda= 2~{\rm fm}^{-1}$. Left panels: Argonne AV18 potential~\cite{Wiringa:1994wb}. Central panels: Nijmegen II potential~\cite{Stoks:1994wp}. Right panels: Chiral N3LO potential of Entem e Machleidt~\cite{Entem:2003ft}.}
\label{fig:AV18-NijmII-N3LO}
\end{figure*}

\section{Numerical results }
\label{sec:num}

Here we provide some details on the numerical evolution. In order to
solve the Wegner's flow equation, we discretize the momentum-space
using a gaussian grid of $N$ mesh-points. This leads us to a system of
$N^2$ non-linear coupled first-order differential equations. The
system can then be solved with an implementation of a fifth-order
Runge-Kutta algorithm with adaptive step. In this work we use $N=200$
points.

Since the potentials we are evolving through the SRG are all regular,
we set an ultraviolet cutoff at $\Lambda = 30~\rm{fm}^{-1}$ which is
beyond the point where regulated potentials vanish.  In principle, one
could take a larger cutoff value provided the number of points in the
grid are enough to ensure the convergence of the Runge-Kutta
algorithm. If the cutoff is too large, one needs too many points in
the grid and the number of coupled equation becomes too large to be
solved in a reasonable time.

We illustrate our points for the Argonne AV18
potential~\cite{Wiringa:1994wb} which fits not only the $NN$ phase-shifts of the Nijmegen data base PWA~\cite{Stoks:1993tb}, but also the
deuteron elastic form factors and has been used quite often
successfully for nuclear structure calculations~\cite{Pieper:2001mp}.  The diagonal
matrix-elements $V (p,p) $ and fully off-diagonal matrix-elements
$V(p,0)$ are depicted at Fig.~\ref{fig:AV18-diag} and
Fig.~\ref{fig:AV18-off-diag} respectively. We compare the initial
potentials and the SRG potentials evolved to $\lambda= 2~{\rm fm^{-1}}$ and
$\lambda= 1~{\rm fm^{-1}}$. As we see, already above $\lambda= 2~{\rm
  fm^{-1}}$ the potentials for the $^1S_0$ and $^3S_1$ waves cross. This is an indication
that Wigner symmetry works very well around that scale. This trend is
also observed in even partial waves such as D and G, where the effect
of SRG evolution becomes less important as the angular momentum is
increased. On the other hand, odd partial waves provide hints of the
Serber symmetry as one sees that the $^1L $ potential is much larger
than the $^3L_C$ combination.

A comparison between the Argonne AV18 potential~\cite{Wiringa:1994wb}, the
Nijmegen II potential \cite{Stoks:1994wp} and the chiral N3LO
potential of Entem and Machleidt~\cite{Entem:2003ft} is presented in
Fig.~\ref{fig:AV18-NijmII-N3LO} for the similarity cutoff $\lambda=
2~{\rm fm}^{-1}$ and for the S- and P-waves. As we see, there is some
degree of universality, as one might expect since these potentials
are phase equivalent, although the strength is distributed
differently, particularly in the N3LO chiral potential case.

As we see, the potential for the $^3S_1$ wave changes rather dramatically when going from
$\lambda=2~{\rm fm}^{-1}$ to $\lambda=1~{\rm fm}^{-1}$ in contrast to
the other waves where apparently a much more stable result is
obtained. This feature is due to the use of the simple generator
$\eta_s=[T,H_s]$, where T is the kinetic energy.  In Wegner's original
formulation~\cite{wegner1994flow} the generator $
\eta=[D_s,H_s]$ was used, where $D_s$ is the diagonal part of $H_s$.  Actually,
as shown by Glazek and Perry~\cite{Glazek:2008pg} the generator
$\eta_s=[T,H_s]$ can produce divergencies in theories with
bound-states, as in the case of the $NN$ interaction in the $^3 S_1- ^3D_1$
channel, limiting how far the transformation can be run. When the SRG
cutoff $\lambda= 1/s^\frac14 $ approaches the momentum scale at which a
bound-state emerges, the strength of the SRG evolved interactions
increases significantly.  This happens because the transformation
tends to move the bound-state eigenvalue to the low-momentum part of
the Hamiltonian's diagonal, forcing the interaction to grow in order
to maintain the bound-state at the right value.  In the appendix we
show that actually in the limit $\lambda \to 0$ the SRG equation
becomes stationary when the SRG potential becomes the reaction matrix
which in the single-channel case becomes
\begin{eqnarray}
\lim_{\lambda \to 0} V_\lambda (p,p) = - \frac{\tan \delta (p)}{p} \, ,
\label{eq:Vto0}
\end{eqnarray}
and a similar equation is obtained in the coupled-channel case. Note that this
diverges when the phase shift goes through $90^o$, a situation which
only occurs for the $^3S_1$-channel for $p \sim 100~{\rm MeV}$.  On
the other hand, there is no divergence problem when using Wegner's
generator $\eta=[D_s,H_s]$, because the bound-state eigenvalue is kept
at the natural momentum scale as the SRG cutoff is lowered suggesting
that actually the generator initially proposed by Wegner has better
infrared properties. In this sense it would be interesting to check
the present results for the Wegner flow case. The advantages of other
SRG generators have been considered recently~\cite{Li:2011sr}.

Finally, we can fine tune the SRG cutoff so that we obtain the best
possible fulfillment of the Wigner symmetry, which is slightly above
$\lambda$. We call this $\lambda_{\rm Wigner}$. This, of course, would
have far reaching consequences for the analysis of finite nuclei on
the basis of symmetry.

\begin{figure*}[tbc]
\begin{center}
\includegraphics[height=6cm,width=7cm]{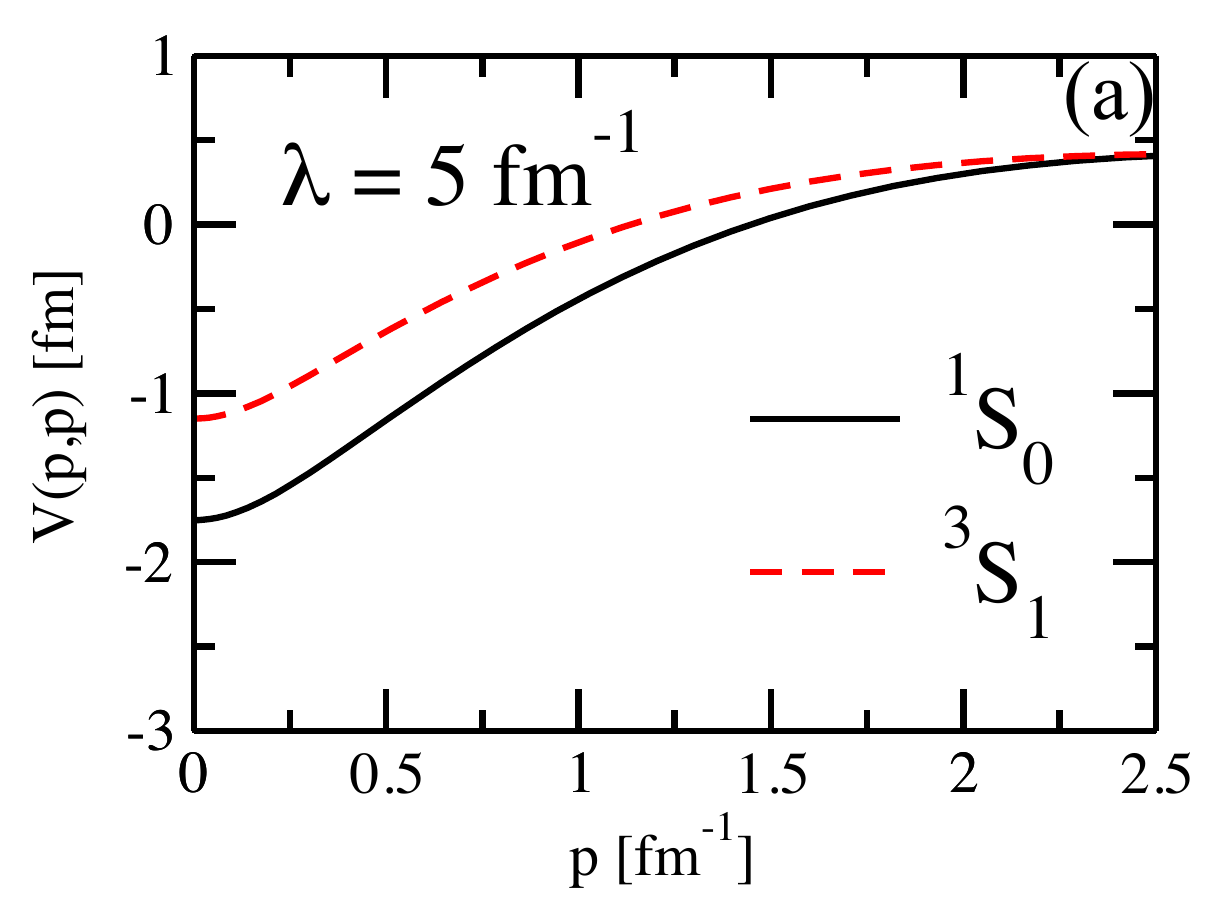}
\includegraphics[height=6cm,width=7cm]{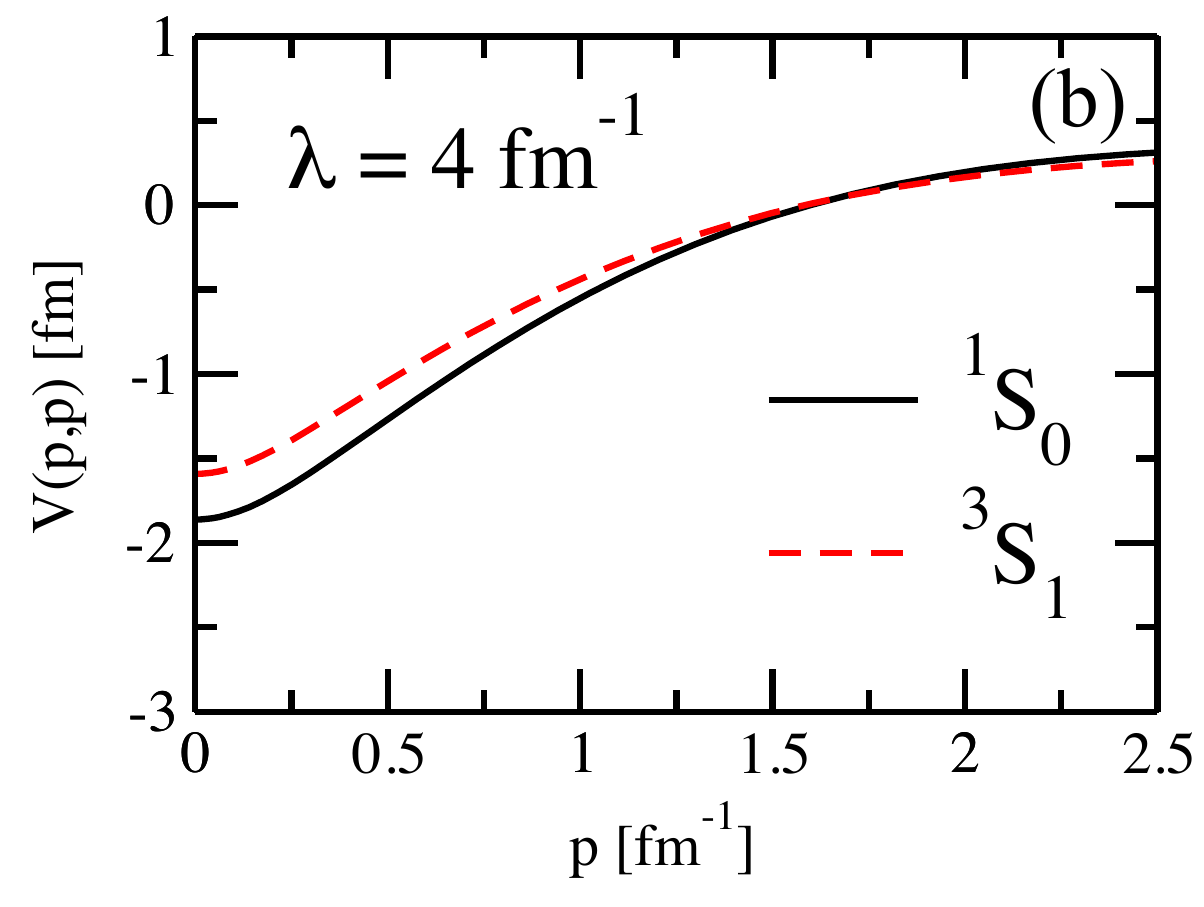} \\
\includegraphics[height=6cm,width=7cm]{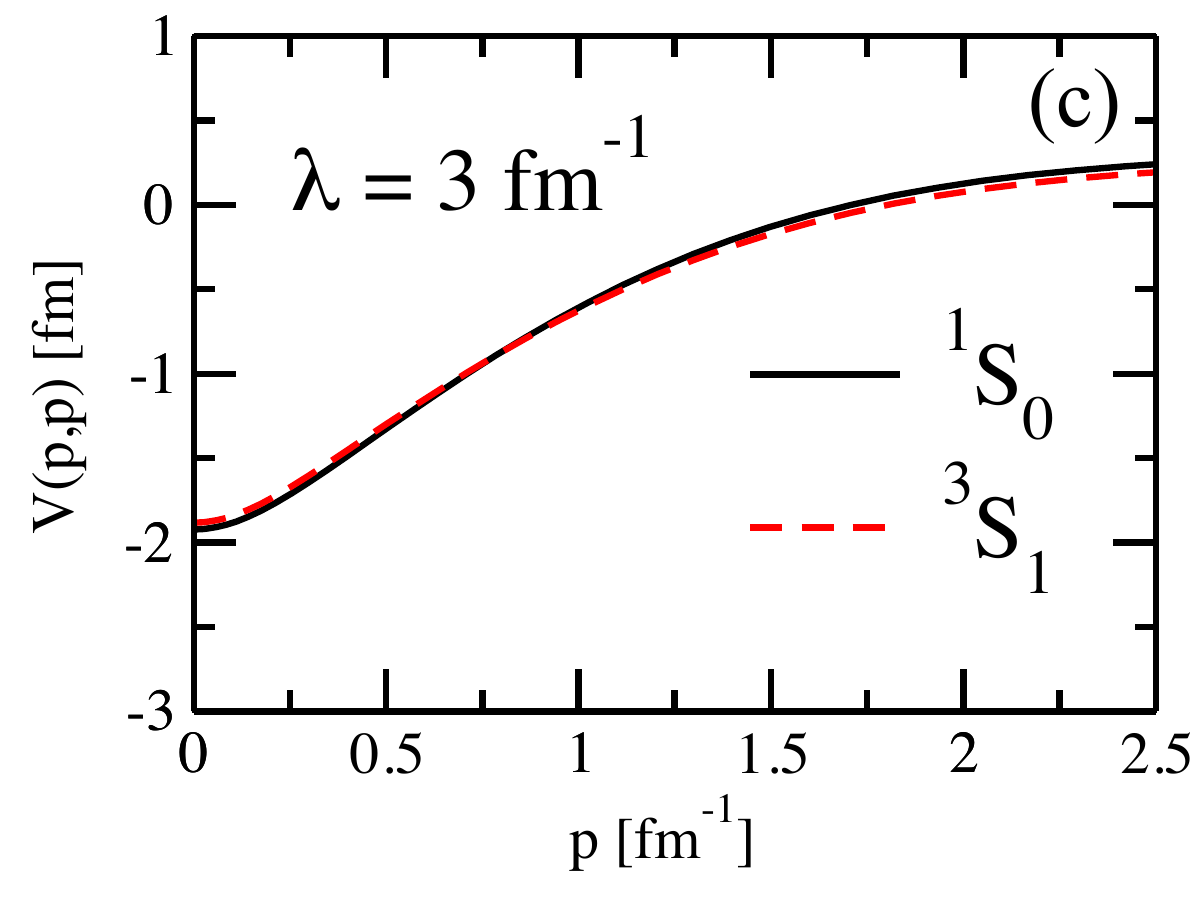}
\includegraphics[height=6cm,width=7cm]{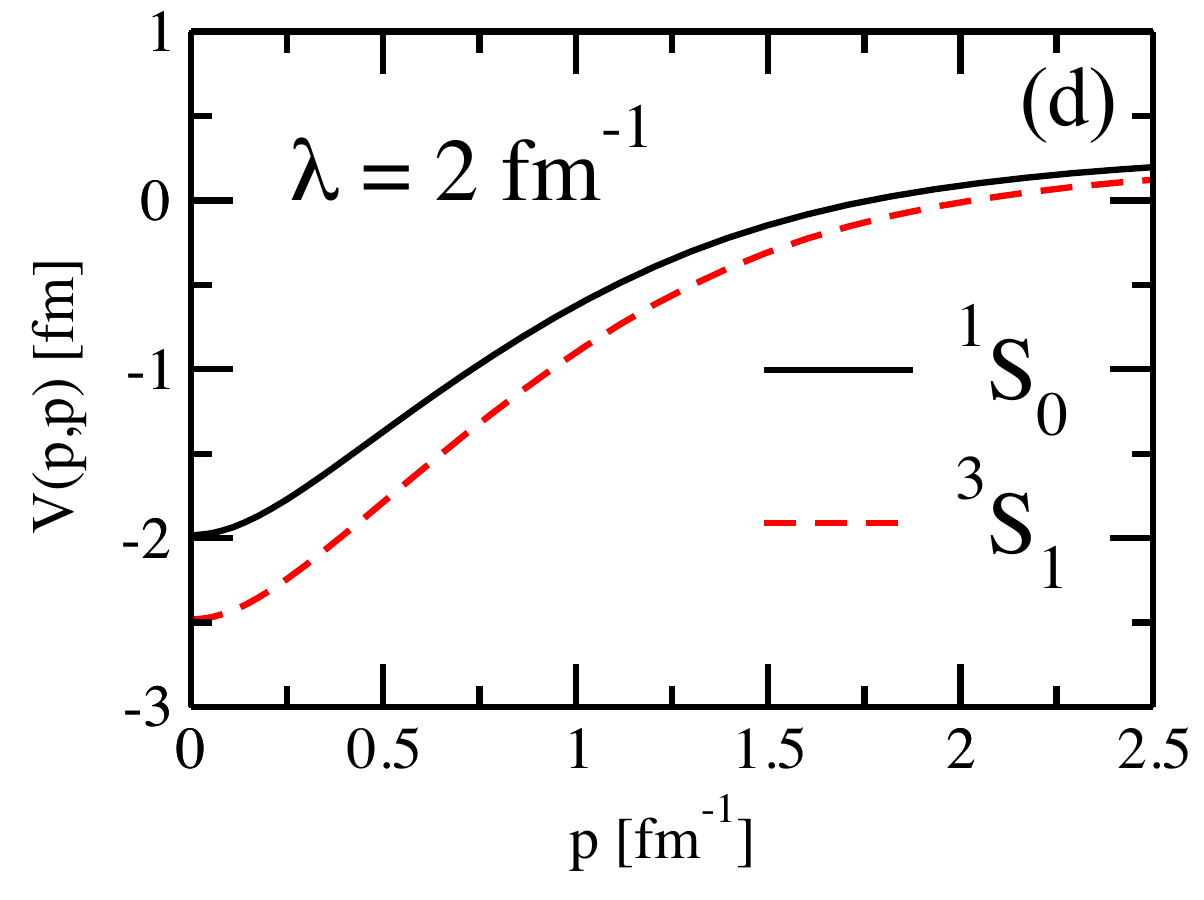}   \\
\includegraphics[height=6cm,width=7cm]{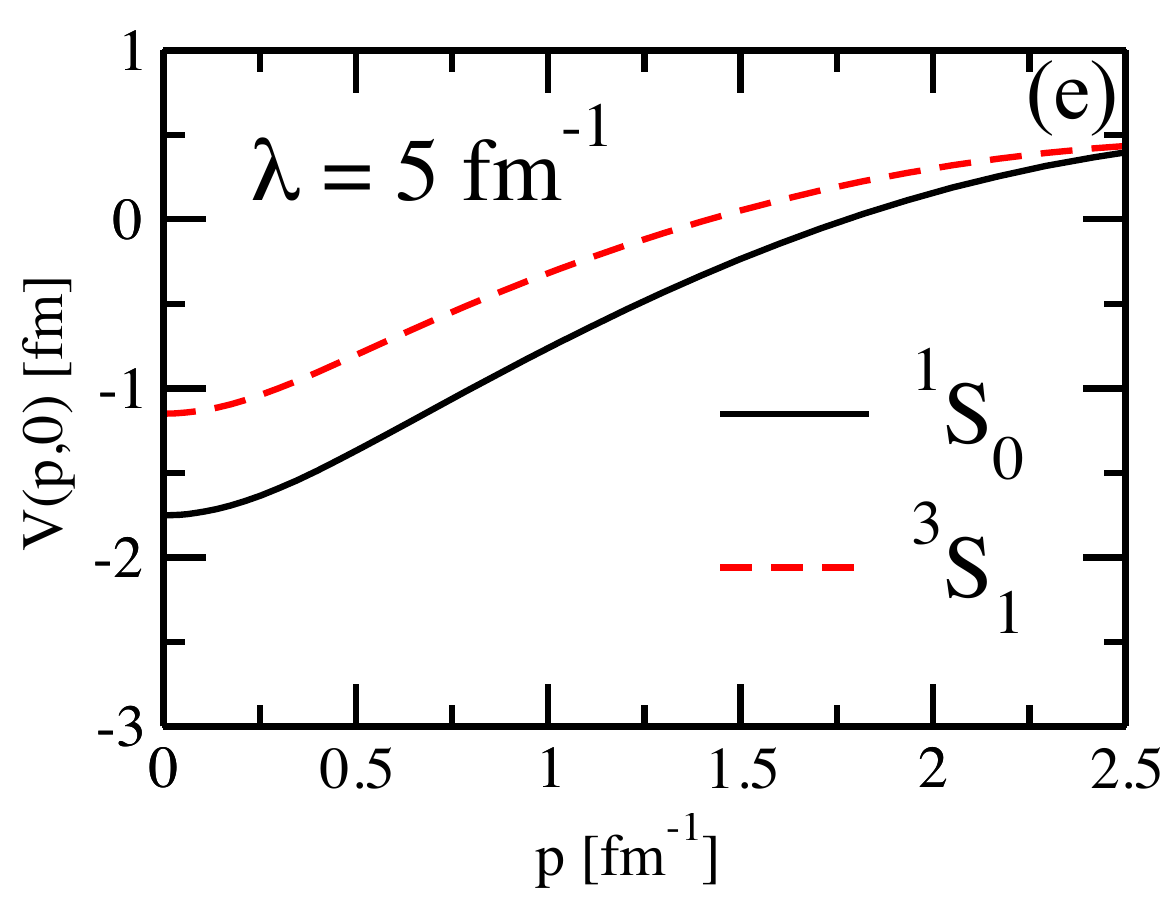}
\includegraphics[height=6cm,width=7cm]{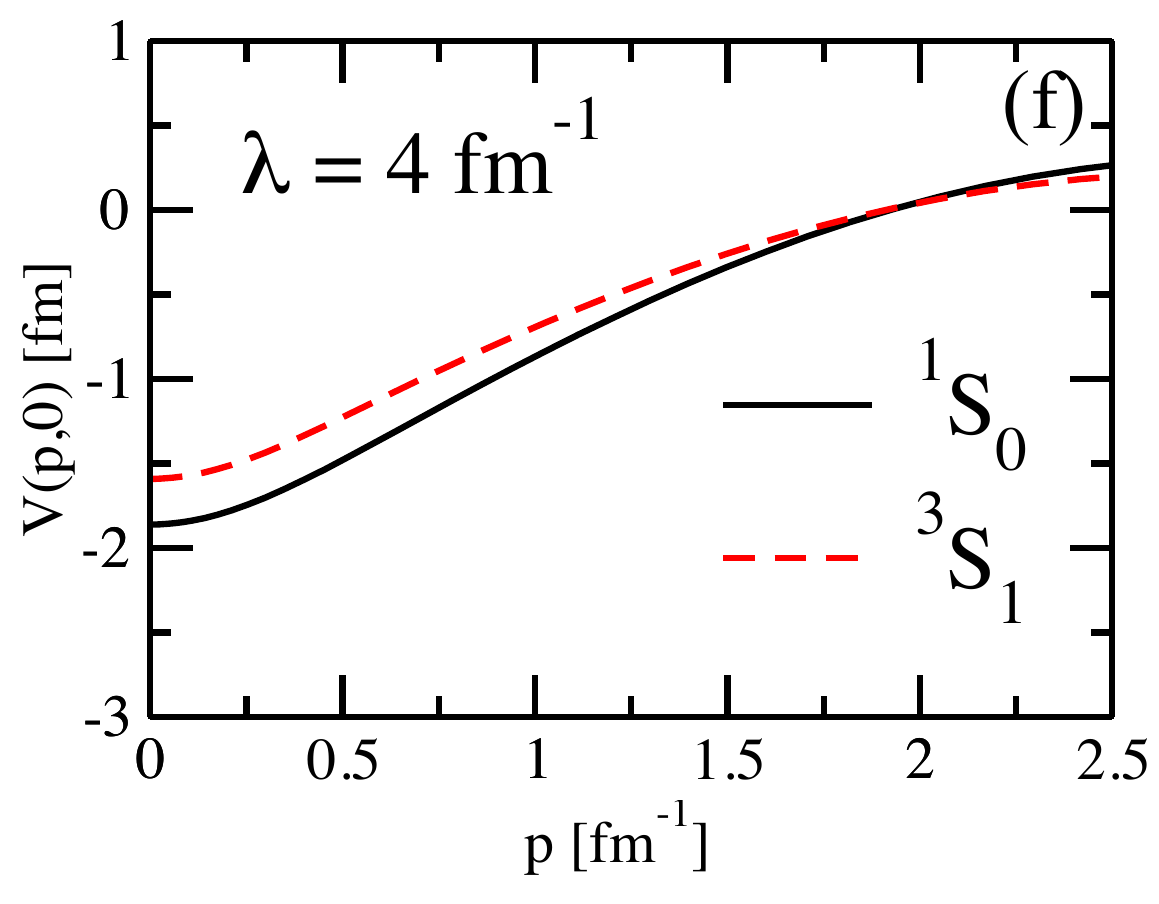}  \\
\includegraphics[height=6cm,width=7cm]{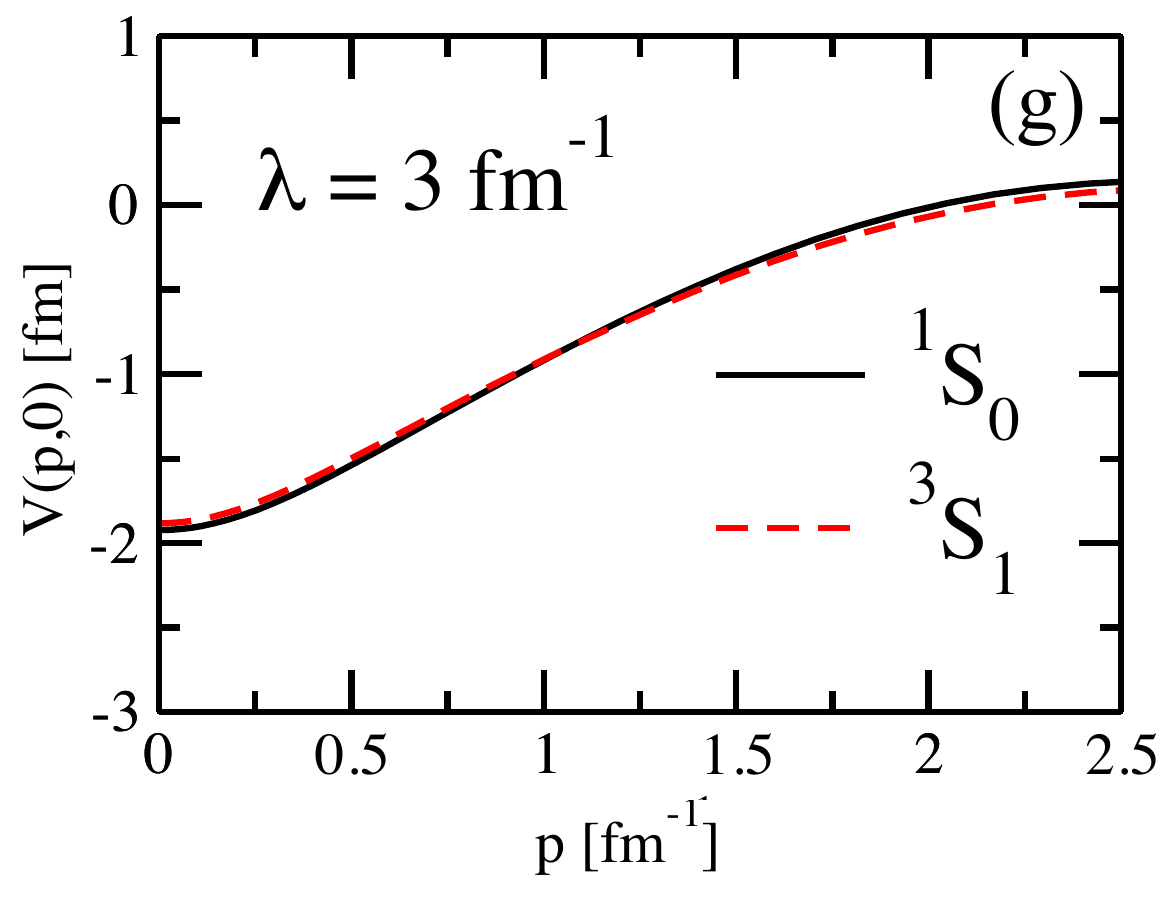}
\includegraphics[height=6cm,width=7cm]{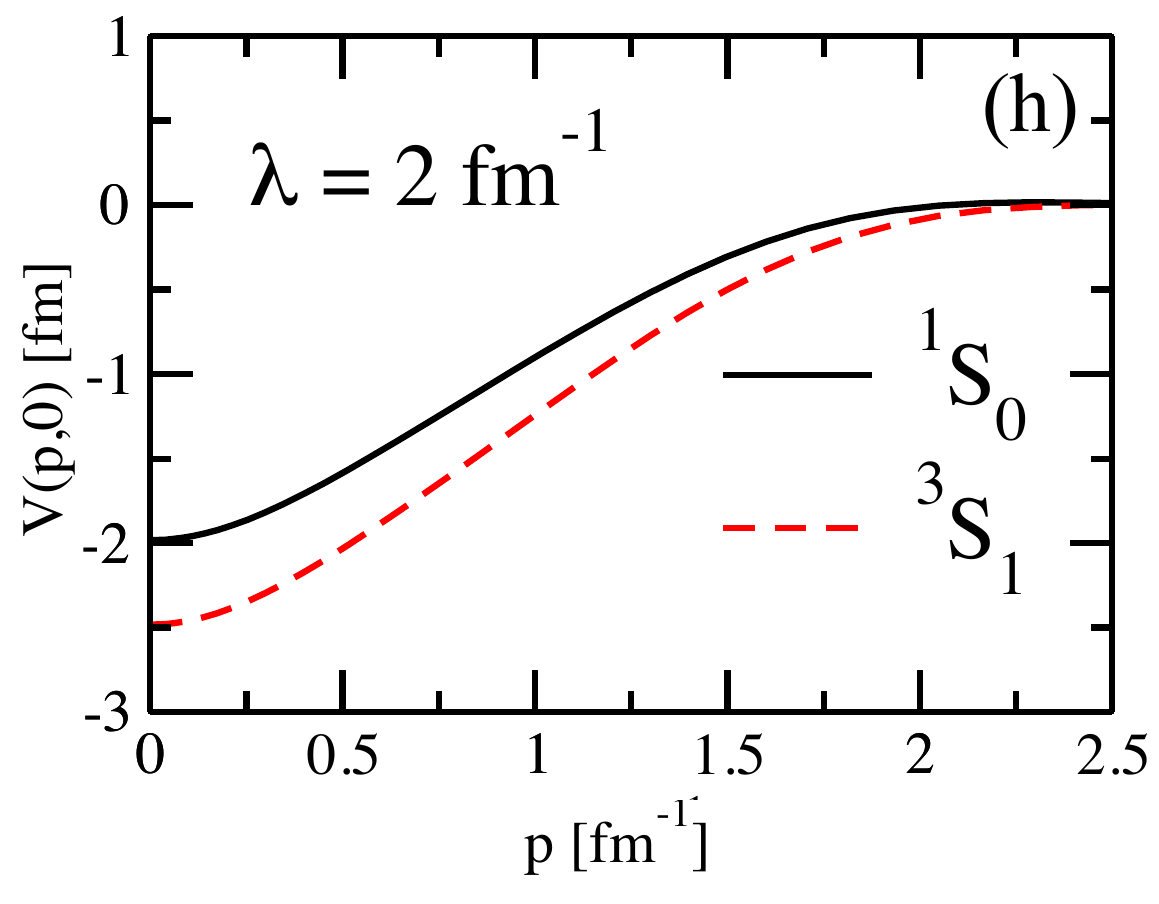}
\end{center}
\caption{(Color online) Comparison between diagonal, $V(p,p)$, and fully off-diagonal, $V(p,0)$,
matrix-elements of the SRG-evolved potentials for the
$S$-waves (in {\rm fm}) as a function of the CM momentum $p$ (in ${\rm fm}^{-1}$),
showing that the Wigner similarity cutoff is $\lambda_{\rm Wigner} \approx 3~ {\rm fm}^{-1}$.
We use the Argonne AV18 potential as the initial condition~\cite{Wiringa:1994wb}.}
\label{fig:AV18-wigner-S}
\end{figure*}

Our results for the $S$-waves are presented in
Fig.~\ref{fig:AV18-wigner-S} where the similarity cutoff runs from
$5~{\rm fm}^{-1}$ to $2~{\rm fm}^{-1}$.  We can see that both
$S$-waves evolve in the same direction, becoming more attractive.  At
$\lambda_{\rm Wigner} = 3~{\rm fm}^{-1}$, the SRG-evolved interaction
in both $S$-waves are practically identical and for $\lambda = 2~{\rm
  fm}^{-1}$ the evolved potential in the triplet channel is stronger
than in the singlet case. As expected, the evolution in triplet state
is faster than in the singlet state. The deuteron pole is at an
imaginary  momentum scale $p_d \sim {\rm i}~0.23~{\rm fm}{^{-1}}$ while the pole
corresponding to the $^1S_0$ virtual bound-state is at a much smaller
momentum scale $p_v \sim {\rm i}~0.04~ {\rm fm}{^-1}$. Thus, when evolving to
similarity cutoffs in the range we are considering, the enhancement in
the strength of the interaction that comes from using the simple
generator $\eta=[T,H(s)]$ is sensible only in the $^3S_1$ channel.  As
we see, at $\lambda_{\rm Wigner}$ the agreement between the $^3S_1$ and
$^1S_0$ SRG-evolved interactions is indeed remarkable both for the
diagonal and the fully off-diagonal elements. It is important to note that the
difference between the $^1S_0$ and the $^3S_1$ phase-shifts evaluated from
the initial unevolved potentials is reproduced at any rate through the SRG evolution, since the unitary SRG transformation changes the interactions (independently in each partial-wave channel) while preserving the corresponding phase-shifts.

\begin{figure*}[tbc]
\begin{center}
\includegraphics[height=5.5cm,width=7.5cm]{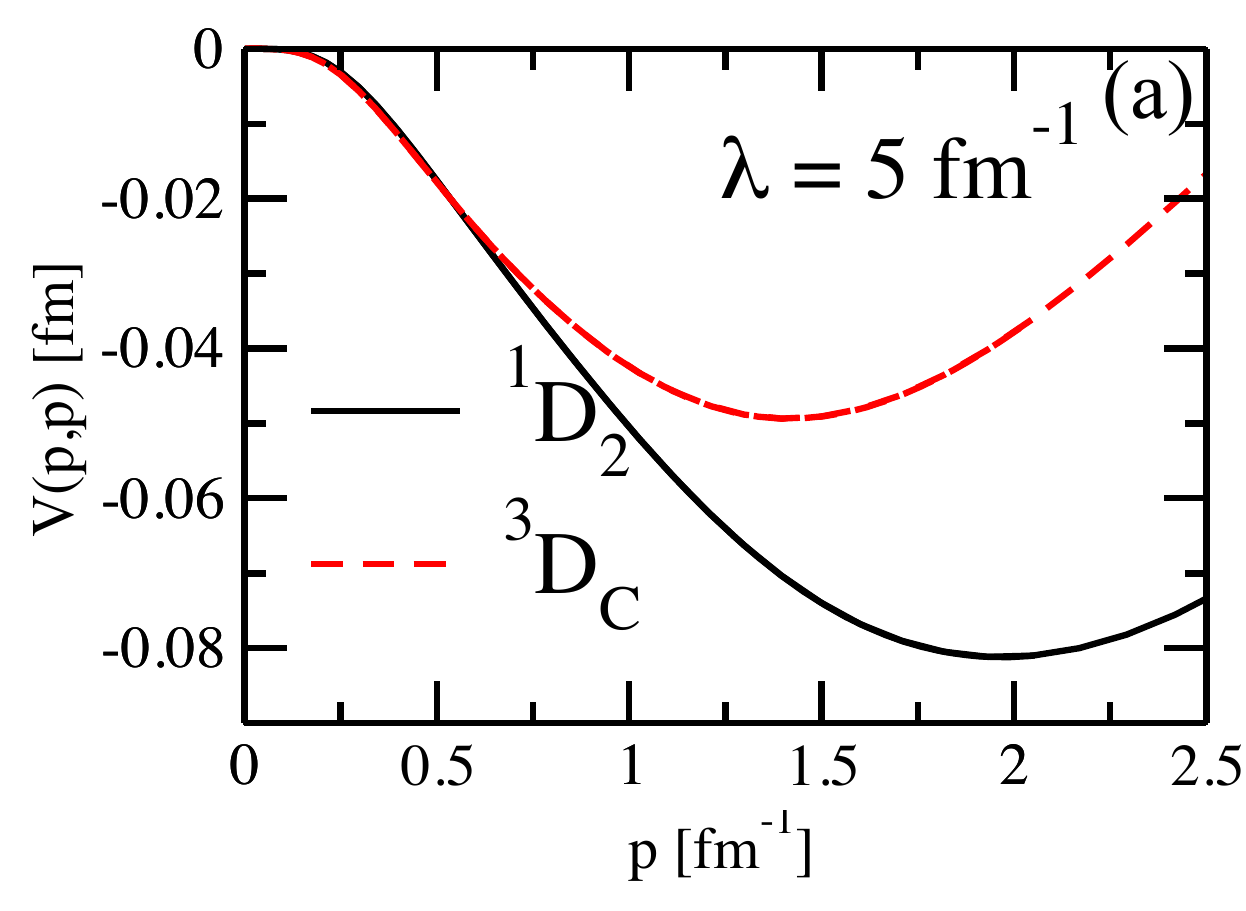}
\hspace{1cm}
\includegraphics[height=5.5cm,width=7.5cm]{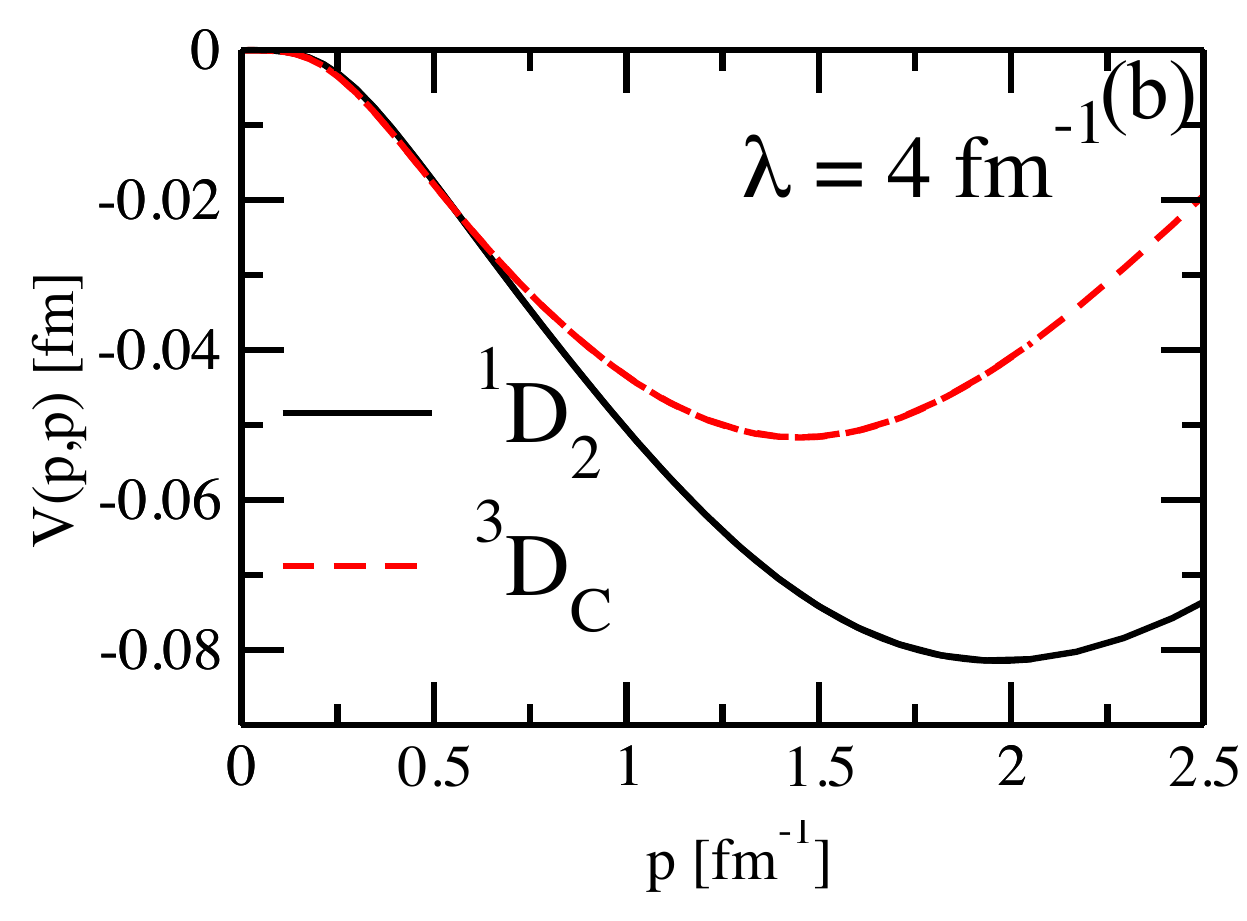} \\
\includegraphics[height=5.5cm,width=7.5cm]{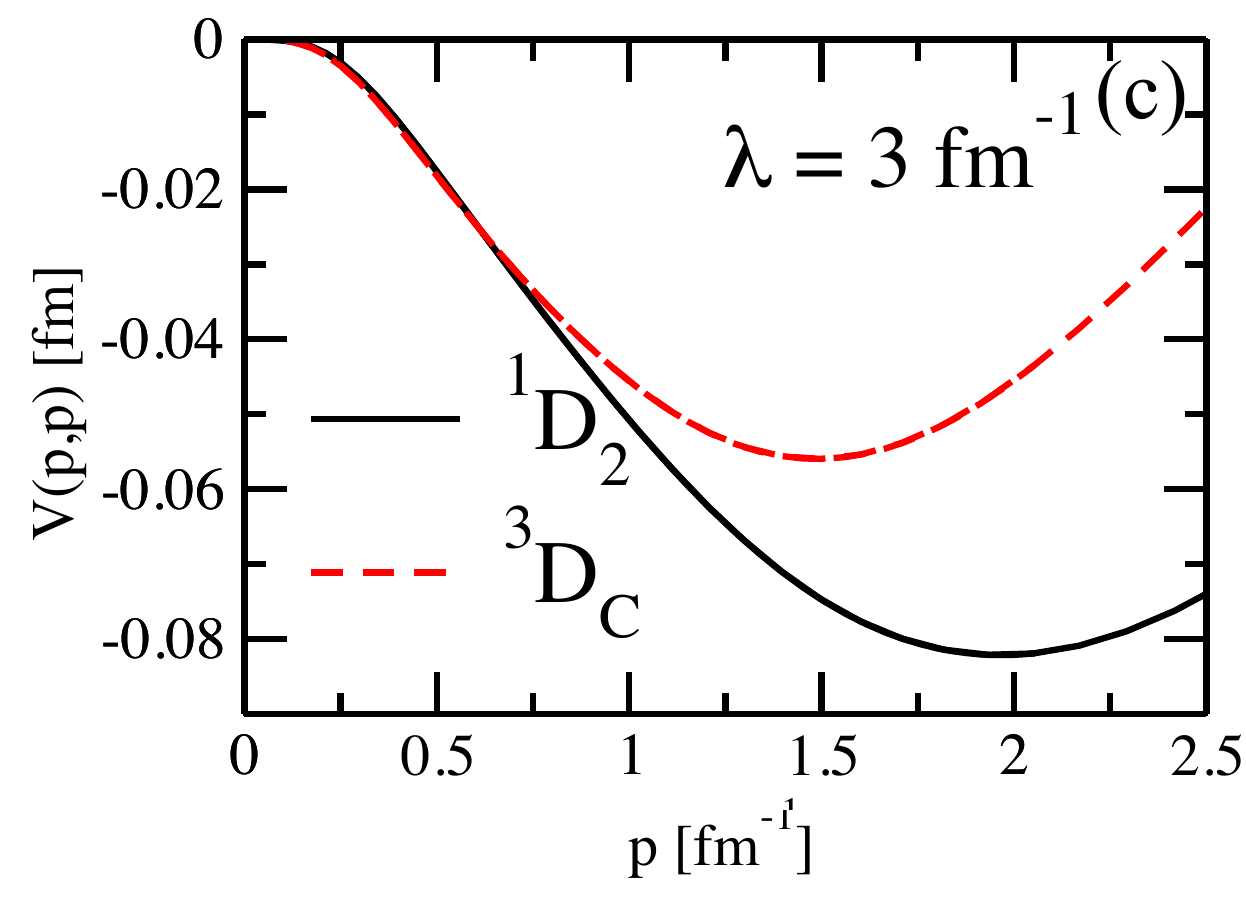}
\hspace{1cm}
\includegraphics[height=5.5cm,width=7.5cm]{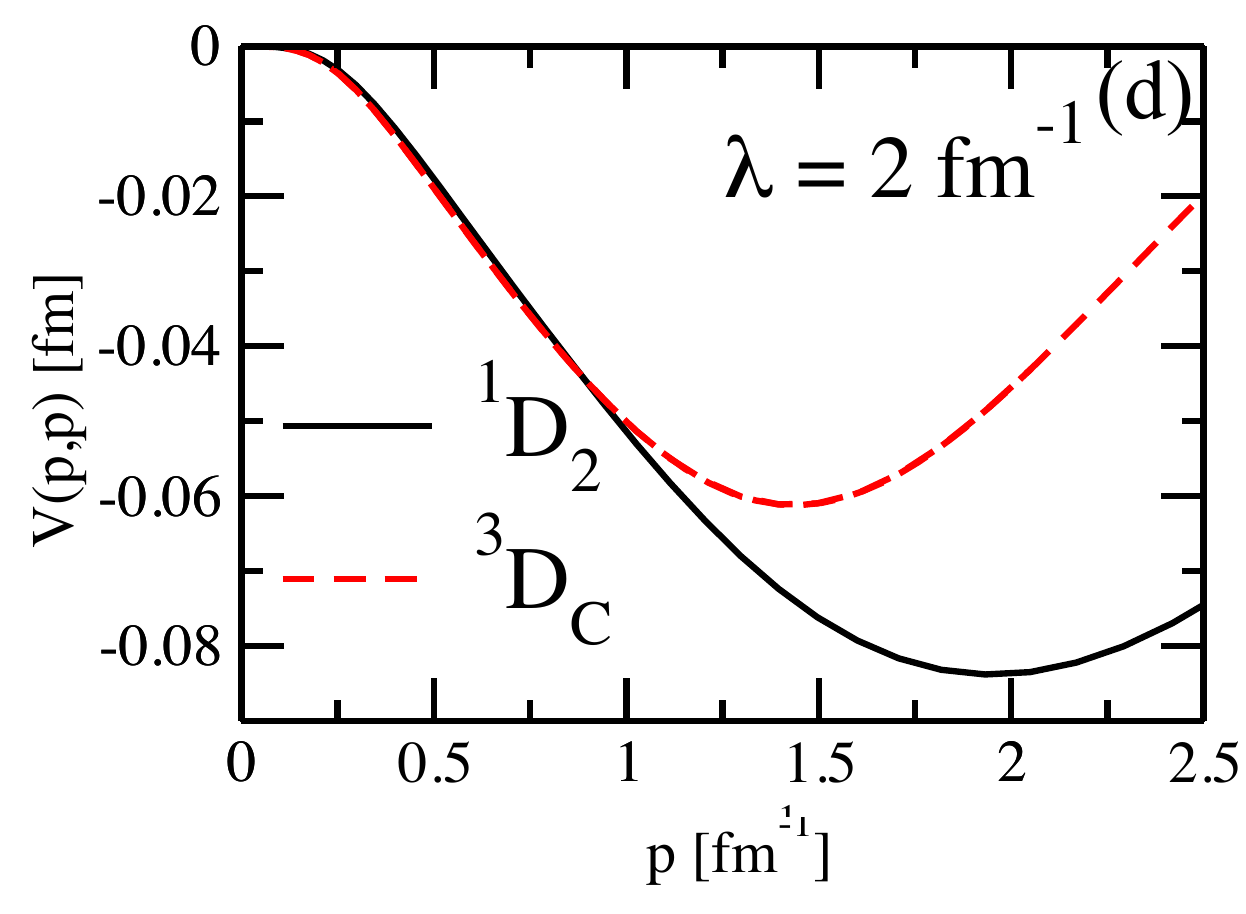}   \\
\hspace{0.5cm}
\includegraphics[height=6.5cm,width=8cm]{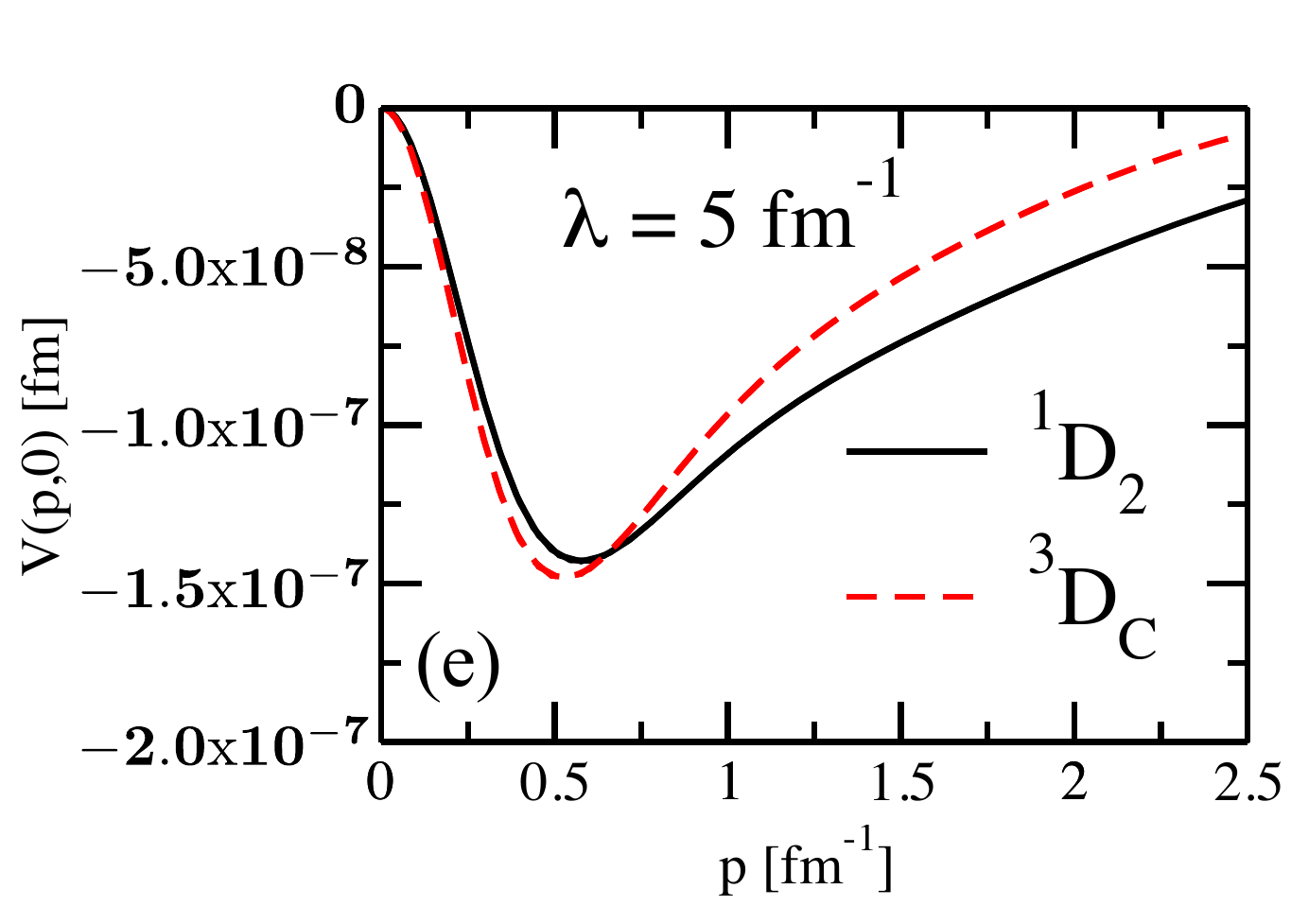}
\hspace{0.5cm}
\includegraphics[height=6.5cm,width=8cm]{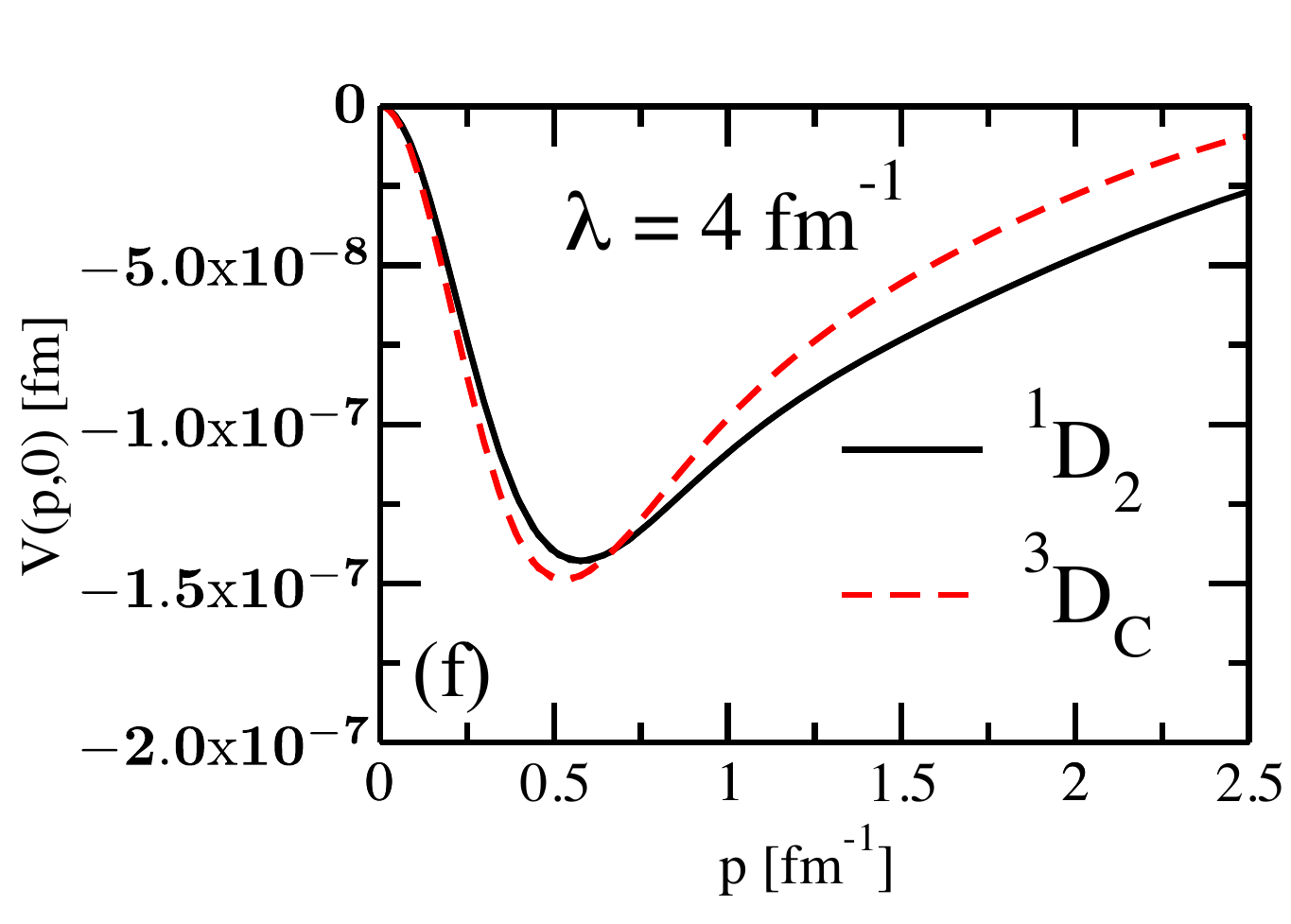}  \\
\includegraphics[height=6.5cm,width=8cm]{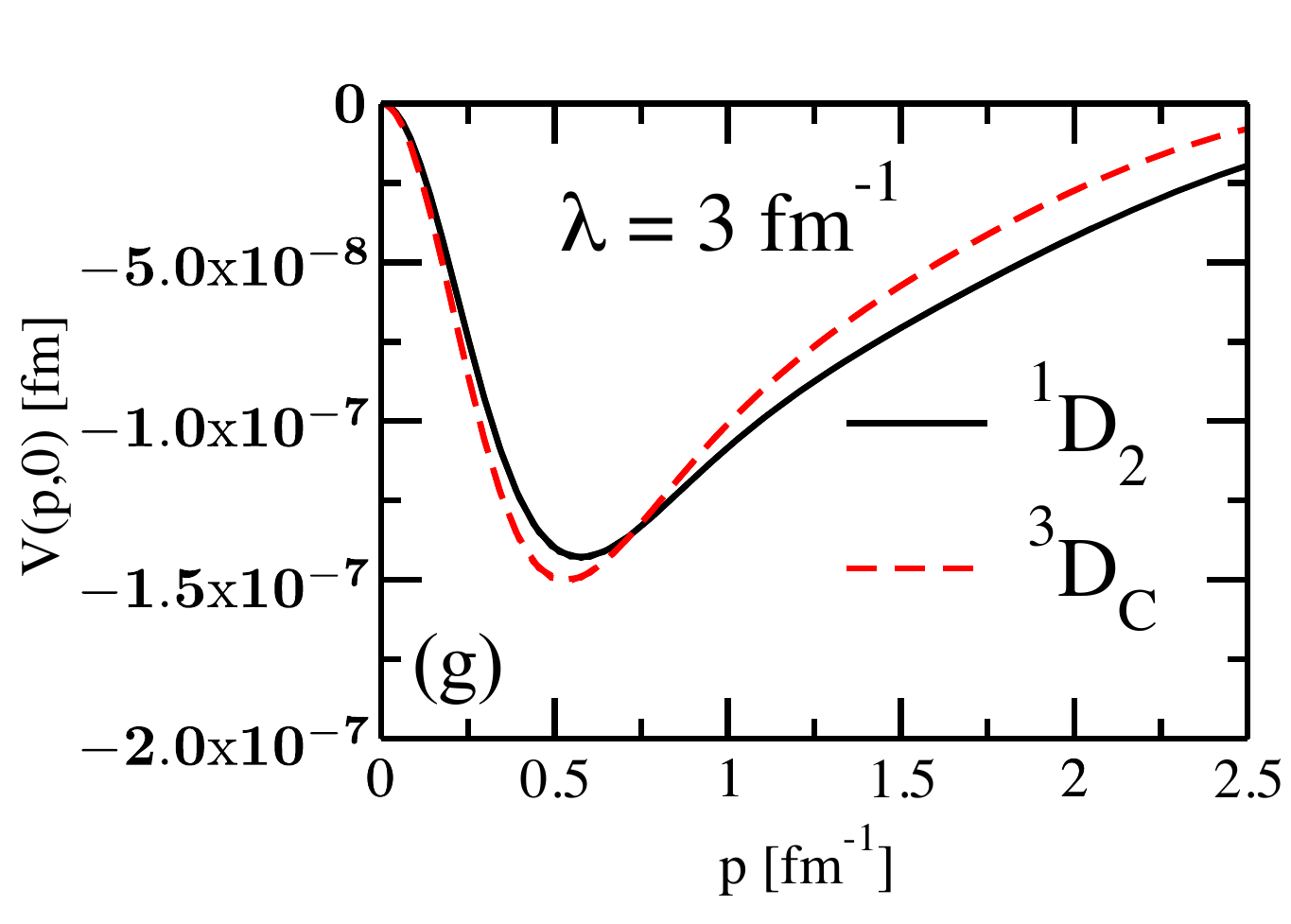}
\hspace{1.0cm}
\includegraphics[height=6.5cm,width=8cm]{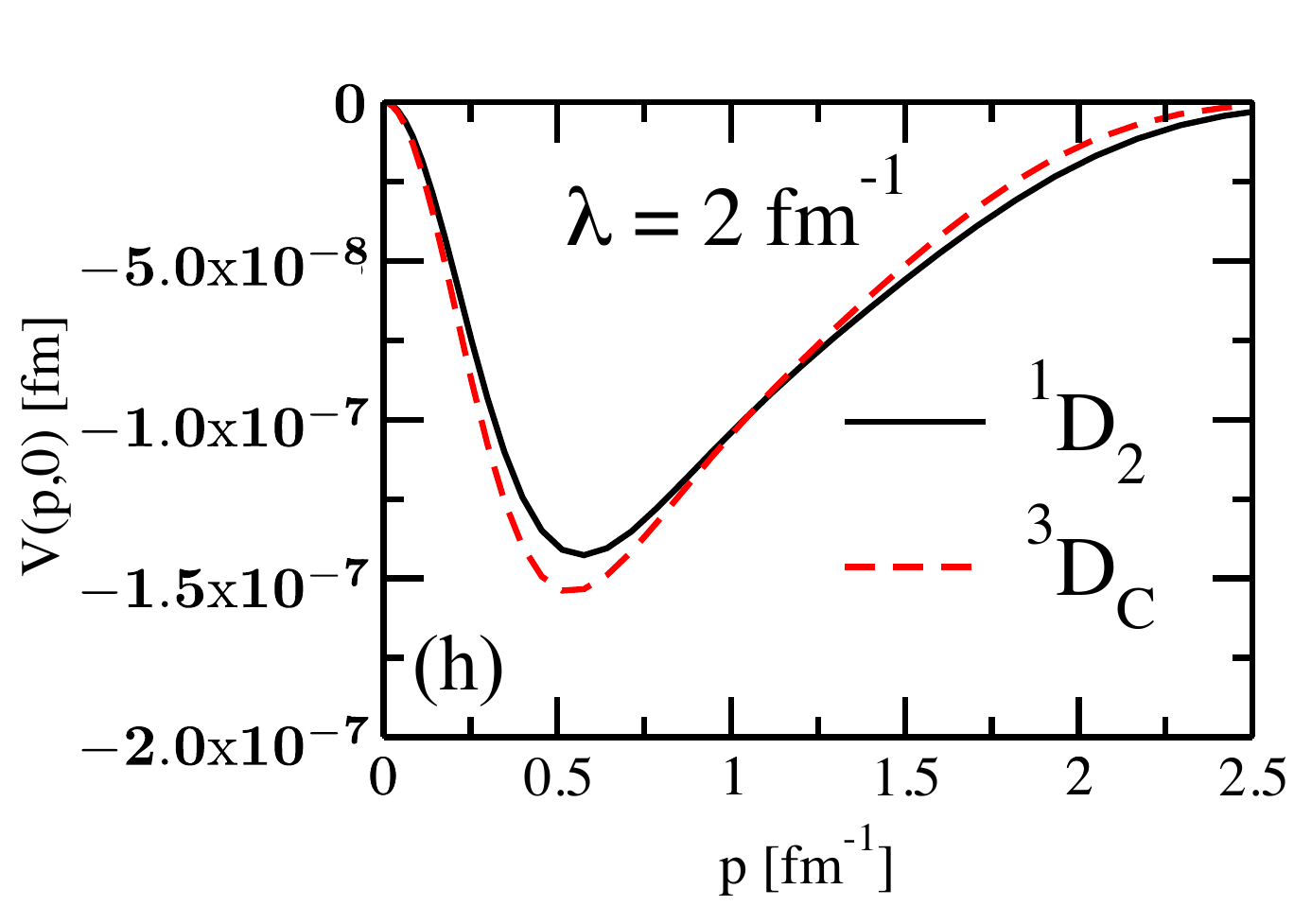}
\end{center}
\caption{(Color online) Comparison between diagonal, $V(p,p)$, and fully off-diagonal, $V(p,0)$,
matrix-elements of the SRG-evolved potentials for the
$D$-waves (in {\rm fm}) as a function of the CM momentum $p$ (in ${\rm fm}^{-1}$).}
\label{fig:AV18-wigner-D}
\end{figure*}

\begin{figure*}[tbc]
\begin{center}
\includegraphics[height=5.5cm,width=7.5cm]{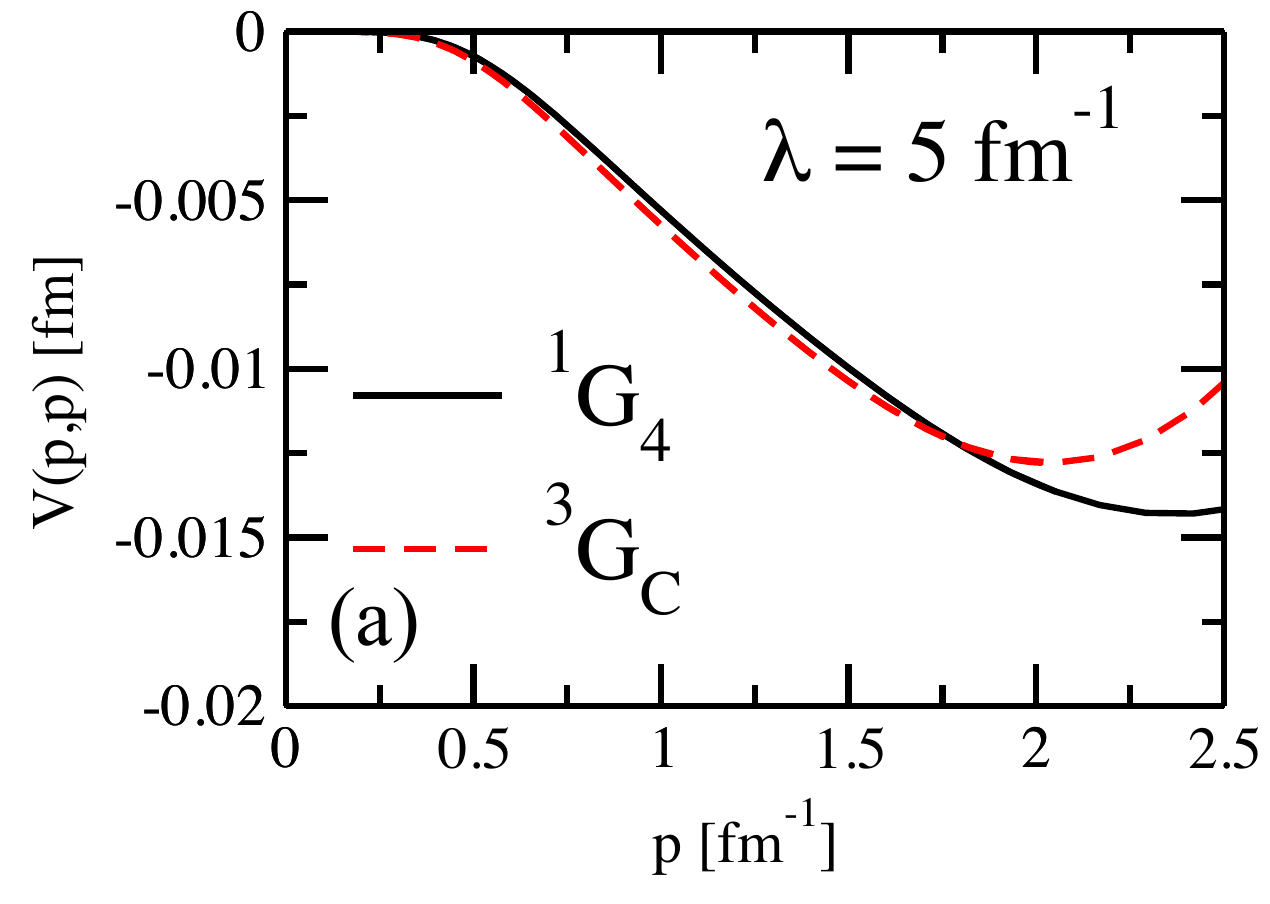}
\includegraphics[height=5.5cm,width=7.5cm]{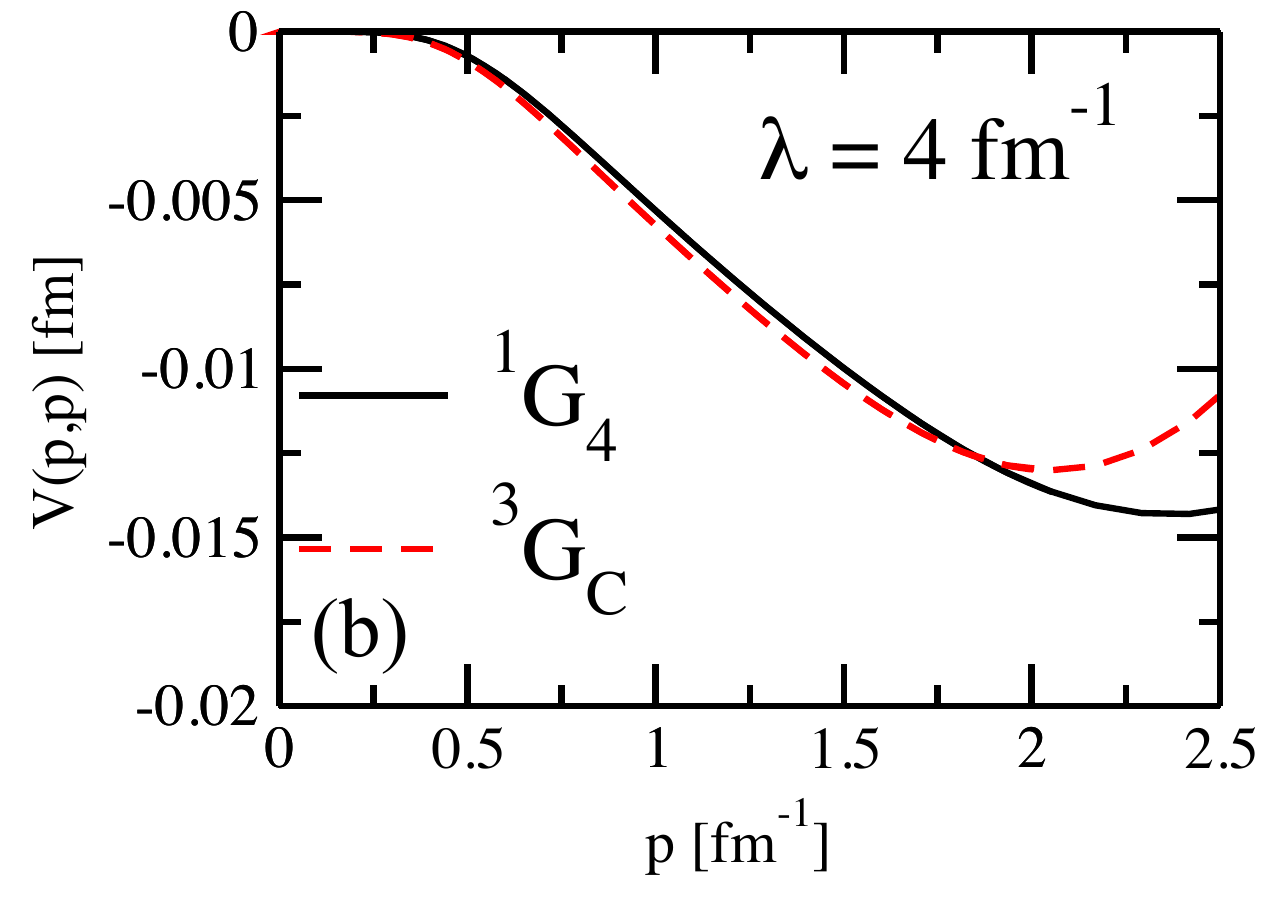} \\
\includegraphics[height=5.5cm,width=7.5cm]{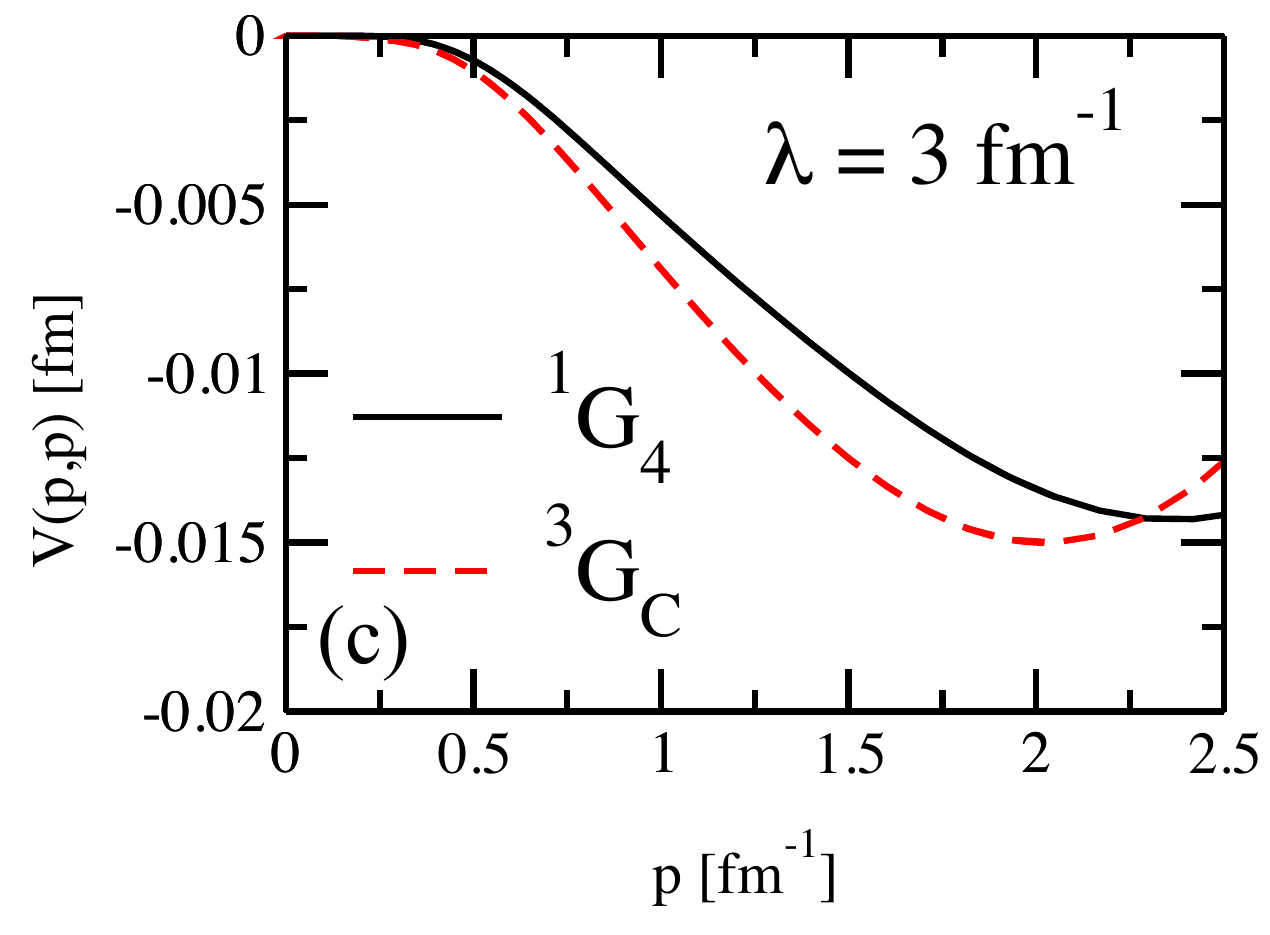}
\includegraphics[height=5.5cm,width=7.5cm]{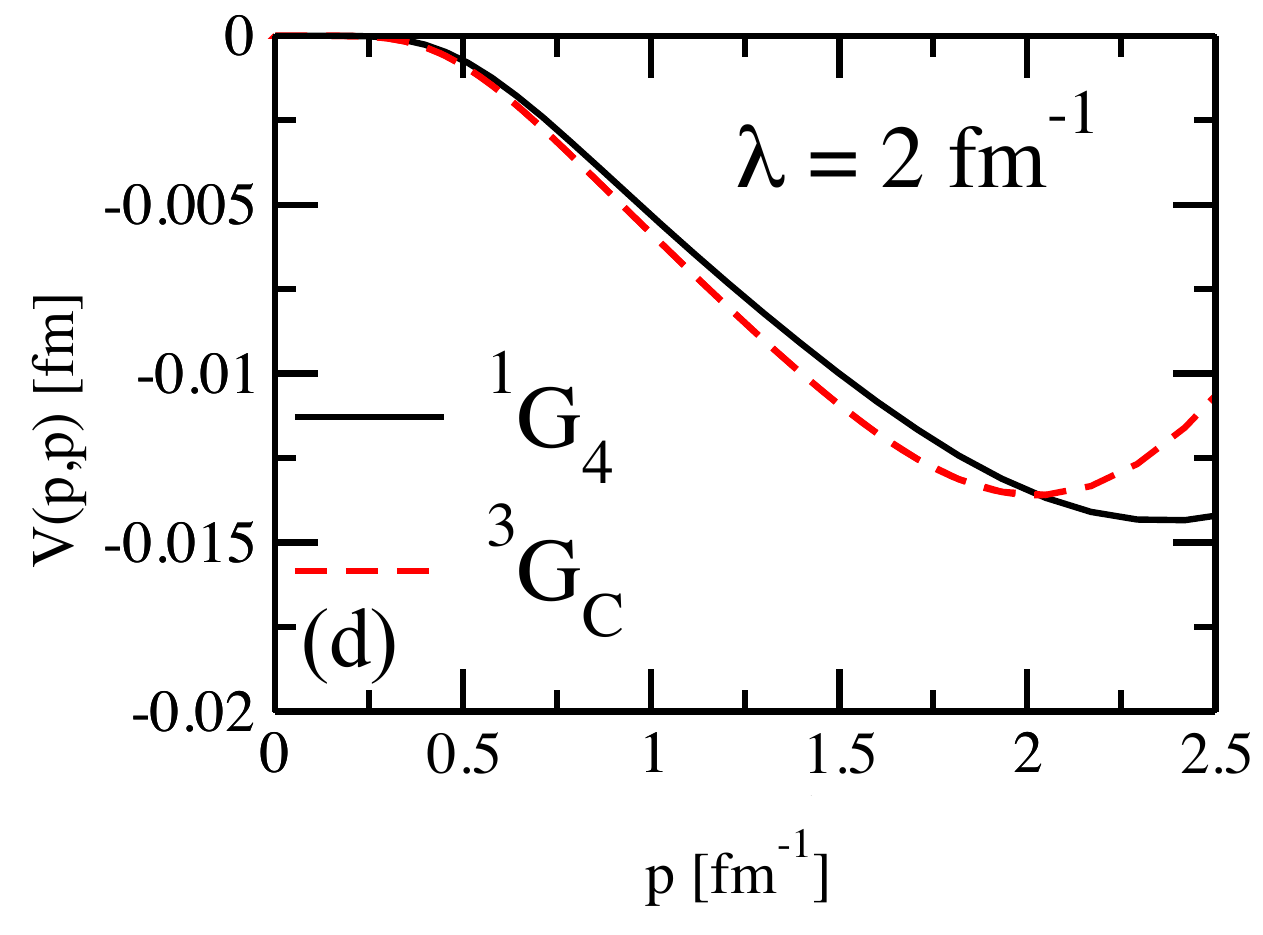}   \\
\includegraphics[height=6.5cm,width=8cm]{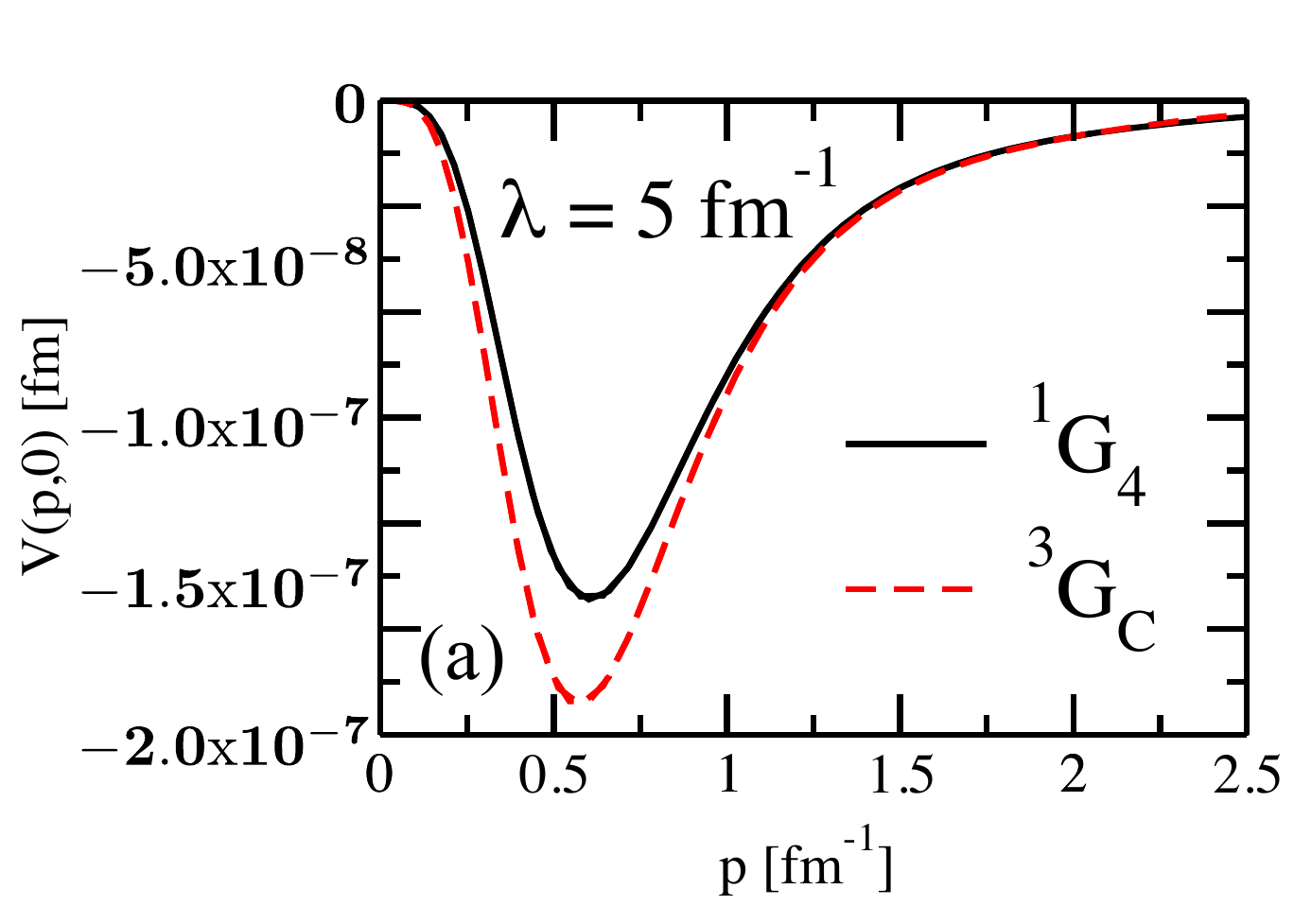}
\includegraphics[height=6.5cm,width=8cm]{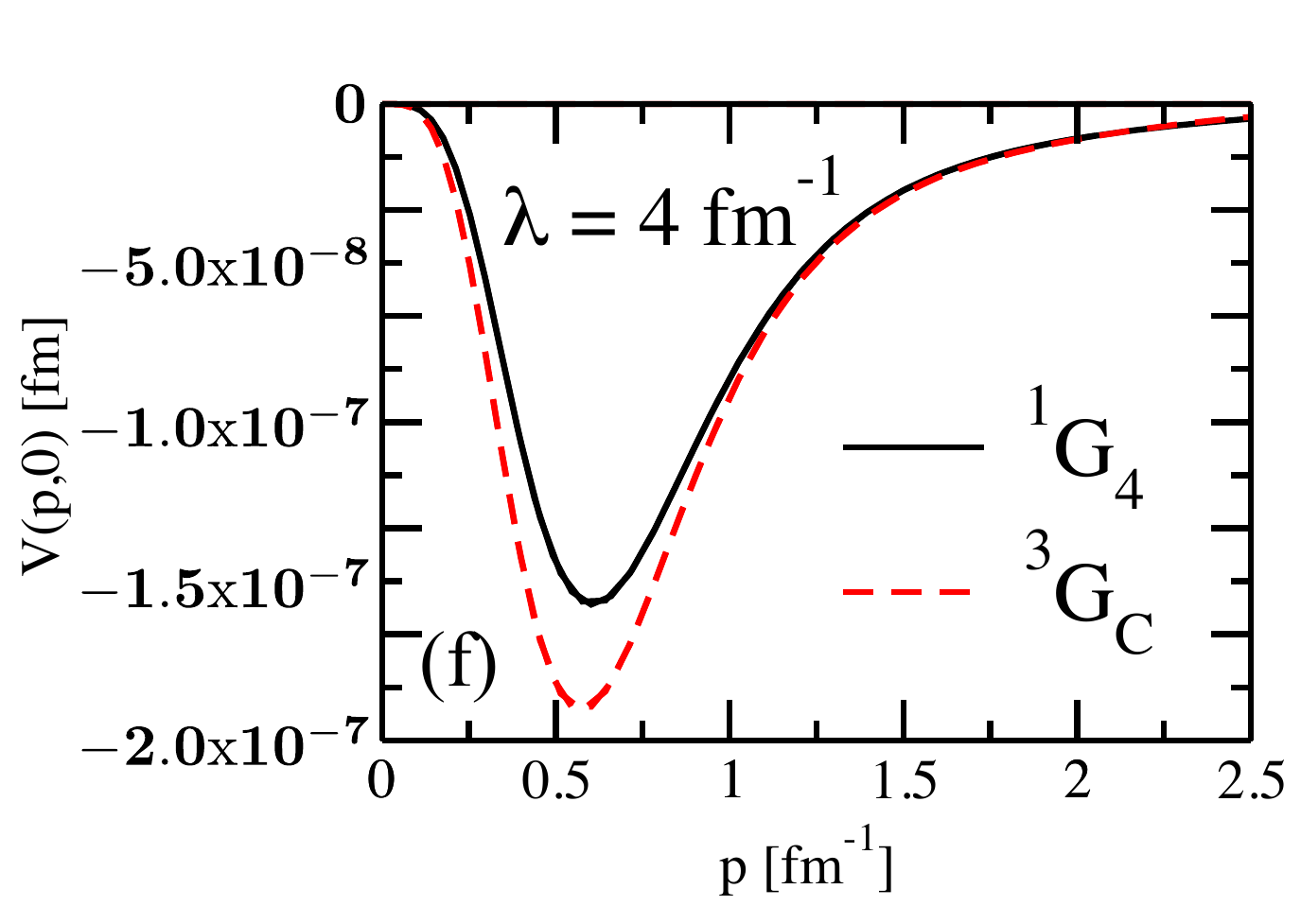}  \\
\includegraphics[height=6.5cm,width=8cm]{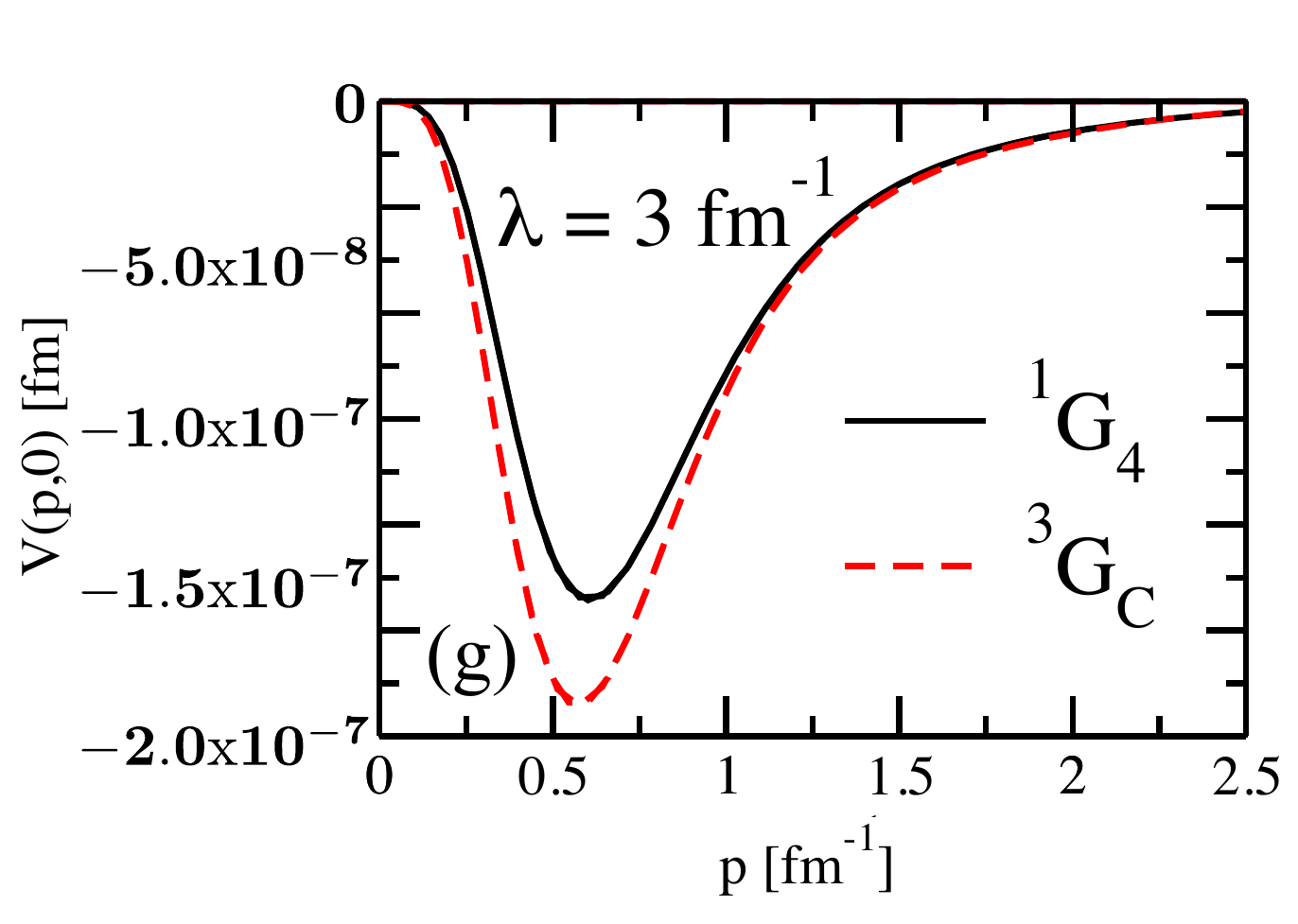}
\includegraphics[height=6.5cm,width=8cm]{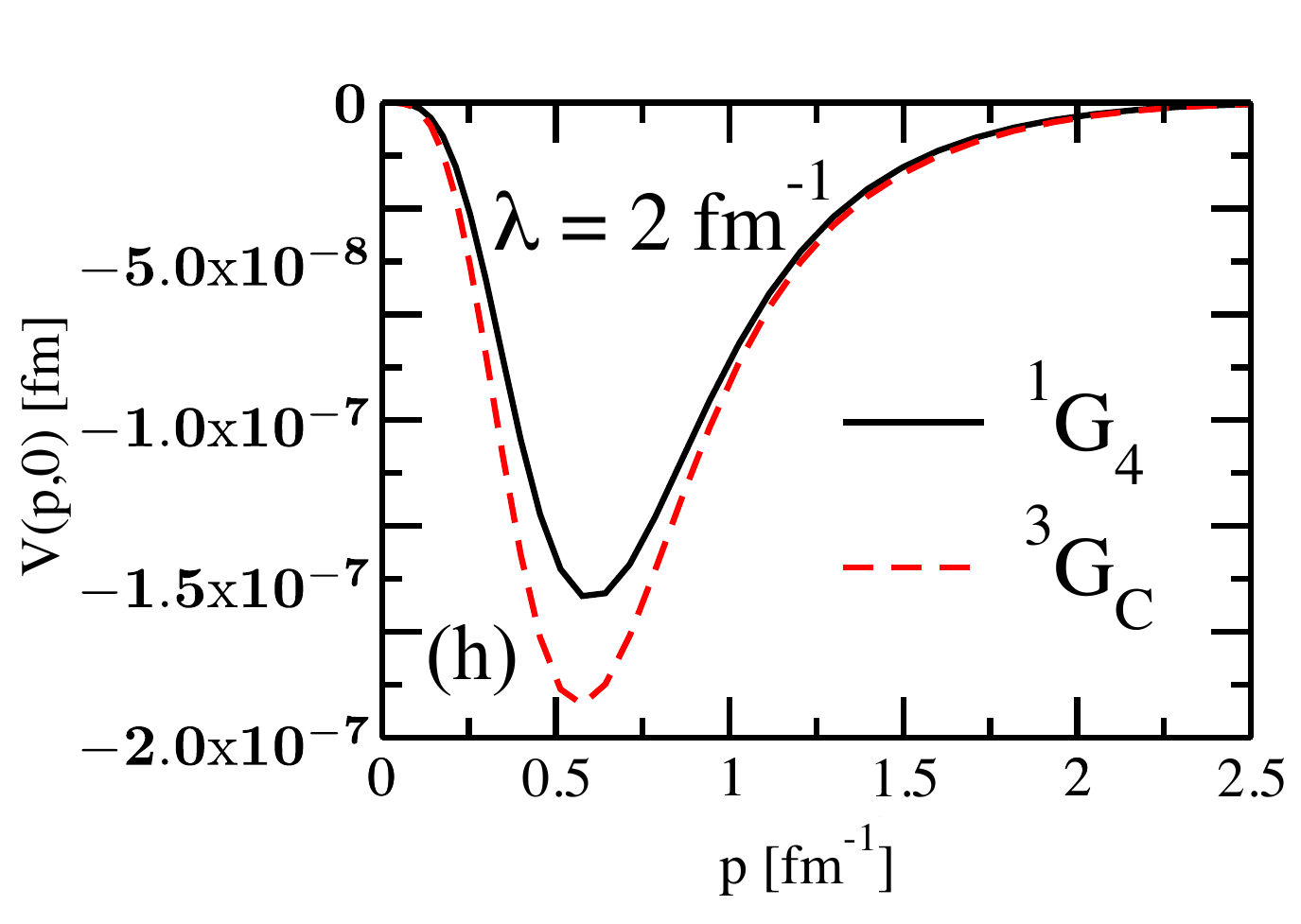}
\end{center}
\caption{(Color online) Comparison between diagonal, $V(p,p)$, and fully off-diagonal, $V(p,0)$,
matrix-elements of the SRG-evolved potentials for the
$G$-waves (in {\rm fm}) as a function of the CM momentum $p$ (in ${\rm fm}^{-1}$).}
\label{fig:AV18-wigner-G}
\end{figure*}

For the $D$-waves the results are displayed in
Fig. \ref{fig:AV18-wigner-D}, where we can see that generally for
$p',p < 0.7 ~{\rm fm}^{-1}$ Wigner symmetry is well fulfilled.  Note
that the better fulfillment of the Wigner symmetry for slightly large
values of $p$ occurs for a similarity cutoff of $2~{\rm fm^{-1}}$. A
somewhat similar situation occurs also for $G$-waves shown in
Fig. \ref{fig:AV18-wigner-G}, where now a better fulfillment of the
Wigner symmetry for a similarity cutoff of $4~{\rm fm^{-1}}$.  It is
interesting to note that the value of the similarity cutoff where we
have the better fulfillment of the Wigner symmetry is not unique,
being different in each of the even angular momentum waves we
considered. Of course, while we do not expect {\it a priori} an exact
fulfillment of $SU(4)$-spin-isospin symmetry it is remarkable how well
it works taking into account that high-quality potentials have not
been designed to implement the symmetry by any means.

With the definitions of the potential we get the simple expressions
\begin{eqnarray}
V_{^1L_C} &=& 2 \pi \int_{-1}^1 dz P_L (z) \left[ V_C - 3 V_S +
  \tau( W_C -3 W_S) \right] \, , \nonumber \\
\label{eq:1Lc} \\
V_{^3L_C} &=& 2 \pi \int_{-1}^1 dz
P_L (z) \left[ V_C + V_S +
  \tau( W_C + W_S) \right] \, , \nonumber \\
\label{eq:3Lc}
\end{eqnarray}
where $\tau= 4 I -3 $.

The non-triviality of the Serber symmetry is realized by noting that
we just change the angle without changing the nucleons. To elaborate
a bit further let us consider the standard Fierz identities. For the
spin-isospin operators one gets
\begin{eqnarray}
P_{1'2'} {\bf 1} &=& \frac{1}{2} \left[  {\bf 1} + \sigma \right] \, ,
\nonumber \\  P_{1'2'} {\bf \sigma} &=& \frac{1}{2} \left[  {\bf 1} - 3\sigma \right] \, ,   \nonumber \\
P_{1'2'} (\sigma_1 + \sigma_2 ) &=& -(\sigma_1 + \sigma_2 ) \, ,  \nonumber \\
P_{1'2'} (\sigma_1 + \sigma_2 ) &=& -(\sigma_1 + \sigma_2 ) \, ,  \nonumber \\
\end{eqnarray}
and similar equations for isospin operators. These identies imply that the following
combinations of scalar functions are respectively even or odd in $z$
\begin{eqnarray}
&& \left[ V_C - 3 V_S + \tau( W_C -3 W_S) \right]\Big|_{-z} \nonumber \\
&&= \left[ V_C - 3 V_S + \tau( W_C -3 W_S) \right] \Big|_{z} \nonumber \\
&& \left[ V_C + V_S + \tau( W_C + W_S) \right]\Big|_{-z} \nonumber \\
&&= -\left[ V_C + V_S + \tau( W_C + W_S) \right] \Big|_{z}
\end{eqnarray}
Therefore, the orbital parity of the integrand and the
Legendre polynomial in Eqs.~(\ref{eq:1Lc}) and (\ref{eq:3Lc}) are {\it
  the same} and do not suggest that any of them vanish. Instead, the
symmetries imply
\begin{eqnarray}
0 &=& V_C + V_S + W_C + W_S  \nonumber \\
V_C - 3 V_S + W_C -3 W_S &=& V_C + V_S -3 W_C -3 W_S  \nonumber \\
\label{eq:wig-ser}
\end{eqnarray}
We remind that the large $N_c$ analysis of Ref.~\cite{Kaplan:1996rk}
yields $V_C,W_S= {\cal O} (N_c)$ whereas $W_C,V_S= {\cal O}
(N_c^{-1})$, suggesting from QCD that Wigner symmetry holds (second
line of Eq.~(\ref{eq:wig-ser}) in the even $L$ partial waves, {\it
  exactly} as we do here for the SRG model. Note also that Serber
symmetry under those conditions implies $V_C+W_S= 0$.  These
conclusions have been analyzed in great detail within several
viewpoints~\cite{CalleCordon:2008cz,CalleCordon:2009ps,RuizArriola:2009bg,Arriola:2010qk}
and are confirmed here within the SRG. In any case, our results
suggest that it must be possible to impose these approximate
symmetries to a $NN$ potential from the very beginning.

\section{Conclusions and Outlook}
\label{sec:concl}

In the present paper we have analyzed the concept of long distance
symmetries as applied to the Similarity Renormalization Group. We have
shown that very similarly to the case of the $V_{\rm low k}$
formulation, the symmetry pattern of Wigner symmetry in partial waves
with even angular momentum as well as the Serber symmetry in odd
partial waves holds for the SRG cutoff around $\lambda \sim 3~{\rm
  fm}^{-1}$.  This is somewhat remarkable since SRG only provides an
exponential decoupling between low- and high-energy modes bringing the
effective interaction to the diagonal form, whereas $V_{\rm low k}$
corresponds to integrate out high-energy components. It is also
noteworthy that the symmetry arises at scales where in few-body
calculations the induced three-body forces become
small~\cite{Jurgenson:2009qs,Jurgenson:2010wy}.

In this work we have not evolved the three-body force through
  the SRG, so that no statements on the three-body SU(4) violating
  terms can be made, even knowing that in the two-body case the SU(4)
  violating terms get small. From this point of view the extension
of the present results to three- and four-body forces and the analysis
of their symmetry structure would be of great interest.

The underlying symmetry pattern unveiled by our SRG analysis appears
intriguing and unexpected from the modern viewpoint of coarse graining
high-quality interactions. From such a perspective this fits somewhat
the vague concept of an accidental symmetry.
Note that here we use the standard concept of accidental symmetry from
quantum mechanics instead of the field theory one: a symmetry which is realized
although not foreseen.

There is a long tradition on the phenomenological consequences
of Wigner symmetry in the properties of nuclei and nuclear matter
(for a review see e.g. Ref.~\cite{1999RPPh...62.1661V}).  A recent
work~\cite{Baroni:2009eh} analyses the SU(4) pattern of pairing
forces within a $V_{\rm low k}$ framework, which quite naturally
follow the symmetry pattern. Our results, in particular the
existence of an SRG scale at which the Wigner symmetry becomes quite
accurate, not only provide a natural explanation for this fact but
suggests to pursue the study further in future work within the
current SRG framework.
  
From a fundamental viewpoint, QCD large $N_c$ based arguments foresee
fulfilling the symmetry with a relative ${\cal O} (1/N_c^2)$ accuracy,
so one does not expect a perfect fulfillment of the Wigner
symmetry. On the other hand, nowhere in the design and optimization of
the modern high-quality interactions which have been successfully
applied to the structure of finite nuclei was the Wigner symmetry
pattern explicitly implemented. In this regard, the accuracy with
which by choosing a suitable SRG scale the symmetry seems to hold
suggests that this is a property of the data themselves which emerges
when the interaction is resolved with a specific length scale and not
so much on the original (bare) potentials used to fit the $NN$
scattering database. Finally, one should recognize that while the use
of symmetries for coarse grained effective interactions is not
mandatory we expect by explicit symmetry considerations additional
simplifications of the complicated nuclear many-body problem. Work
along these lines is in progress.

\begin{acknowledgments}
Computational power provided by FAPESP grants 2011/18211-2 and 2010/50646-6.
The work of V.S.T. was supported by FAEPEX/PRP/UNICAMP and FAPESP,
S.S. was supported by Instituto Presbiteriano Mackenzie through Fundo
Mackenzie de Pesquisa and E.R.A. by the Spanish DGI and FEDER funds
with grant  FIS2011-24149, Junta de Andaluc{\'\i}a grant FQM225,
and EU Integrated Infrastructure Initiative Hadron Physics Project
contract RII3-CT-2004-506078.
\end{acknowledgments}


\appendix

\section{Wigner symmetry for $NN$}
\label{eq:su4}

For completeness we remind here some features of the Wigner $SU(4)$
spin-isospin symmetry. It consists of the following 15-generators
\begin{eqnarray}
T^a &=& \frac12 \sum_A\tau_A^a  \, ,  \\
S^i &=& \frac12 \sum_A \sigma_A^i \, ,  \\
G^{ia} &=& \frac12 \sum_A \sigma_A^i \tau_A^a \, ,
\end{eqnarray}
where $\tau_A^a$ and $\sigma_A^i$ are isospin and spin Pauli matrices
for nucleon $A$ respectively, and $T^a$ is the total isospin, $S^i$
the total spin and $G^{ia}$ the Gamow-Teller transition operator.  The
quadratic Casimir operator reads
\begin{eqnarray}
C_{SU(4)} &=& T^a T_a + S^i S_i + G^{ia} G_{ia} \, ,
\end{eqnarray}
and a complete set of commuting operators can be taken to be
$C_{SU(4)}$, $T_3$ and $S_z, G_{z3}$.  The fundamental
representation has $C_{SU(4)}=4$ and corresponds to a single nucleon
state with a quartet of states $p\uparrow$, $p\downarrow$,
$n\uparrow$, $n\downarrow$, with total spin $S=1/2$ and isospin
$T=1/2$ represented ${\bf 4}= (S,T)=( 1/2,1/2)$. For two-nucleon
states with good spin $S$ and good isospin $T$ Pauli principle
requires $(-)^{S+T+L}=-1$ with $L$ the angular momentum, thus
\begin{eqnarray}
C_{SU(4)}^{ST} = \frac12\left(\sigma + \tau + \sigma \tau\right) +
\frac{15}2 \, ,
\end{eqnarray}
where $\tau= \tau_1 \cdot \tau_2 = 2 T(T+1)-3 $ and $\sigma= \sigma_1
\cdot \sigma_2 = 2 S (S+1)-3 $.

One has two SU(4) supermultiplets, which Casimir values are
\begin{eqnarray}
C_{SU(4)}^{00} &=& C_{SU(4)}^{11} = 9  \, , \\
C_{SU(4)}^{01} &=& C_{SU(4)}^{10} = 5  \, ,
\end{eqnarray}
corresponding to an antisymmetric sextet ${\bf 6}_A=(0,1) \oplus
(1,0)$ when $L=$ even and a symmetric decuplet ${\bf 10}_S=(0,0)
\oplus (1,1)$ when $L=$odd. According to $LSJ$ quantum numbers we have
the following supermultiplets
\begin{eqnarray}
(^1S_0, ^3S_1 ) \,, \,  (^1P_1, ^3P_{0,1,2})
\,, \,  (^1D_2, ^3D_{1,2,3}) \dots
\end{eqnarray}
When applied to the $NN$ potential, the requirement of Wigner symmetry
for {\it all} states, implies
\begin{eqnarray}
V_T &=&W_T=V_{LS}=W_{LS}=0 \, , \nonumber \\ W_S &=& V_S=W_C \, ,
\end{eqnarray}
so that the potential may be written as
\begin{eqnarray}
V = V_C + (2C_{SU(4)}^{ST}-15) W_S  \, .
\end{eqnarray}
The particular choice $W_S=0$ corresponds to a spin-isospin
independent potential, but in this case no distinction between the
${\bf 6}_A$ and ${\bf 10}_S$ supermultiplets arises.

\section{The infrared cut-off limit of the SRG}
\label{sec:appa}

In this appendix we show that when the kinetic energy is taken as the
generator of the SRG transformations the SRG evolved potential becomes the
standard reaction  matrix, i.e.
\begin{eqnarray}
V_s(p,p) \to R (p,p;p)
\end{eqnarray}
when there are no bound-states. To avoid unnecessary mathematical
complications, we will analyze the problem in the discretized form
like the gauss integration points used in the numerical
calculation. The SRG equation in the basis where $T$ is diagonal
becomes
\begin{eqnarray}
\frac{d V_{ik}}{ds}= - (E_i - E_k)^2 V_{ik}+ \sum_k ( E_i +E_k - 2 E_l ) V_{il} V_{lk}
\end{eqnarray}
The discrete representation has many advantages. One can see that
along evolution in the discrete representation one has an infinite number
of constants of motion $ \frac{d}{ds}{\rm Tr} (V_s^n)=0$ due to the
commutator structure of the SRG equation ~\footnote{Mathematically,
  such a property is ill defined in the continuum limit since even for
  $n=1$ one has $ {\rm Tr} (V_s)= \int_0^\infty p^2 V_s(p,p) =
  \int_0^\infty r^2 dr V(r,r)$ which for a local potential $V(r,r') =
  V(r) \delta (r-r')$ diverges as the momentum cutoff. Also, the
  trace of a commutator ${\rm Tr}[A,,B] $ only vanishes when both
  ${\rm Tr}(AB)$ and ${\rm Tr}(BA)$ are finite, as the choice $A=p$
  and $B=x$ clearly illustrates, since $[p,x]= - i\hbar$ and hence
  ${\rm Tr}[p,x]= - i\hbar {\rm Tr} ({\bf 1}= \infty$.}.

The fixed-point solution implies $\frac{d V_{ik}}{ds}$ which requires that
$[[T,V],V]=0$  so that $V$ and $[T,V]$ become diagonal in the same basis, not necessarily the one where $T$ is diagonal,
 i.e.
\begin{eqnarray}
V_{\alpha \beta} &=& v_{\alpha} \delta_{\alpha \beta} \\ 0
&=&\sum_\gamma (T_{\alpha \gamma} V_{\gamma \beta} - V_{\alpha \gamma}
T_{\gamma \beta})=  T_{\alpha \beta}( V_{\beta}- V_{\alpha} )
\end{eqnarray}
The second line requires that for $\alpha \neq \beta $ then $T_{\alpha
  \beta}=0$ which means that $T$ is diagonal also. Therefore $V_{ij}=
v_i \delta_{ij}$ , i.e. $V$ is diagonal in the basis where $T$ is also
diagonal. If we write now the Lippmann-Schwinger (LS) equation for the
reaction matrix
\begin{eqnarray}
R(p',p;k) &=& V(p',p) \nonumber \\
&+& \frac{2}{\pi} \dashint_0^\infty  q^2 dq \frac{V(p',q)
  R(q,p;k)}{k^2-q^2}
\label{eq:LS}
\end{eqnarray}
With this normalization the phase-shift in the one-channel case reads
\begin{eqnarray}
R(p,p;p) = - \frac{\tan \delta (p)}{p}
\end{eqnarray}
Note that the SRG actually implies that the phase-shift is constant
along the evolution, so that one may take any $V_s$ in
Eq.~(\ref{eq:LS}). The discrete version of this equation for the half off-shell
$R$-matrix, $R_{ij}= R(p_i, p_j ; p_i$ reads
\begin{eqnarray}
R_{ij} = V_{ij} + \sum_{k \neq j}
\frac{2}{\pi} \,  \Delta q \, q_k^2 \frac{R_{ik} V_{kl}}{p_i^2-q_k^2}
\end{eqnarray}
where the principal value corresponds to exclude the point in the sum.
Thus, for a diagonal potential we get
\begin{eqnarray}
R_{ij} = V_{i} \delta_{ij}
\end{eqnarray}
Turning to the phase-shift and going to the continuum limit we thus
obtain the assertion, so that Eq.~(\ref{eq:Vto0}) is obtained.  The
proof also holds for the continuum limit, provided a momentum cutoff
in the integrals is supplemented.

The infrared solution suggests looking for perturbations around it.
Actually we will see now under what conditions are these perturbations
stable. If we write
\begin{eqnarray}
V_{ik}(s) = V_i \delta_{ij} + \Delta V_{ij}
\end{eqnarray}
so that we get to first order in the perturbation
\begin{eqnarray}
\Delta V_{ik}'(s) =  (E_i - E_j) ( E_i + V_i - E_j - V_j) \Delta V_{ij}(s)
\end{eqnarray}
which yields the solution
\begin{eqnarray}
\Delta V_{ij}(s) &=& \delta_{ij} \Delta V_{ii} (\infty) \\
&+& \Delta V_{ij}
(\infty) e^{-s(E_i - E_j) ( E_i + V_i - E_j - V_j)} \nonumber
\Delta V_{ij}(\infty)
\end{eqnarray}
The diagonal part is constant as required by the property ${\rm Tr}
V(s)={\rm const.}$.  To identify this contribution we use again the LS
equation and find that  $\Delta V_{ii}=0$
and $\Delta V_{ij} (\infty)=
R_{ij}$ for $i \neq j$. Thus, we get
\begin{eqnarray}
\Delta V_{ij}(s) = (1-\delta_{ij}) R_{ij} e^{-s(E_i - E_j) ( E_i + V_i
  - E_j - V_j)}
\end{eqnarray}
which clearly shows that the fixed-point is stable provided
$(E_i-E_j)V_i-V_j) > 0 $. Departures from it measure some
off-shellness of the potential. Going to the continuum limit we get
for $p' \neq p$
\begin{eqnarray}
\Delta V_s (p',p) = R (p',p) e^{-s(E_p - E_{p'}) ( E_p + R(p,p) -
  E_{p'} - R(p',p'))}
\end{eqnarray}
where for large $p$ we obtain $R(p,p) \to -\delta (p)/p = V (p,p)$
which means that at high-momentum the kinetic energy dominates and the
fixed-point is stable. If there are no bound-states, we have that
$\delta (p) $ never becomes $\pi/2$ (Levinson's theorem states
that $\delta(0)-\delta(\infty)=n_B \pi$, with $n_B$ the number of
bound-states). However, for a pole at $p=p_0$ we get
\begin{eqnarray}
R(p,p) = \frac{1}{p_0 (p-p_0) \delta'(p_0)} + {\rm reg.}
\end{eqnarray}
where ${\rm reg.}$ means regular contributions, so that for $p > p_0 > p'$ and
$\delta'(p_0)> 0$ or $p'> p_0 > p$ and $\delta'(p_0)< 0$ becomes
stable and unstable otherwise. In the $^3S_1$ channel one has
$\delta'(p_0) < 0$ which means that the corrections increase
dramatically for $p > p_0 > p'$.

\bibliography{srg-symmetries}

\end{document}